\documentclass[twocolumn]{aastex7}

\usepackage{xspace}
\usepackage{shortbold}
\usepackage{pifont}
\usepackage{amsmath}
\usepackage{booktabs}
\usepackage{verbatim}
\DeclareMathOperator*{\argmax}{arg\,max}
\DeclareMathOperator*{\argmin}{arg\,min}

\newcommand{\hmi}{{{\it SDO}/HMI}\xspace}
\newcommand{\aia}{{{\it SDO}/AIA}\xspace}
\newcommand{\gong}{{{\it NSO}/GONG}\xspace}
\newcommand{\hmip}{HMIP\xspace}
\newcommand{\method}{{\tt SSLOS}\xspace}

\newcommand{\blos}{$B_{\textrm{los}}$\xspace}
\newcommand{\bperp}{$B_\perp$\xspace}
\newcommand{\br}{$B_\textrm{R}$\xspace}
\newcommand{\bp}{$B_\phi$\xspace}
\newcommand{\bt}{$B_\theta$\xspace}
\newcommand{\brpt}{$B_{R\phi\theta}$\xspace}
\renewcommand{\alphaB}{$||\BB||$\xspace}
\newcommand{\gauss}{Mx\,cm$^{-2}$\xspace}

\begin{document}

\title{Learning to Estimate Photospheric Vector Fields from Line-of-Sight Magnetograms}

\author{David Fouhey}
\affiliation{New York University, Courant Institute of Mathematical Sciences, New York, NY}
\affiliation{New York University, Tandon School of Engineering, New York, NY}
\email{david.fouhey@nyu.edu}

\author{KD Leka}
\affiliation{NorthWest Research Associates, Boulder, CO}
\email{leka@nwra.com}

\begin{abstract}
Solar photospheric line-of-sight magnetograms are easier to estimate than full vector magnetograms since the line-of-sight component (\blos) can be obtained from total intensity and circular polarization signals, unlike the perpendicular component (\bperp), which depends on harder-to-measure linear polarization. Unfortunately, the line of sight component by itself is not physically meaningful, as it is just one component of the underlying vector and one whose relationship to gravity changes pixel-to-pixel. To produce an estimate of the radial component (\br) a common ``correction'' is often applied that assumes the field is radial, which is nearly always false. This paper investigates recovering full vector field information from \blos by building on the recent SuperSynthIA approach that was originally used with Stokes vectors as input for simultaneous inversion and disambiguation. As input, the method accepts one or more line-of-sight magnetograms and associated metadata; as output, our method estimates full vector field in heliographic components, meaning that the physically-relevant vector components are returned without need for further disambiguation steps or component transforms. We demonstrate the ability to produce good estimates of the full vector field on unseen examples from both HMI and GONG, including examples that predate the Solar Dynamics Observatory mission. Our results show that learning is not a replacement for a dedicated vector-field observing facilities, but may serve to unlock information from past data and at the very least, provide more accurate \br maps from \blos than are created using the simple viewing angle assumption.

\end{abstract}

\section{Introduction} 
\label{sec:intro}

Photospheric magnetograms have long served as key data products for
solar physics research and space weather applications, serving topics as diverse as
modeling the solar corona~\citep{wiegelmann2021solar}, active-region evolution and 
dynamics~\citep{LekaBarnes2007,cheung2012method,gombosi2018extended,Lionello2014,schuck2022origin}, and downstream 
operational space-weather facilities \citep{ffc3_2,WSA_ENLIL_2003}.
Since Hale's discovery of magnetic fields in the Sun \citep{Hale08}, 
the community has built and deployed a variety of
instruments to investigate the magnetic properties of the solar plasma. 
For photospheric magnetograms, the typical approach taken captures Zeeman-polarized light which
is then inverted to recover estimates of the magnetic field.
While a full discussion of inversion techniques is beyond the scope here (interested readers are directed to discussions by~\cite{Stenflo94} and \cite{delToroIniesta_RuizCobo_2016}), the relevant
point here is that (using standard Stokes notation), one can use Stokes $I$ (total
intensity) and $V$ (circular polarization) to obtain good information about the
projection of the magnetic vector along the line of sight to the observer (\blos),
but the component perpendicular to the line of sight
(\bperp) requires Stokes $Q$ and $U$ that represent the linear
polarization and its direction (up to a $180^\circ$ inherent ambiguity in direction described by \cite{Harvey69}). Additionally, 
the Q and U components have far less sensitivity to the magnetic field and thus significantly lower
signal-to-noise ratio (SNR) as discussed by \cite{Stenflo94}. 

As such, while the full vector enables more physical interpretation and importantly 
a direct estimate of the radial component of the field (\br), it is more difficult 
to obtain and inherently suffers from greater noise. Accordingly, line-of-sight magnetograms are 
``workhorses'' that solar physics and especially space-weather applications rely upon. For instance, those 
produced by the National Solar Observatories Global Oscillations Network Group \citep[``\gong'';][]{gong,Harvey1996,Harvey1998} are presently the United States' `operational'' source of solar magnetic field data.

\definecolor{losDiagOrange}{HTML}{ED7D31}
\definecolor{losDiagPurple}{HTML}{44546A}
\definecolor{losDiagBlue}{HTML}{4472C4}

\begin{figure}[t!]
    \centering
    \includegraphics[width=\linewidth]{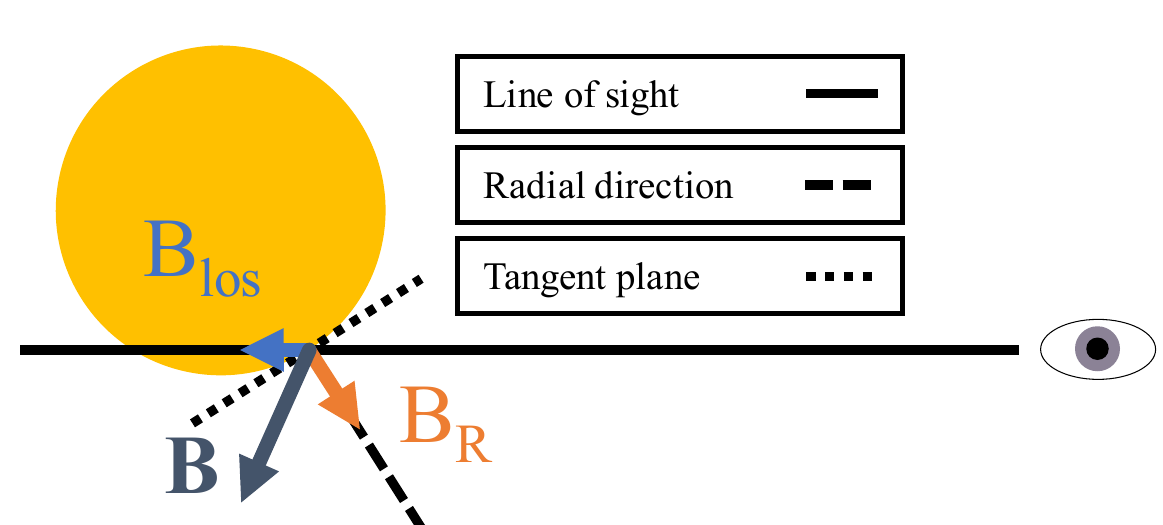}
    \caption{{\bf A schematic illustrating why \blos cannot be easily converted to physically meaningful data}, where the paper is the ecliptic plane. Many applications need \textcolor{losDiagOrange}{\br}, the radial component of the magnetic field \textcolor{losDiagPurple}{$\BB$}, but many important instruments only provide the projection of the field onto the line of sight \textcolor{losDiagBlue}{\blos}. When the field is inclined away from radial, \blos cannot be straightforwardly related to \br: in this example, \blos has the wrong sign. Similarly, if the field were flipped around the $B_R$ angle and pointed at the viewer, it would be a substantial over-estimate.
    Diagram following~\cite{leka2017evaluating}.
    }
    \label{fig:diagram}
\end{figure}

Typically, the \blos data is ``corrected''to \br by assuming that the field is entirely radial (oriented parallel to the gravity vector). If true, then \br can be recovered from \blos by taking the cosine of the viewing angle $\mu$ and treating \blos $/\mu$ as \br.
Unfortunately, the $\mu$-correction is incorrect for fields with {\it any}
inclination away from radial, often catastrophically so \citep{leka2017evaluating}. 
This effect is illustrated in Figure~\ref{fig:diagram}, where the the magnetic field's tilt
leads \blos to have a different sign than \br. No scaling can fix this and, in
fact, scaling only exacerbates the error.

In this paper, we investigate a machine learning technique for recovering both \br and the full vector $\BB$, 
from line of sight magnetograms. We use an ML-based approach SuperSynthIA~\citep{SuperSynthIA}, but instead of using the Stokes vector as input, we use observations of \blos.

This investigation serves both scientific and practical purposes. Scientifically, the question is how well the vector field could be recovered from limited information, given access to a large repository of vector data and precisely what information might be needed. Our experiments show that the ML methods are both clearly far superior to the $\mu$-correction and, importantly, can also identify {\it which} areas are most uncertain by virtue of the training process; yet the ML methods still have limitations that underscore the need for continued vector-capable instruments. On the other hand, today's information regarding the global vector field is now precariously limited. The resulting system could provide a tool that  provides physically-informative enhanced information from the ground-based \gong network.

\begin{figure*}[t!]
\centering
\includegraphics[width=\linewidth]{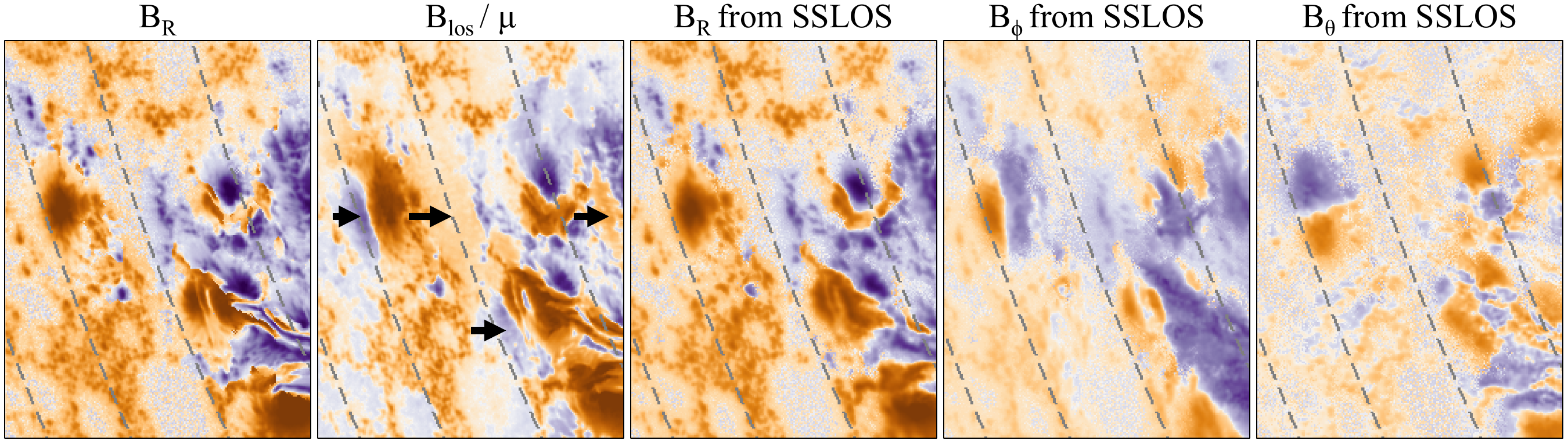} 
\caption{{\bf Overview.} The radial component of the magnetic field \br is the preferred quantity of interest, but many instruments only provide the magnetic field projected on the line of sight, \blos. \blos is often heuristically corrected to \br by accounting for viewing angle $\mu$ and assuming that the field is radial. We show 130\arcsec x 100\arcsec~cutouts from NOAA AR~13664 close to the limb, 2024 May 10, 22:48 TAI, two hours before a X5.8 flare. The $\mu$-corrected \blos has false polarity inversion lines and several other incorrect polarities (all indicated with arrows $\rightarrow$). Our learned correction improves over the $\mu$-correction for \br and also estimates a rough (albeit not insufficiently sheared) \bp and \bt. 
Colormap: -3000 \includegraphics[width=30pt,height=7pt]{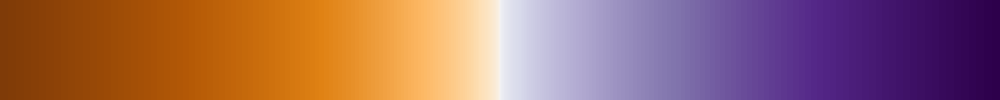} 3000 Mx cm$^{-2}$ with $\mu$ isocountours in dotted lines (every 0.1 here, and throughout the paper unless otherwise specified). Please note that this and all subsequent figures, are best viewed in color on a screen.
}
\label{fig:teaser}
\end{figure*}

 In brief, as input our method ``SuperSynthIA for line-of-sight'' or \method, takes one or more \blos magnetograms as well as auxiliary coordinate information. As output we aim to produce a vector field (i.e., \br, \bp, and \bt). To demonstrate the approach, we train \method to mimic the pipeline output from the Solar Dynamics Observatory / Helioseismic and Magnetic Imager \citep{Pesnell2012,scherrer2012helioseismic}; ``\hmip'' thus refers to the {\tt hmi.B\_720s} series deploying the ``pipeline'' version of VFISV~\citep{borrero2011vfisv}, plus the minimum-energy~\citep{metcalf1994resolving} disambiguation in well-measured regions (we generally choose the ``random'' solution for non-annealed pixels), at the default $720$\,s cadence \citep{Hoeksema2014}. As input, we show that \method can be trained to use \blos from essentially any \blos input; we demonstrate \method on effectively two extremes:  \hmi and \gong.  We demonstrate that \method produces good vector data from either, albeit unsurprisingly not with the quality of a true vector magnetogram pipeline that employs the full Stokes vector.  We show one such an example in Figure~\ref{fig:teaser} in comparison with the standard $\mu$-correction: our approach produces a demonstrably better estimate of the radial component, while also providing the \bp and \bt component. Note that Figure~\ref{fig:teaser} and all subsequent figures are best viewed on a screen.

Our approach and experiments build off of lessons learned building SuperSynthIA~\citep{SuperSynthIA}, an approach for simultaneous Stokes inversion and disambiguation. While the precise formulation differs slightly as described below, key lessons from SuperSynthIA inform the creation and evaluation of \method. Methodologically, our approach consists of regression-by-classification, in which an approximate discrete guess is made and then followed by adjustments. This approach enables a number of properties that are critical for downstream applications. 
We present the details of the methodology in Section\,\ref{sec:method}, and describe the experiments 
regarding input and evaluation metrics in Section\,\ref{sec:data}.  Overall evaluation under various scenarios is presented in Section\,\ref{sec:results}.

\section{Method}
\label{sec:method}

The goal of this work is to generate an estimate of the photospheric vector field, with uncertainties, from a line of sight magnetogram. Generating a plausible vector magnetic field with a machine learning method is difficult because any vector field result that best satisfies standard ML criteria (e.g., MSE or $\ell_1$ residual) will be at odds with the characteristics observational (or numerical model) data used as the ground truth.
For instance, models trained with the mean squared error over the three vector components often produce vector magnetic fields of $\zeroB$ in low-signal regions, which does not preserve the unsigned magnetic flux. The undesired behavior arises because the MSE is minimized by the mean. In areas with poor signal-to-noise ratio, the plane-of-sky inclination angle is near-$90^\circ$ (with no directional preference, meaning uniform azimuthal distribution) due to the lower SNR levels in the linear polarization and the positive-definite character of the noise~\citep{borrero2011inferring,leka2022identifying}. These angles lead to a symmetric distribution over $\BB$, specifically a uniform distribution over a great circle with an orientation in heliographic components determined by disk location. Under this distribution, $\zeroB$ is the optimal answer for the MSE. Several other common losses are minimized by $\zeroB$. For instance, the $\ell_1$ loss is minimized by the median of a distribution, which is the same as the mean for symmetric distributions. A Huber loss would also be similarly minimized by $\zeroB$. A uniform-in-3D distribution would also have the same problem -- the expected vector is also $\zeroB$. In each case, training on more data does not help, because the optimal answer itself is the problem. There are, of course, solutions such as conditional GANS~\cite{Isola2017} or diffusion~\cite{ho_denoising_2020}. These, however come at the cost of mode-collapse and difficulty producing distributions and uncertainty quantification in the case of conditional GANs, as well as typically a need to resample in the case of diffusion models.

Before moving to the details of the approach, we specify some notation and magnetogram-specific details. First, we refer to the vector magnetic field throughout as $\BB$, which is generally referred to in heliographic components (\br, \bp, \bt) following the convention described in~\citep{Hoeksema2014}. The length, or norm, of $\BB$ is denoted $||\BB||$. Scalars are denoted in roman face (e.g., $a$), all vectors other than $\BB$ in bold lowercase (e.g., $\aB$), and matrices in bold uppercase (e.g., $\AB$). To avoid double subscripts, we refer to the $i$th component of a vector $\vB$ as $\vB[i]$. Throughout we only work with magnetograms where magnetic field strength $||\BB||$ and fill factor (and/or scattered light) $\alpha$ are {\it not} disentangled in the Stokes inversion process. Thus, {\it e.g.}, the radial field data shown here, \br, really ought to be referred to as $\alpha B_R$ -- the product of radial field strength $B_R$ and fill factor $\alpha$. For notational clarity, we drop $\alpha$ throughout, but continue to use the more correct units of \gauss. We do not address the estimation of $\alpha$ from \blos data, but past work suggests that fill factor can be estimated fairly well~\citep{Higgins2022,SuperSynthIA,leka2022identifying}.

\subsection{Overview}
\label{sec:method_intro}

Our goal is to produce vector magnetograms using \blos data as input.  In other words, \method accepts a {\it reference} observation or \blos magnetogram with the aim to produce the vector magnetogram ($\BB$) that would have been produced at the reference time. Optionally, \method also accepts {\it auxiliary} \blos magnetograms that come from before or after the reference observation time. We hypothesized that these increase the signal-to-noise ratio of the inputs, especially at the limb.

In turn, we will introduce the outputs produced, the objective for learning, followed by a few key steps used for post-processing. We then discuss the inputs used and the machine learning architecture. Finally, we will discuss implementation details.

\subsection{Outputs and Learning Objective}
\label{sec:outputs}

We produce vector magnetic fields by decomposing the field vector into a norm and a unit vector that represents the angles, and applying recent innovations from~\cite{SuperSynthIA}. Given a magnetic field vector $\BB \in \mathbb{R}^3$, we independently factor it into a non-negative norm $n = ||\BB||_2$ and unit vector direction / angle vector $\aB \in \mathbb{R}^3$ with $||\aB||_2 = 1$. We predict the norm and direction separately and re-combine them as  $n \aB$. The factorization helps a model express a belief that the norm is well-defined but the angle is not. We then treat each subproblem (norm and angle) as a regression-by-classification problem in which we estimate a discrete bin that gives an approximate answer that is then refined. The regression-by-classification provides an easy way to generate uncertainty in a single function evaluation pass, which is useful for downstream applications, and also prevents the network from splitting the difference between bimodal distributions.

We will first illustrate the idea for the norm for a single pixel, which
then simply extends across all pixels. We start with an overcomplete set of $K$ bins 
$\sB_n \in \mathbb{R}^K$ covering the space of possible norms with an upper limit of 5000 \gauss (matching a common upper-bound for Milne-Eddington inversions \citep{Centeno2014} but which is arguably artificial \citep{OkamotoSakurai2018,CastellanosDuran_etal_2020}).
The primary output of the network is a
logit vector $\zB_n \in \mathbb{R}^K$ giving the un-normalized log-likelihoods
of each bin, which can be converted to a probability using a standard softmax function
$\sigma(\zB)[i] = \exp(\zB[i]) / \sum_{j} \exp(\zB[j])$. The bin value with the
highest probability $\sB_n[\argmax(\zB_n)]$  provides a good first
approximation.

To improve results, the network also produces a per-pixel continuous correction $\deltaB_n \in \mathbb{R}^K$ of the bins, resulting in a final prediction of $(\sB_n + \deltaB_n)[\argmax(\zB_n)]$. 
While the network predicts corrections for all bins for computational convenience,  one can see this as first predicting a most likely bin $i_n = \argmax(\zB_n)$, using the discrete nominal bin value, and then applying a correction. The soft-label formulation in~\cite{higgins2021fast} would also likely work equally well, but here we apply a continuous correction to use the same approach for both the norm and the angles, since the soft label case of unit vectors is a bit more involved \citep{Ladicky14b}. 
Corrections are often close to zero, leading to some amount of bias towards the bin values; if the bias towards bin values needs to be removed, $\deltaB_n$ can instead replaced with appropriately-sampled random noise (described in Appendix~\ref{sec:app_dither}). Since the bulk of the work involved is in finding the right bin, the corrections are small and so randomizing them has limited impact on performance. We hypothesize that there is insufficient signal in the \blos data to reliably produce an estimate with the few-dozen \gauss / few-degrees precision between bins.

To train the network to predict $\zB_n$ and $\deltaB_n$, one uses paired samples with known ground-truth norm $y_n$. To train $\zB$, one identifies the index $l_n \in \mathbb{Z}$ of the bin that is closest to $y_n$ (or $l_n = \argmin_i |y_n - \sB_n[i]|$) and trains the network's prediction of $\zB_n$ by minimizing the negative log-likelihood $\textrm{NLL}(\zB,l_n) = -\log(\sigma(\zB)[l_n])$. The continuous correction $\deltaB_n$ is trained with the mean-squared error $((\sB_n + \deltaB_n)[l_n] - y_n)^2$, which means that the network learns a correction only for points falling into that bin. 

Angular predictions follow largely identically apart from ensuring that unit vectors are normalized to norm unity throughout. The network predicts logits $\SB_a \in \mathbb{R}^{K \times 3}$, as well as a 
continuous correction $\DeltaB_a \in \mathbb{R}^{K \times 3}$; the set of corrected predictions $(\SB_a + \DeltaB_a)$ is re-normalized to the unit sphere. Throughout, we use a heliographic coordinate frame: the vector $[1,0,0]$ corresponds to a purely radial vector, and $[0,1,0]$ to a poloidal vector. While the choice means that the coordinate frame changes from pixel to pixel, results from SuperSynthIA suggest that networks can handle the changing coordinate frame easily.

Putting things together, at each pixel \method produces four outputs: $\zB_n \in \mathbb{R}^K$ giving logits over the norm bins, $\zB_a \in \mathbb{R}^K$ giving logits over the angular bins, $\delta_n \in \mathbb{R}^{K}$ a corrected vector for each norm bin, and  $\DeltaB_a \in \mathbb{R}^{K \times 3}$ giving a correct angle for each angular bin. 
During training, the final loss per pixel depends on the ground-truth norm $y_n$ and norm bin $l_n$, as well as ground-truth angle $\yB_a$ and angle bin $l_a$, and is a combination of negative log-likelihood losses to encourage the logits $\zB_n, \zB_a$ to make the right bin indices likely and MSE losses that encourage the corrections $\deltaB_n, \deltaB_a$ to make the bin values plus corrections closely approximate the ground-truth. In total, the loss at a pixel is
\begin{equation}
\label{eqn:loss}
\begin{split}
\textrm{NLL}(\zB_n,l_n) + 
((\sB_n + \deltaB_n)[l_n] - y_n)^2 + \\
 \textrm{NLL}(\zB_a,l_a) +
||s((\SB_a + \DeltaB_a )[l_a]) - s(\yB_a)||^2_2,
\end{split}
\end{equation}
where $s(\xB) = \xB / ||\xB||_2$ projects $\xB$ to the unit sphere. Note that minimizing the squared Euclidean distance between $s((\SB_a + \DeltaB_a)[l_a])$ and $s(\yB_a)$ is equivalent to maximizing the cosine of the angle between the two vectors. 
During training, Equation~\ref{eqn:loss} is minimized over all pixels in the training set.

\begin{figure*}
\includegraphics[width=\linewidth]{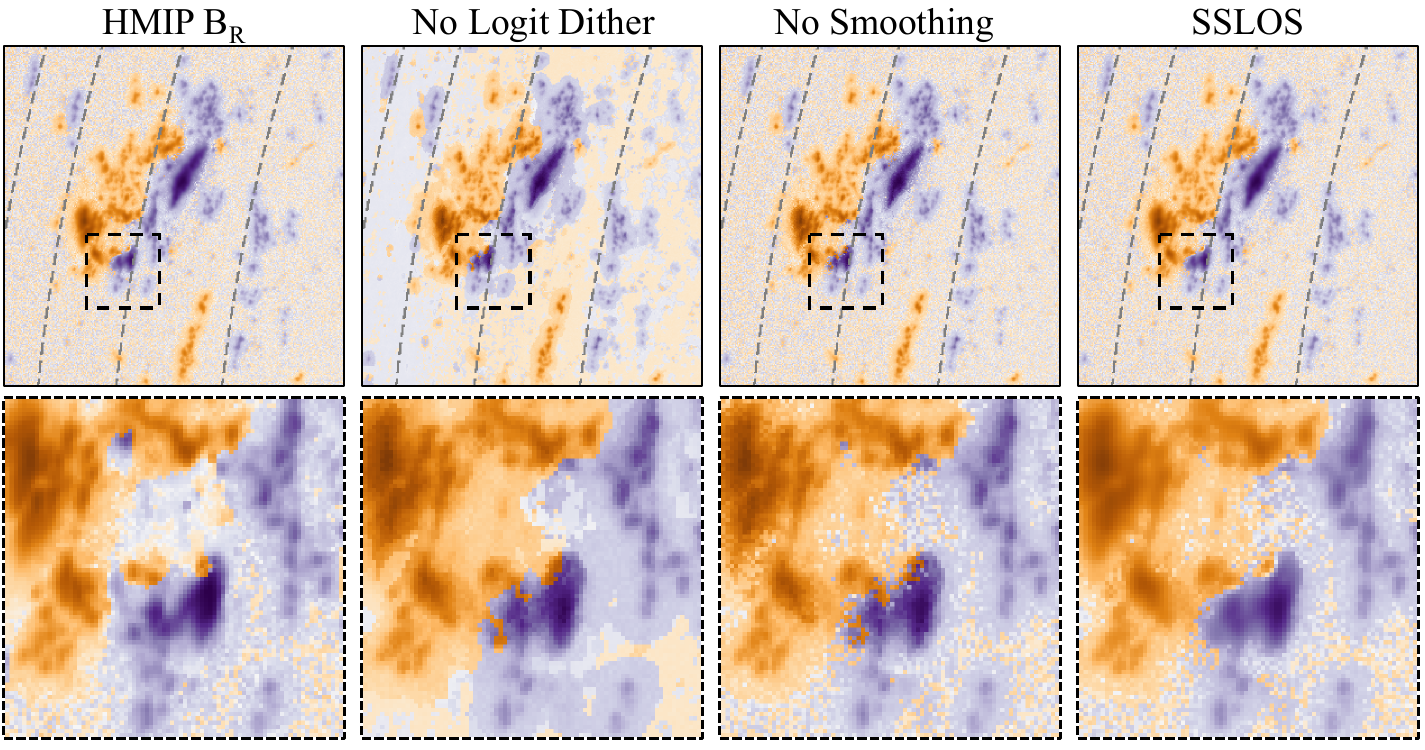}
\caption{{\bf Comparison of \method with ablations (variants of the method with modules/steps removed).} Without logit dither, there are large regions with the same orientation in low polarization regions. Without smoothing, there are regions of uncertainty and ``flip-flopping'' near polarity inversions. Smoothing helps mitigate the artifact. While the \hmip \br appears to have smaller values in parts of the region as compared to \method, this is due to a disagreement in orientation: \method and \hmip agree on the total length as seen in Figure~\ref{fig:totalflux}.}
Data:  NOAA AR 1570 captured 2016 August 7, 13:12 TAI. Colormap: -2000 \includegraphics[width=24pt,height=7pt]{PuOrSqrt.png} 2000 Mx cm$^{-2}$. 
\label{fig:ablation}
\end{figure*}

\begin{figure*}
\includegraphics[width=\linewidth]{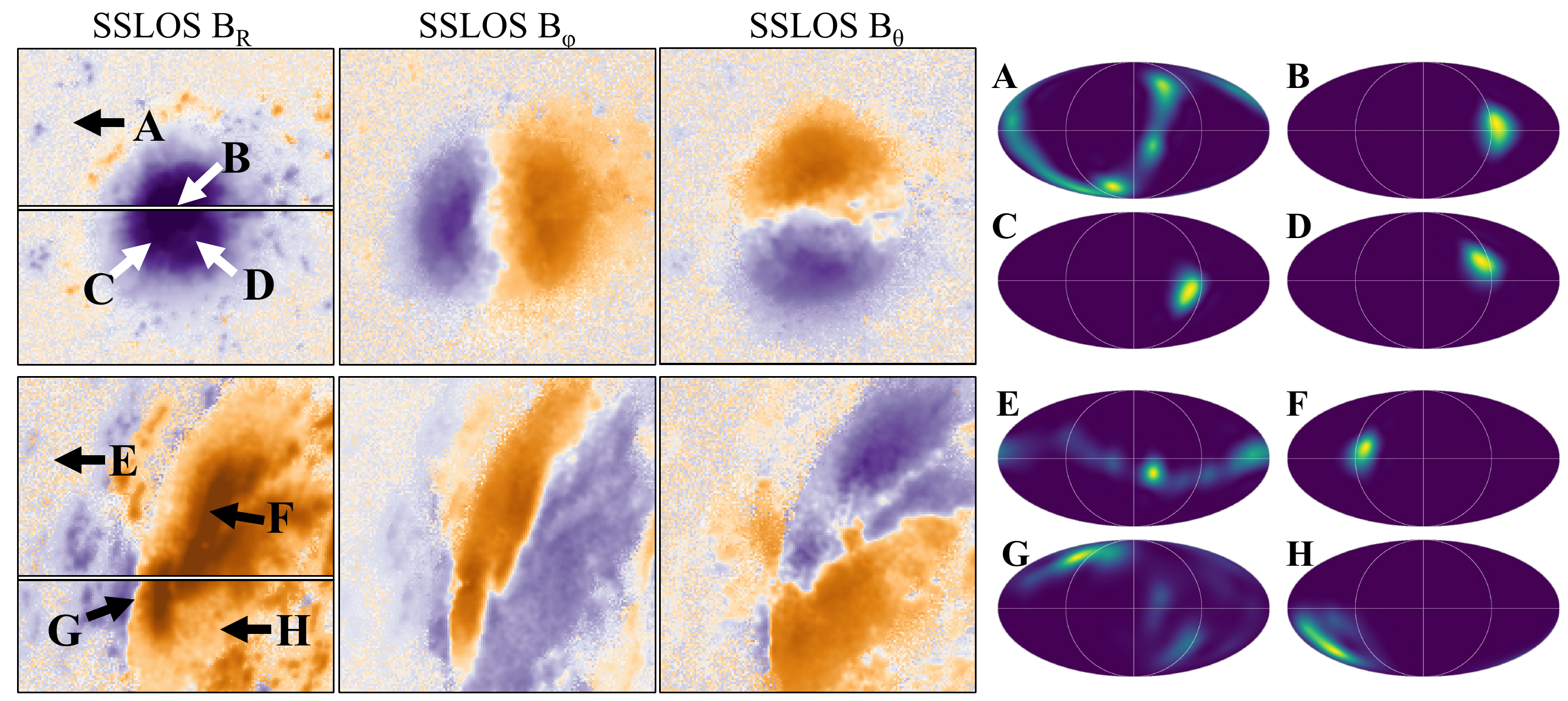} \\
\includegraphics[width=\linewidth]{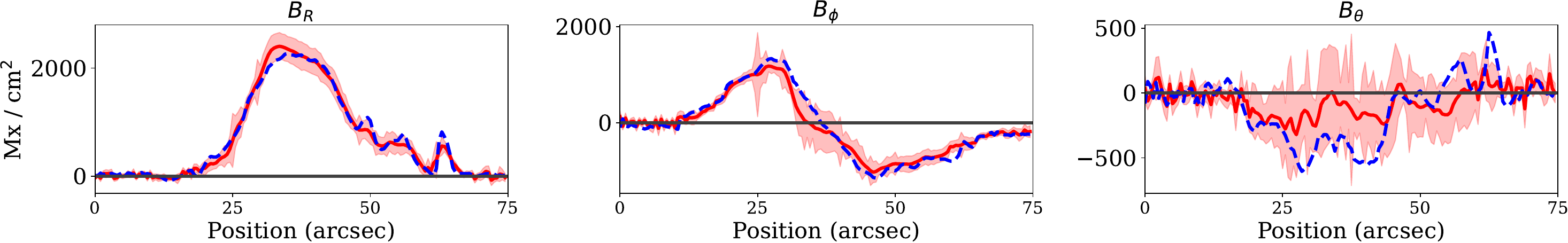} \\
\includegraphics[width=\linewidth]{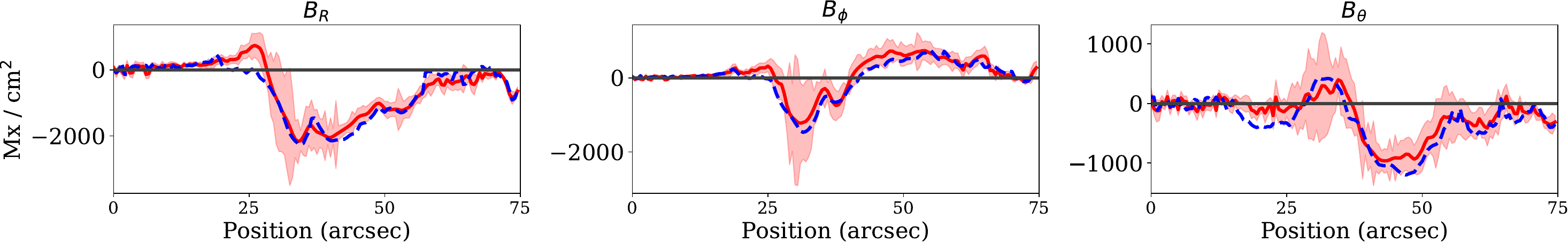} \\
\caption{{\bf Distributions produced by \method.} We show two examples (both $75\arcsec \times 75\arcsec$): a simple active region near disk center (NOAA 12418 2015 September 17, 12:36TAI) and a more complex region on the limb (NOAA 12488; 2016 January 28, 18:24 TAI). {\bf (top)} we show the inferred three components as well as the inferred distribution over angles at four locations per sample. We visualize the angular distribution on a Mollweide projection with a power-law applied to the colormap (min \includegraphics[width=20pt,height=6pt]{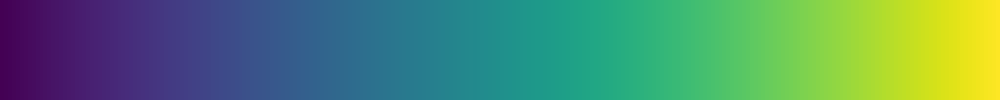} max probability) to enhance contrast. At the low polarization region {\bf (A)}, \method is uncertain about direction and produces the great circle corresponding to the Milne-Eddington plus disambiguation method. Within the region ({\bf B,C,D}), it is more confident. At another low-polarization area ({\bf E}), \method is again uncertain, but produces a different distribution (with \br near zero). At {\bf (F)}, it is confident. Near the polarity inversion line {\bf (G)}, the results are very uncertain, and at ({\bf H}), there is considerable spread.
\\
{\bf (Bottom)} Line plots through the samples of each component. We plot \textcolor{red}{\method in red solid lines} (along with shaded confidence intervals) and \textcolor{blue}{\hmip in blue dashed lines}. 
Colormap: -2000 \includegraphics[width=20pt,height=6pt]{PuOrSqrt.png} 2000 \gauss. Mollweide coordinate system: $B_R=0$ along the vertical center; $B_\phi = 0$ along the horizontal center line; $B_\theta=0$ along center ring. 
}
\label{fig:uncertainty}
\end{figure*}

\subsection{Postprocessing}
\label{sec:inference_postprocessing}

At test time, our networks produce logits over the norm ($\zB_n$) and orientation ($\zB_a$) of the vector field, as well as continuous corrections ($\delta_n$, $\Delta_a$). We apply three steps to improve
results; these ``finalizing'' steps produce marginal improvements in terms of per-pixel statistics compared to directly taking the network's output, but serve to mitigate issues that may impact downstream use cases. 

The first postprocessing step is logit dithering, following~\cite{SuperSynthIA}. 
In low-signal regions, models trend to have a slight preferences that lead to large splotches of identical values. As discussed by~\cite{SuperSynthIA}, one bin tends to have a slightly higher likelihood, which we hypothesize is exacerbated by the Milne Eddington plus disambiguation formulation, which distributes low-signal data on a location-dependent subspace. 
Logit dithering {\it slightly} randomizes decisions about which bin index $i_a$ is selected. Rather than select the most likely bin (i.e., $\argmax_i \zB_a[i]$), one selects from noised-logits, or $\argmax_i (\zB_a + \epsilonB_a)[i]$. We draw each element of $\epsilonB_a$ from a triangular distribution with zero mean and variance of unity. Since samples from this distribution only can take values from within $-\sqrt{6}$ to $\sqrt{6}$, the dithering does not  alter clear-cut decisions where one direction is preferred by a large margin, but randomizes cases with high uncertainty.

The second post-processing step is angular smoothing. Occasionally, \method produces isolated pixels in active regions that disagree sharply with their neighbors. Since each active region has a large number of pixels, these flipped pixels appear frequently enough (some in most  complex active regions) that it creates a need for smoothing.
The smoothing was not needed in SuperSynthIA since the mapping from Stokes vectors to $\BB$ is relatively well defined; here, the difficulty of estimating the vector data from \blos increases the chance of these intermittent mis-estimations, and they are mitigated with the same motivation as the ``current-minimization'' term in the minimum-energy disabmbiguation optimization function, meaning a somewhat ad-hoc approach to remove occasional arguably unphysically-sharp gradients.
We construct a smoothed version at each pixel by finding the angular bin that maximizes agreement with predictions at the surrounding $W \times W$ window. We operationalize this as maximizing the number of  predictions in the window that are within $45^\circ$. We break ties by the 
the average angular distance to the pixels within the window. The described operation can be implemented quickly by a direct search over the bins. We then update the pixel's selected bin if both: (a) 
the predicted norm $||\BB||$ is over 500 \gauss (to avoid smoothing in low-signal regions),
and (b) the smoothed angle has a negative dot product with the predicted bin (and is thus substantially different and worth updating). The update changes the average error by a trivial amount, but reduces the number of small isolated disagreeing predictions. We set $W{=}7$px for \hmi (${\approx}3.5\arcsec$) and $W{=}3$px for \gong (${\approx}7.5\arcsec$).

The final post-processing step is a simple Gaussian filter that is applied per-component ({\it e.g.}, separately \bt, \bp, \br). The predictions often have small magnitude speckling, or spatially inconsistent solutions. These may create higher than expected spatial derivatives, which may cause issues in downstream analysis.
We create a smoothed estimate of each component's prediction with a Gaussian ($\sigma = 1$px or 0.5\arcsec~for \hmi and $0.5$px or 1.25\arcsec~for \gong). We only filter high norm ($||\BB||$) pixels to avoid smoothing out the low polarization pixels. The final per-pixel estimate of the component is unsmoothed if $||\BB||$ is less than 500 \gauss. To avoid sharp transitions, the data is fully smoothed only above $750$ \gauss; from $500$ to $750$, the final output is a linear combination, with the weight linearly transitioning from fully unsmoothed (below $500$) to fully smoothed (above $750$). 

We show the impact of these steps in Figure~\ref{fig:ablation}. Dithering has close to no impact in regions with good polarization signal, but a large one in low polarization regions. 
Large regions of uniform orientation are turned into more uniformly distributed angles, an effect seen both in the large field of view as well as the bottom right of the zoom-in.
Although the field is weaker in these areas, there are many pixels in them, and so the total effect to any global use-case may be quite large. The logit dithering provides a simple method to mitigate the artifact and produce regions with more uniformly distributed angles. Smoothing helps helps avoid small pixel-to-pixel changes in polarity.

To summarize: at inference, time, one infers a norm bin $i_n = \argmax \zB_n$, producing a final estimate of the norm $(\sB_n + \delta_n)[i_n]$. One samples a noise vector $\epsilonB_a$, to pick the angle 
angle bin $i_a = \argmax(\zB_a + \epsilonB_a)$, and then optionally filters the resulting vectors. The final output is 
\begin{equation}
(\sB_n + \delta_n)[i_n] \cdot s( (\SB_a + \DeltaB_a)[i_a]),
\end{equation}
or the angle prediction scaled by the norm prediction. The output is then Gaussian filtered per-component.

\subsection{Uncertainties}
\label{sec:Q_uncertainties}

One advantage of the series of approaches developed in~\citep{higgins2021fast,Higgins2022,SuperSynthIA} is that they can produce uncertainties. Practically speaking, these uncertainties help identify pixels that are likely to be incorrect, and enables uncertainty propagation for downstream tasks. Additionally, these uncertainties demonstrate what the network knows it knows. Recall that the method simultaneously produces a distribution over unit vectors (giving the direction of the field), as well as a norm (giving the strength). This distribution can be reported and visualized a few different ways: (a) per-component uncertainties, which can then be used for error propagation; (b) explicit depictions of the directional distribution; and (c) scalar quantities capturing the uncertainty in norm and angular distribution. These quantities follow directly from the definitions of variance, and we give explicit formulas in Appendix~\ref{sec:app_uncertainty}.

We show the structure of the uncertainty in Figure~\ref{fig:uncertainty} in two ways. First, we plot the inferred probability density over the direction (shown with a Mollweide projection). Quiet-Sun predictions are (correctly) not uniformly distributed over the unit sphere but instead a great circle. These structured predictions show that the model has correctly learned the Milne-Eddington Stokes Inversion plus disambiguation approach -- near disk center, the radial component is parallel to the viewing direction, and so low polarization data will have $B_R \approx 0$; likewise at the limb at the equator, $B_\phi \approx 0$. Angular uncertainty is also higher near polarity inversion lines. Second, we additionally line plots of the the three components across a cut. In these cuts, even when \hmip disagrees with \method, \hmip is often within the confidence intervals produced by \method.

\subsection{Network Architecture}
\label{sec:architecture}

The \method architecture accepts one or more \blos magnetograms as input plus meta-information about disk location consisting of per-pixel $\mu$, latitude, longitude, and a disk mask (all computed by Sunpy; ~\cite{sunpy_community2020}). Typically, \method accepts two auxiliary magnetograms, leading to an input of height $H$ and width $W$ and $7$ channels.

The network follows our past work in this area~\citep{higgins2021fast,Higgins2022,SuperSynthIA} and follows a U-Net~\citep{ronneberger2015unet} which can be thought of as a parametric function that maps from an input in $\mathbb{R}^{H \times W \times 7}$ to an output in $\mathbb{R}^{H \times W \times (6K)}$ where $K$ is the number of bins: $K$ for the norm bins probabilities, $K$ for the norm bin corrections, $K$ for the angle bin probabilities, and $3K$ for angle bin corrections. The function consists of interleaved $3 \times 3$ convolutions (which contain the parameters) and pointwise nonlinearities (which have no parameters). Spatially, the network contracts (while increasing the number of feature channels) and then re-expands (while collapsing the number of feature channels), which facilitates integration of information across large regions.  We believe that additional work may reveal better architecture designs (e.g., as done by~\cite{liu2022convnet}), but we keep a simple design to emphasize our contributions (the problem, formulation, and learning objective) rather than the orthogonal challenge of designing good networks.

\subsection{Implementation Details}
\label{sec:implementation}

For both the norm and the angles, we define $K = 80$ bins automatically to ensure good coverage of the output space. For the bins for norms $\sB_n$, we use a grid of points from $0$ to $5000$ \gauss whose square roots are distributed evenly (i.e., like $[1,4,9,16]$), which helps ensure that each higher-field strength bin has sufficient samples for learning (similar to SuperSynthIA's use of k-means clustered values for bins). 
The maximum distance between the bins is ${\approx}125$\gauss at $5000$ \gauss, roughly matching the uncertainties in \hmip for field strength. On average, points sampled from $1$ to $5000$ \gauss are within $21$ \gauss of
a bin. We use a Fibonacci lattice to define the set of angular bins $\SB_a$, which automatically defines a set of points with good coverage over the unit sphere, near-equidistance spacing between points, and without concentrations at the poles. With 80 bins, each unit vector is ${\approx}9^\circ$ away from a bin on average, again similar to the uncertainties in angles from \hmip.

To align our training data with conventional recipes for training these models, we found it helpful to scale our inputs to make the typical input range be ${\approx}10^0$. We divide all \blos magnetograms by $1000$, longitude by $180$, latitude by $90$, and the disk mask is set to $0$/$1$. 

We train the U-Net architectures using AdamW~\citep{loshchilov2017decoupled}. Full training details appear in Appendix~\ref{sec:app_trainingdetails}, but here we note a few critical optimization hyperparameter choices. We set the learning rate (i.e., step size in optimization of the objective) to $10^{-3}$ and decayed it three times by a factor of $10$ at a fixed schedule. We set weight decay (a form of  regularization performed during optimization that penalizes the squared norm of the weights) to $3 \times 10^{-7}$ and AdamW's $\epsilon$ parameter to $10^{-4}$  following~\cite{higgins2021fast}. We found that the higher-then-usual value of $\epsilon$ helps reduce instabilities in the later stages of training when the low-polarization pixels have converged nearly perfectly. The full $4096 \times 4096$ disk does not fit into memory, so we broke each disk into $1024 \times 1024$ tiles and treated these as separate training samples. Off-disk pixels are set to the zero vector $\zeroB$. We trained each mode for approximately $1.5$ million steps with a batch size of five, which took a little under four days on one NVIDIA A100 GPU with 80GB memory. The model can also trained, at the cost of longer wall-clock time, on a less capable GPU using either data parallelism across multiple GPUs or gradient accumulation.

When training with \gong data, we use transfer learning and data augmentation. We initialize \gong models with models that have been trained to map from \hmi \blos to \hmi vector magnetograms. While the instruments are different, this transfers knowledge of aspects such as statistics that indicate plage-vs-active regions as well as the prior distribution over $\BB$. We also apply a moderate data augmentation: for each real \gong sample, we include: three additional samples where the image has been rescaled by a factor sampled from a uniform distribution from $0.9$ to $1$ (one sided to prevent the disk from escaping the image) and shifted by an offset in each direction that is sampled uniformly from 5\% to 7\% of the image; and four additional samples that have been rescaled and where the flux in both input and output has also been scaled by a factor sampled uniformly from $[0.75,1.25]$.

\section{Experimental Setup}
\label{sec:experiment}

Before we describe our experiments, we discuss data preparation strategies, what data are used, and how we evaluate the methods.

\subsection{Data Preparation}
\label{sec:data_prep}

For \hmi data, we use \blos data from the {\tt hmi.M\_720s} data series and vector data from the {\tt hmi.B\_720s} data series. 
We use the randomized option for disambiguation, which applies ME0 disambiguation~\citep{metcalf1994resolving} with full annealing in regions with good \bperp signal and random disambiguation otherwise, which avoids large-scale biases. When using a model that uses multiple temporally-adjacent auxiliary \blos observations, we rotate the observations to the reference timestamp, accounting for differential rotation with Sunpy~\citep{sunpy_community2020}. 
By aligning the magnetograms, the network does not spend its capacity learning
the solar differential rotation.

For \gong data, we use \blos data from the networked-merged daily magnetogram, or the \gong~{\tt mrbzi} series.
Our definition of vector data here is 
vector data from \hmip re-mapped to the \gong grid with a scaling and translation. A rigid transformation is appropriate since \hmi's Fe I 6173\AA~line and \gong's Ni I 6768\AA~line are believed to have similar formation heights, with a difference of ${\approx}25$km \citep{fleck2011formation}. The height difference corresponds to far less than an \hmi pixel and an even smaller fraction of a \gong pixel. 
The transformation is done using {\tt affine\_transform} from SciPy~\citep{SciPy2020}, consisting of smoothing and re-sampling; while this approach is suitable for a proof-of-concept version, down-sampling vector magnetograms is non-trivial due to structures becoming unresolved, and a future effort might consider resampling the Stokes vectors and re-inverting \citep[see ][for further discussion]{leka2012modeling}.

The \hmi to \gong transformation is obtained by matching solar disks: after an alignment of disk masks, we cross-correlate the absolute value \gong $|B_{\rm los}|$ and \hmi \alphaB. 
Since some \gong disks have artifacts close to the limb, the cross-correlation score is computed only within a \gong disk mask that has been eroded by 1\,px,  5 times. Nothwithstanding that 
$|B_{\rm los}|$ and \alphaB are not identical quantities, any expected statistical offset is small compared to the \gong spatial resolution.  Empirically, we find our approach to produce good alignment. For consistency, we also transfer metadata from \hmi to \gong using the scaling and translation transformation. Due to the worse resolution of \gong, no rotation is applied to auxiliary observations (if used).

\begin{figure*}
    \centering
    \begin{tabular}{@{}c@{}c@{}@{}}
    \includegraphics[width=0.5\linewidth]{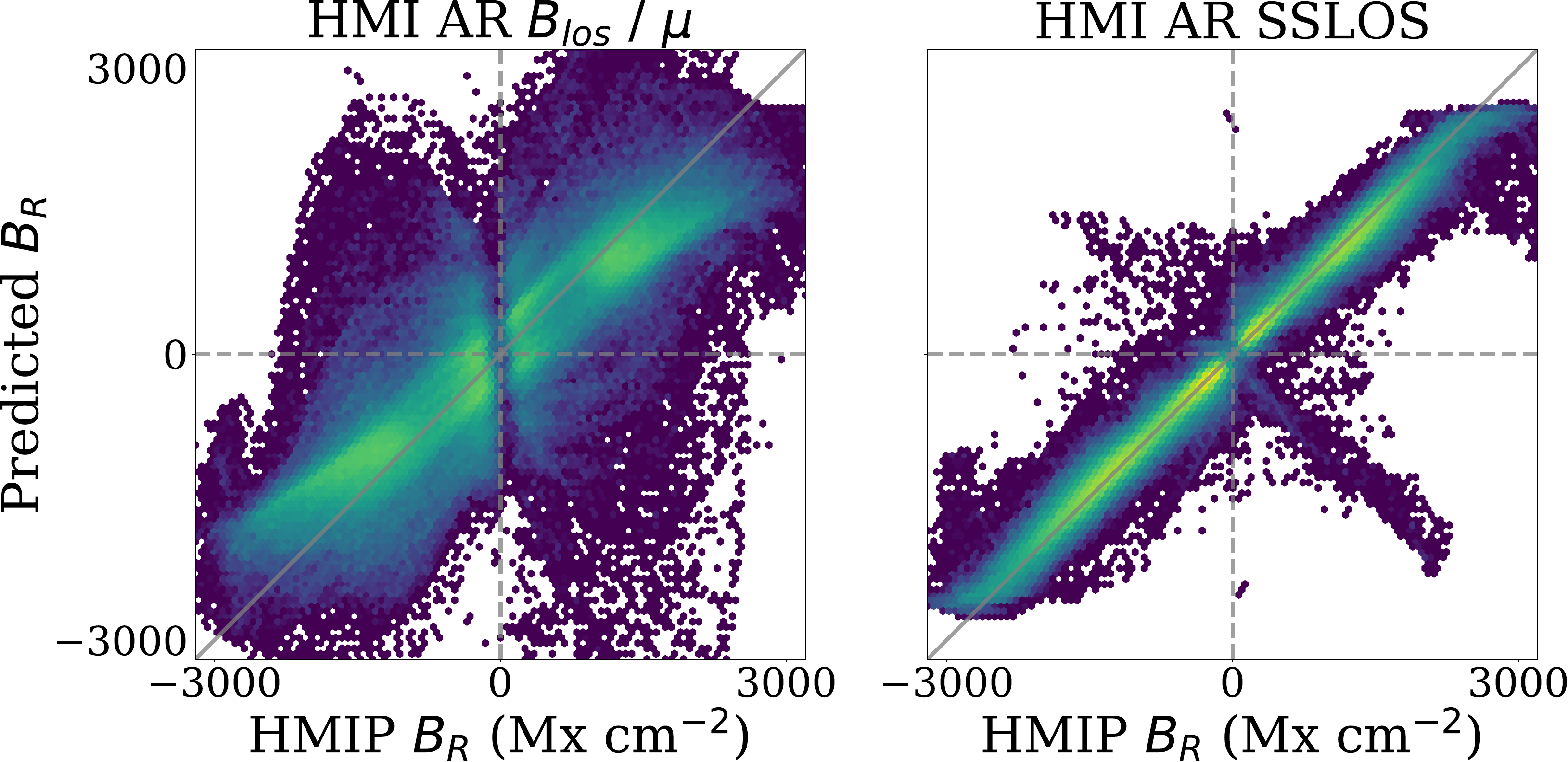} & 
    \includegraphics[width=0.5\linewidth]{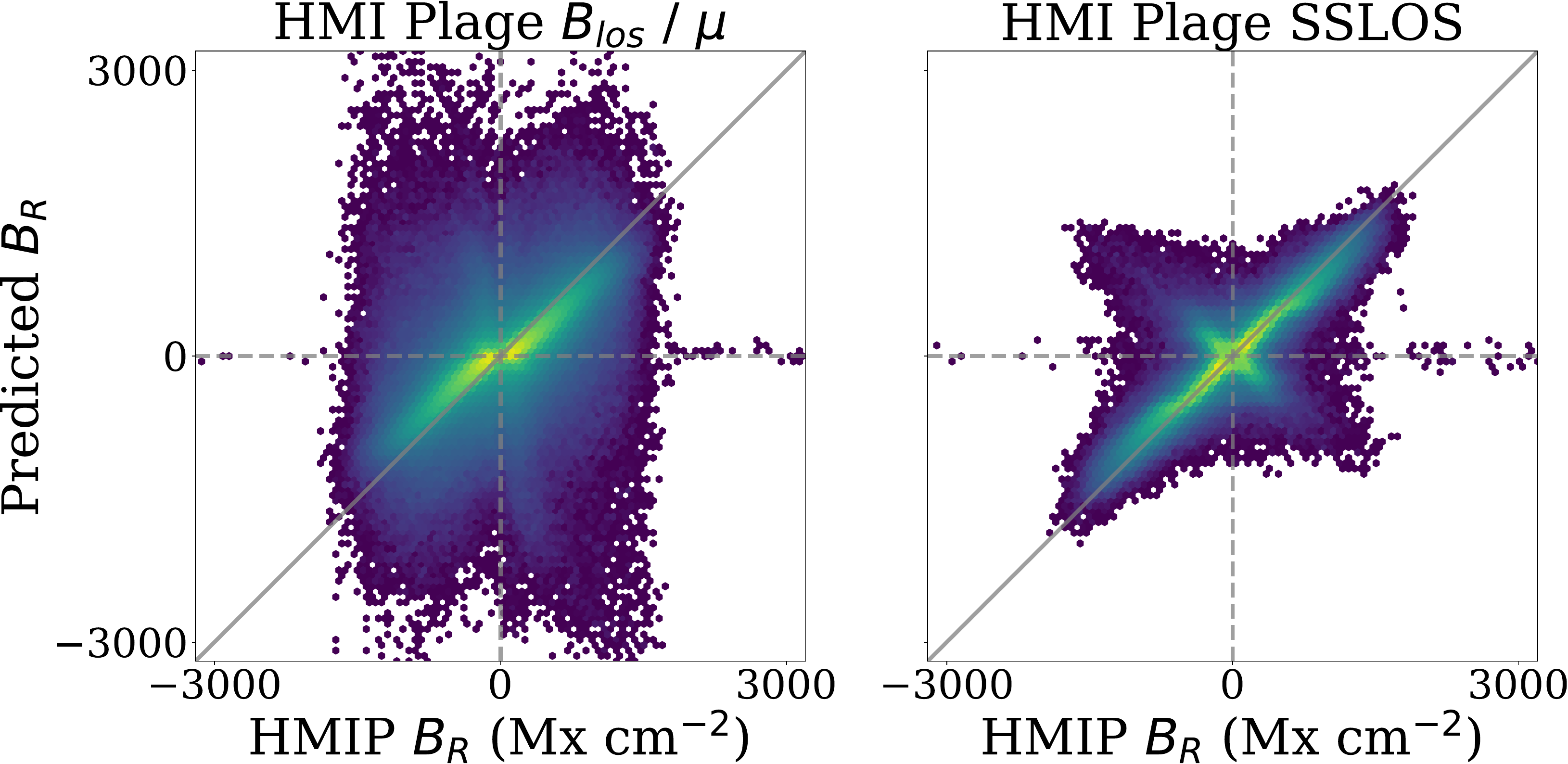} \\
    \includegraphics[width=0.5\linewidth]{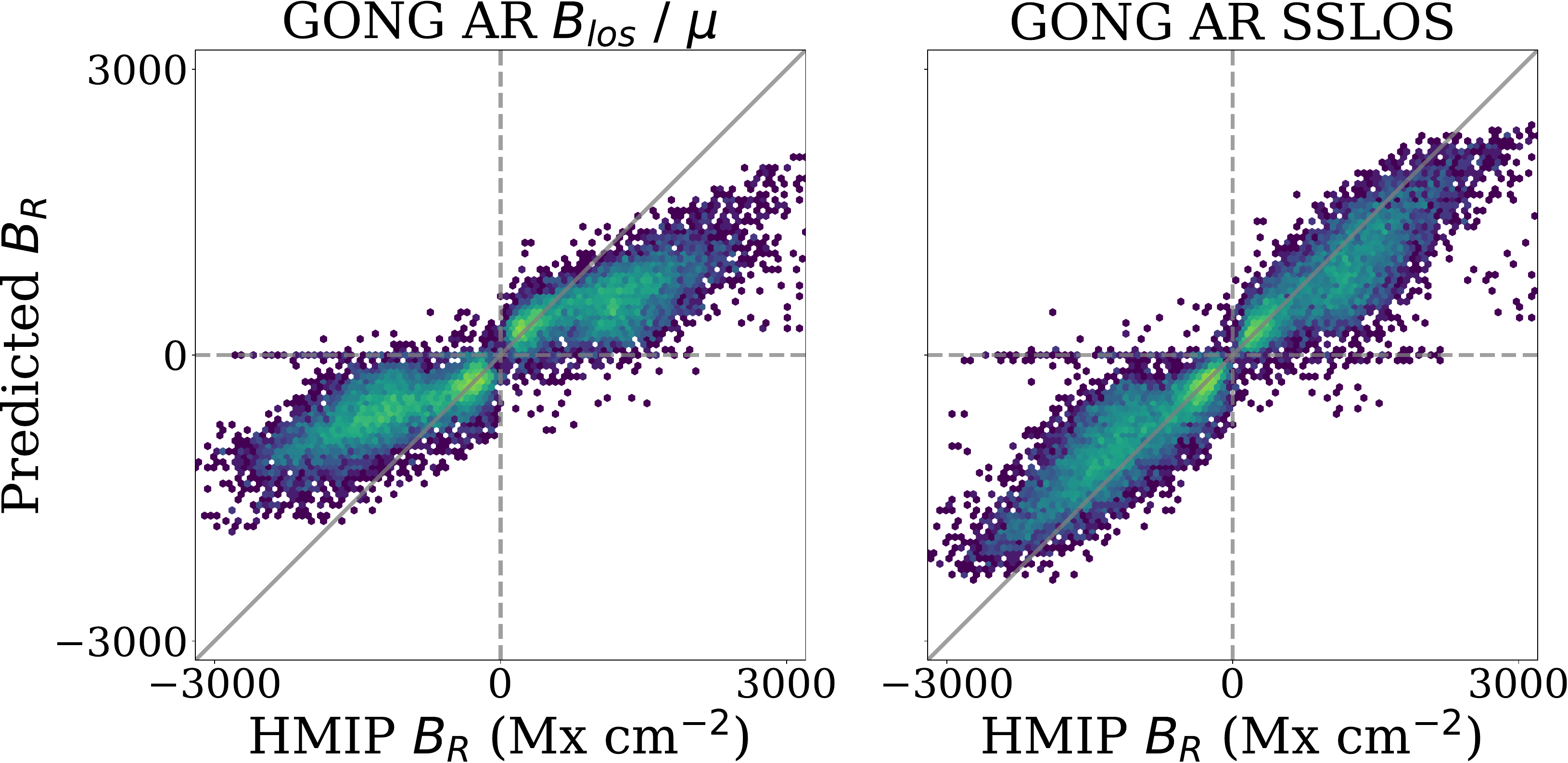} & 
    \includegraphics[width=0.5\linewidth]{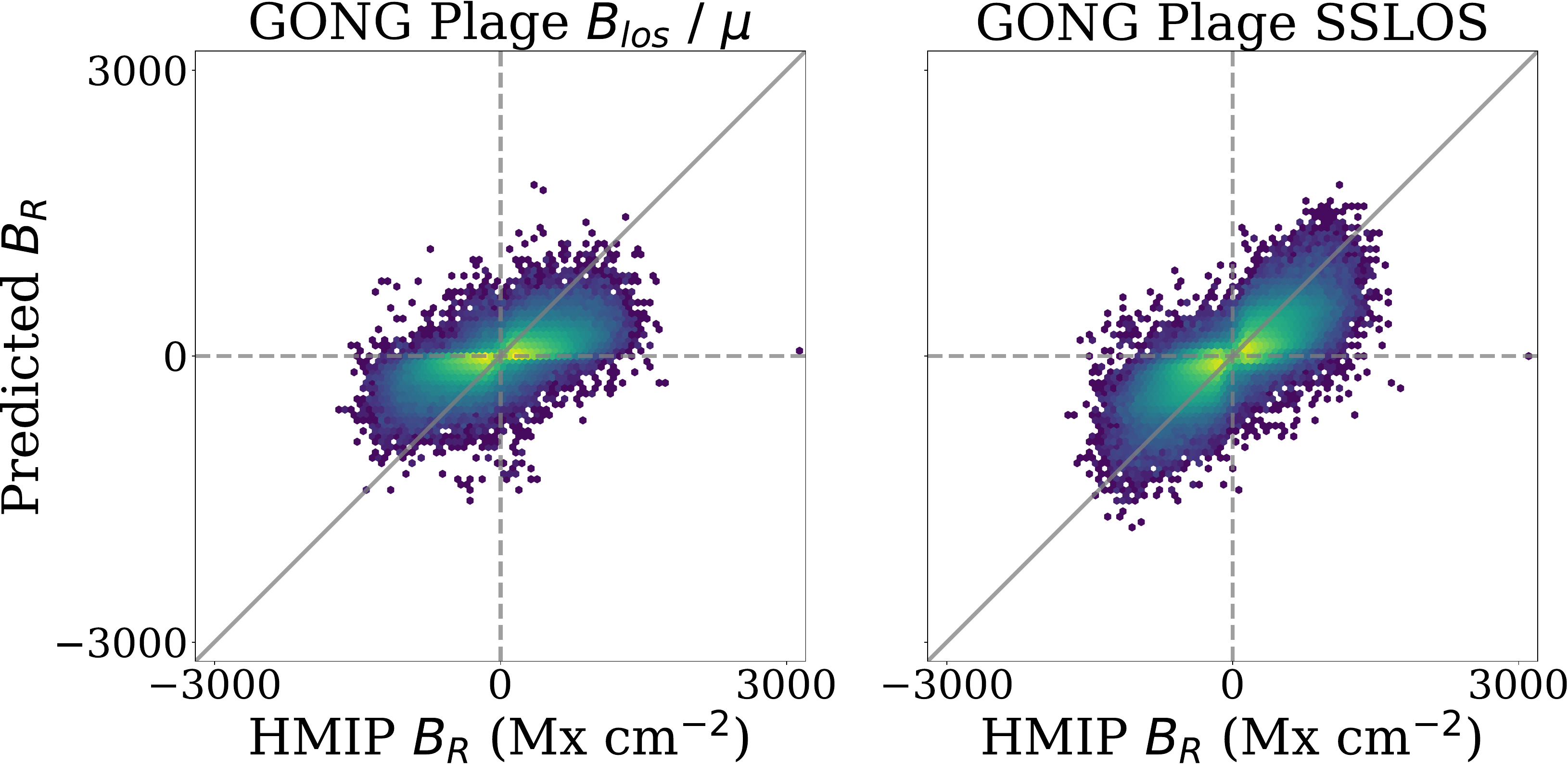} \\
    \end{tabular}
    \caption{{\bf Bivariate log-histograms of \br in active regions and plage using \hmi and \gong \blos data}. We compare results from $\mu$-corrected \blos and \method. We plot a $y=x$ line and $y=0$ and $x=0$ lines in red. Content in the top-left or bottom right quadrants are polarity errors. 
    The proposed method improves over the standard $\mu$-correction as seen by the tighter alignment with the axis and substantial reduction in polarity confusions. In plage with \hmi \blos, the proposed method shows slight concentration at the $y=-x$ diagonal. These errors are on pixels with low field strength; the $\mu$-correction has a similar amount of polarity confusion, but it is spread out over more values since the $\mu$-correction does not get the magnitude right. Colormap: empty bins are colored white; non-empty bins follow lowest \includegraphics[width=20pt,height=6pt]{viridis.png} highest density. Colormaps are logarithmically colored, and normalized identically per-condition (e.g., GONG AR, HMI Plage) in this and the \bp/\bt plots in Figure~\ref{fig:bivariabeBpt}.}
    \label{fig:bivariateBr}
\end{figure*}

\subsection{Data}
\label{sec:data}

\begin{deluxetable}{l@{~~}l@{~~}l}[t]
\caption{Parameters defining regions for evaluation, as well as the segments used.}
\label{tab:region_defs}
\tablehead{
Region & Source & Limit or Value 
}
\startdata
Plage Regions & {\tt conf\_disambig} & $\ge 60$ \\
              & {\tt Ic\_noLimbDark\_720s} & $> 0.8$ \\
              & \alphaB & $\ge 150$ \gauss \\
Active Regions & {\tt conf\_disambig} & $\ge 60$ \\
               & {\tt Ic\_noLimbDark\_720s} & $< 0.8$ \\
              & \alphaB & $\ge 150$ \gauss \\
Limb          & $\mu = \cos(\theta)$  & $\le 0.5$ \\            
\enddata
\end{deluxetable}

Our primary evaluation set consists of data from the period of time that \hmi and \gong have co-observed up to June 30, 2023. Our test period starts 1 July 2015 and ends 30 June 2017. The adjacent six months to the test set are ignored, to allow nearly seven Carrington rotations of time for the Sun to reset and thus prevent train-test leakage (as discussed by \cite{Galvez2019}). The validation period is the union of 1 July 2014 through 31 December 2014 and 1 January 2018 through 30 June 2018. The remaining portion of the {\it Solar Dynamics Observatory} mission up to June 30, 2023 is available for training. 

We generate a training set by sampling data at a fixed cadence (29 hours and 24 minutes to ensure coverage of times of day and thus SDO observer velocities) from a given starting point. For each year in the training set, we start at midnight on July 1 and sample with the fixed cadence for one year. For more coverage of active years, we repeat the sampling starting at noon, July 1 for years 2010 -- 2013, 2021, and 2022). We finally add samples for the first two months of the {\it SDO} mission (May 1 - June 30, 2010).

For our validation and testing set, we sample 50 timestamps randomly from the validation and testing periods, respectively.

\begin{deluxetable*}{lcccccccccccc}
    \caption{{\bf $B_R$ results (\hmi) for Mean Absolute Error (MAE), Median Error (ME), and Percent within 200 Gauss (\% $< 200$).} \blos / $\mu$ is catastrophically wrong in active regions, with few pixels accurately measured; rescaling (\blos/$\mu$ Fit) helps only a little. 
    Learning-based methods ($\ell_1$ regression and \method) do substantially better, with $\ell_1$ regression slightly outperforming \method. However, $\ell_1$ regression catastrophically underestimates flux as shown in Table~\ref{tab:totalFlux}.}
    \label{tab:hmi}
\tablehead{
 & \multicolumn{3}{c}{All Pixels} 
 & \multicolumn{3}{c}{Active Regions (AR)}
 & \multicolumn{3}{c}{Limb ARs}
 & \multicolumn{3}{c}{Plage Regions}
 \\
 &  MAE & ME & $\%<200$&
 MAE & ME & $\%<200$ &
 MAE & ME & $\%<200$ &
 MAE & ME & $\%<200$
 }
\startdata
\blos / $\mu$  & 34 & 0.0 & 99.5 & 449 & 44.5 & 30.1 & 1099 & 52.1 & 11.4 & 119 & -2.2 & 88.9 \\
\blos / $\mu$ Fit  & 34 & 0.0 & 99.5 & 441 & 30.7 & 31.8 & 851 & 84.6 & 16.9 & 103 & 0.2 & 92.1 \\
$\ell_1$  & 32 & 0.0 & 99.8 & 126 & -7.3 & 83.8 & 252 & 16.9 & 56.9 & 48 & -0.4 & 98.1 \\
\method  & 42 & 0.1 & 97.8 & 134 & 2.1 & 81.6 & 248 & 8.5 & 61.2 & 54 & -0.2 & 95.4 \\
\enddata
\end{deluxetable*}

\subsection{Evaluation Metrics}
\label{sec:eval_metrics}

\begin{figure*}[t!]
    \centering
    \includegraphics[width=\linewidth]{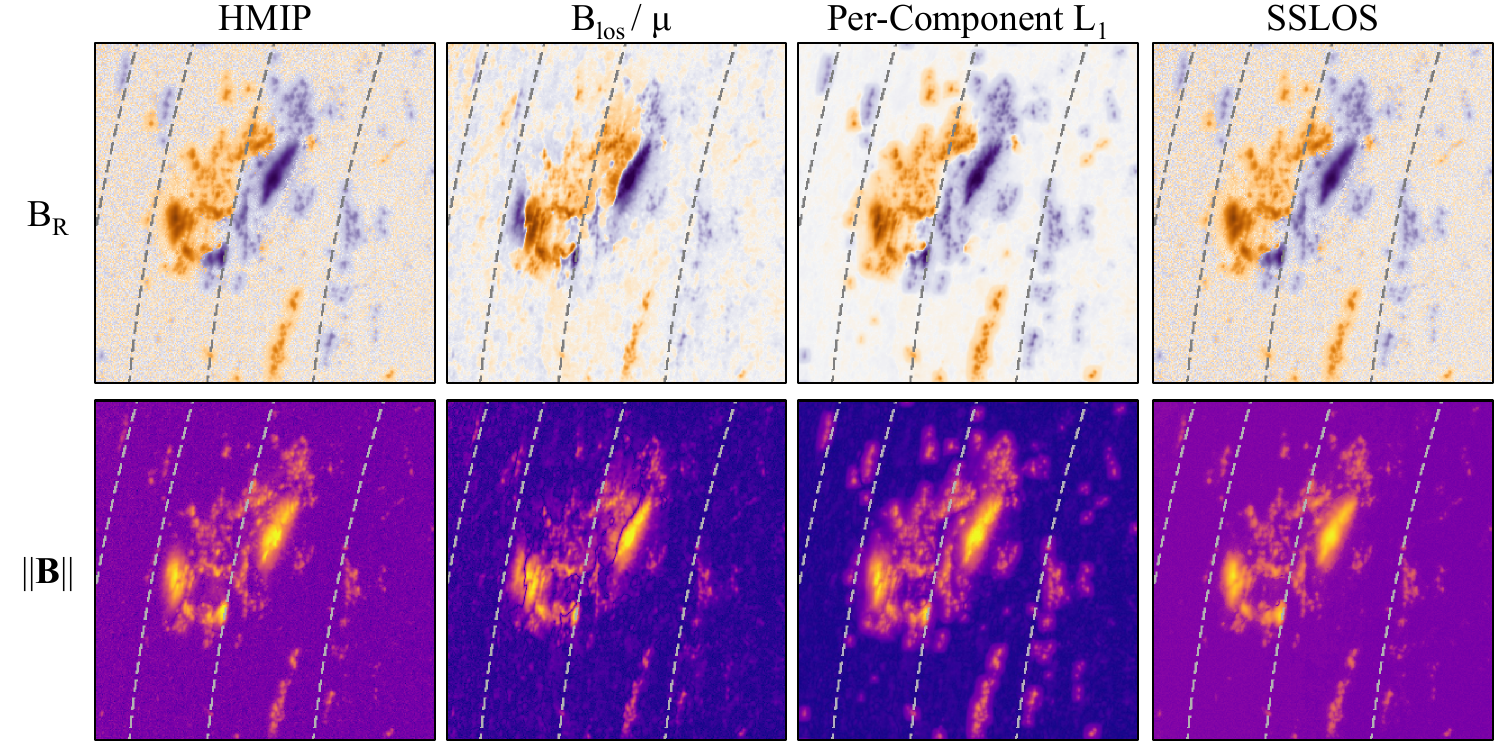}
    \caption{{\bf Comparison of \br and $||\BB||$, using the same data (NOAA AR 12570) as Figure~\ref{fig:ablation}.} We show the reference \hmip data, \blos / $mu$, the per-component $\ell_1$ regression, and \method. Per-component $\ell_1$ regression and \blos / $\mu$ systematically underestimate the total magnetic flux density. \blos $/ \mu$  assumes that \bp and \bt are zero, and so is an underestimate and has false zero-flux density lines through sunspots. Per-component $\ell_1$ regression does well in active regions, but zeros-out the flux density in regions with low polarization signal. This is because the per-component $\ell_1$ minimizing solution for these pixels is, in fact zero. \method faithfully handles both the active regions as well as the low polarization regime.}
    \label{fig:totalflux}
\end{figure*}

\begin{deluxetable}{l@{~~}c@{~~}c@{~~}c@{~~}c@{~~}c@{~~}c}[t]
\caption{Evaluation of total flux density \alphaB (i.e., $(B_R^2 + B_\phi^2 + B_\theta^2)^{1/2}$). Both \blos $/\mu$ and $\ell_1$ substantially underestimate the total magnetic flux over all pixels; $\ell_1$ recovers in the active regions, but is still substantially less accurate. Trends are similar when evaluation is restricted to limb active regions.}
\label{tab:totalFlux}
\tablehead{
 & \multicolumn{3}{c}{Active Regions} & \multicolumn{3}{c}{Plage Regions} \\
 & MAE & ME & \%\,$<200$ & MAE & ME & \%\,$<200$
}
\startdata
\blos / $\mu$  & 1659 & -1342 & 6 & 277 & -195 & 52\\
\blos / $\mu$ Fit  & 1516 & -1410 & 0 & 276 & -205 & 48\\
$\ell_1$ Reg.   & 185 & -116 & 67 & 66 & -49 & 98\\
\method  & 144 & -102 & 78 & 42 & -26 & 99\\
\enddata
\end{deluxetable}

The performance validation is defined in terms of the metrics used to quantify performance and the selection of pixels to which the metrics are applied.  

Three metrics are reported based on a set of $N$ ground-truth values $\{y_i\}$ and $N$ predictions $\{\hat{y}_i\}$. The first is the mean absolute error (MAE), or $\frac{1}{N} \sum_{i=1}^N |y_i - \hat{y}_i|$. The second is the Percent-Good-Pixels (PGP), or percent of pixels within a threshold $t$, or $\frac{1}{N} \sum_{i=1}^N H( |y_i - \hat{y}_i| < t)$ where $H(\cdot)$ is a $1$ if its argument is true and $0$ otherwise. We set $t = 200$ \gauss. Finally, we report the median signed error, or $\textrm{median}(\{\hat{y}_i - y_i\})$, as this can often catch systematic under/overestimates.

The vast majority of the Sun has low polarization signal at the resolutions discussed here, hence any summary statistics over the full disk are driven largely by the handling of low-signal ``quiet-Sun'' regions.  For example, fixing a phantom polarity inversion line that appears in a $100\arcsec$ sunspot might only influence ${\approx}10$K pixels of the ${\approx}11$M pixels observed by \hmi, accounting for just 0.1\% of the pixels. Indeed, in Appendix \ref{sec:app_scale}, we show that some changes that radically improve estimates in active regions have no discernible impact on full-disk summary statistics.

We perform evaluation in four regions: all on-disk pixels, plage regions, active regions, and near-limb active regions. The precise thresholds are described in Table~\ref{tab:region_defs} and are similar to~\cite{higgins2021fast}. Briefly, we define active regions and plage as pixels with sufficient total field strength and signal in $B_\perp$ that are are either dark (active regions) or bright (plage).

\section{Results}
\label{sec:results}

In this section, our main goal is to understand the overall accuracy of \method.  In turn, we answer how well we can recover \br and $||\BB||$ when input \blos data are from \hmi and \gong, how well we recover the components perpendicular to \br (parallel to the local surface), and finally demonstrate and evaluate \method on examples from historic data. 

These experiments are necessary, but not sufficient, to show the value of a learning-based system. Downstream users must also ask about the data systematics of machine learning methods, just as they must ask about any instrument pipeline's systematics. We show spatiotemporal systematics in Appendix~\ref{sec:app_CTL} and \ref{sec:app_oscillations}.

\begin{figure}
\includegraphics[width=\linewidth]{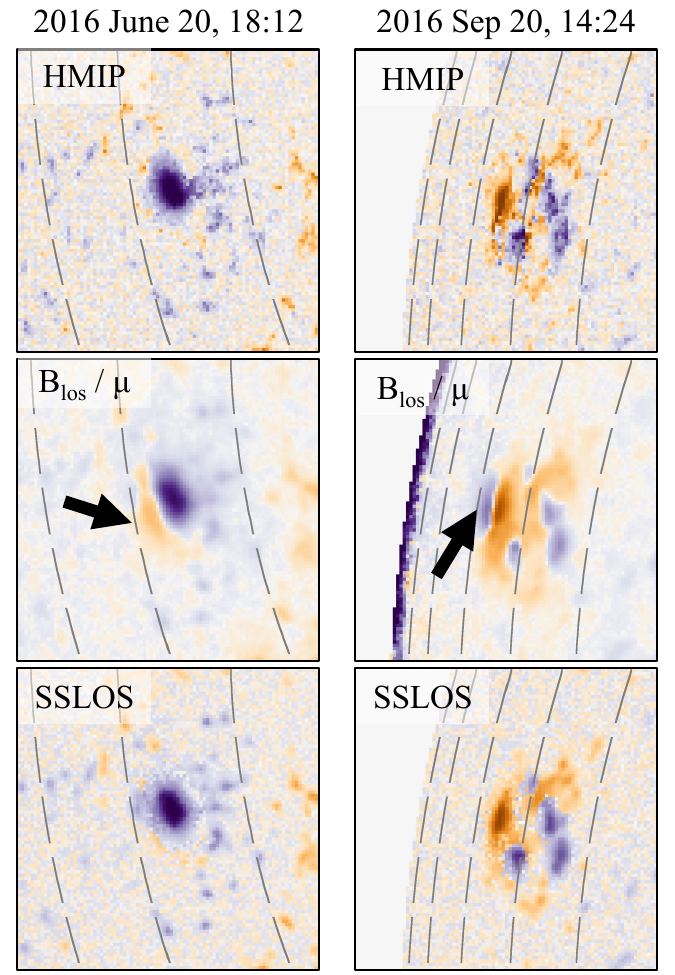}
\caption{{\bf Selected results for \br for \gong.} We show both a simple sunspot on left (NOAA AR 12553), and a more complex region on right (NOAA AR 12595). In both cases, we point to false polarity inversion lines with arrows in the \blos $/\mu$ results.  While more easily seen on the limbs, these false inversion lines appear closer to disk center.  Colormap: \br colormap: -2000 \includegraphics[width=24pt,height=7pt]{PuOrSqrt.png} 2000 Mx cm$^{-2}$.
}
\label{fig:gong}
\end{figure}

\begin{deluxetable*}{lccccccccccccccc}
\caption{{\bf Results recovering \hmi \br from \gong \blos.} We report \blos / $\mu$, as well as a variant that does a fit (to control for any instrumental scaling needed), as well as \method. \method does substantially better on active regions compared to the correction.}
\label{tab:gong}
\tablehead{
 & \multicolumn{3}{c}{All Pixels} 
 & \multicolumn{3}{c}{Active Regions (AR)}
 & \multicolumn{3}{c}{Limb ARs}
 & \multicolumn{3}{c}{Plage Regions}
 \\
 &  MAE & ME & \% $<$200 &
 MAE & ME & \% $<$200 &
 MAE & ME & \% $<$200 &
 MAE & ME & \% $<$200 
 }
\startdata
\blos / $\mu$  & 36 & -0.1 & 99.1 & 559 & 46.9 & 28.4 & 573 & 59.4 & 22.5 & 196 & -9.5 & 62.9 \\
\blos / $\mu$ Fit  & 35 & -0.1 & 99.2 & 382 & 12.5 & 33.2 & 510 & -59.2 & 24.0 & 174 & -5.0 & 68.5 \\
Proposed  & 41 & -0.1 & 99.4 & 320 & 17.3 & 44.7 & 397 & -5.7 & 34.1 & 150 & -3.6 & 73.1 
\enddata
\end{deluxetable*}

\subsection{Recovering \hmi \br and $||\BB||$}
\label{sec:Q_HMIBrBB}

We begin by analyzing how well \method can recover the \hmi radial field and total flux density from the \hmi line of sight data. This experiment tests performance under ideal conditions: there is no co-alignment or change of gridding. 

We compare \method to a number of alternatives. The first is the $\mu$-correction (``\blos $/ \mu$''). The second is (``\blos $/ \mu$-Fit''), which multiplies the $\mu$-correction by an empirically determined scaling factor to ensure that a simple calibration constant is not responsible for differences (e.g., when doing studies with \gong). The scaling factor is determined by minimizing the MAE between the \blos $/ \mu$ result and ground-truth \hmip \br. The final is $\ell_1$ Regression, which minimizes the sum of the per-component distances, or $|\hat{B}_R - B_R| + |\hat{B}_\phi - B_\phi| + |\hat{B}_\theta - B_\theta|$.  Minimizing the $\ell_1$ error minimizes the average MAE for each component, independently without consideration of the total norm.

Qualitative comparisons of \method to \blos / $\mu$ appear in Figures~\ref{fig:teaser} and~\ref{fig:totalflux}. Quantitative results are given in Table~\ref{tab:hmi} with bivariate histograms presented in Figure~\ref{fig:bivariateBr}. Quantitative results for $||\BB||$ are presented in Table~\ref{tab:totalFlux}.  Compared to \blos / $\mu$, \method's MAE is ${\approx}3.5\times$ smaller in active regions. The gain is driven both by improved correction to false polarity inversion lines and substantial overestimates by \blos / $\mu$  for regions with  the correct polarity but that are inherently inclined. 
Since the $\mu$ correction assumes that the field perpendicular to the radial component is zero, the $\mu$-correction also does worse in recovering of the total flux density $||\BB||$. 
Even with an additional post-hoc factor added by \blos / $\mu$-Fit, \method's MAE is ${\approx}12\times$ smaller in active regions, as shown in Table~\ref{tab:totalFlux}.

\begin{figure*}
\centerline{
\includegraphics[width=0.33\textwidth,clip, trim = 10mm 10mm 22mm 24mm]{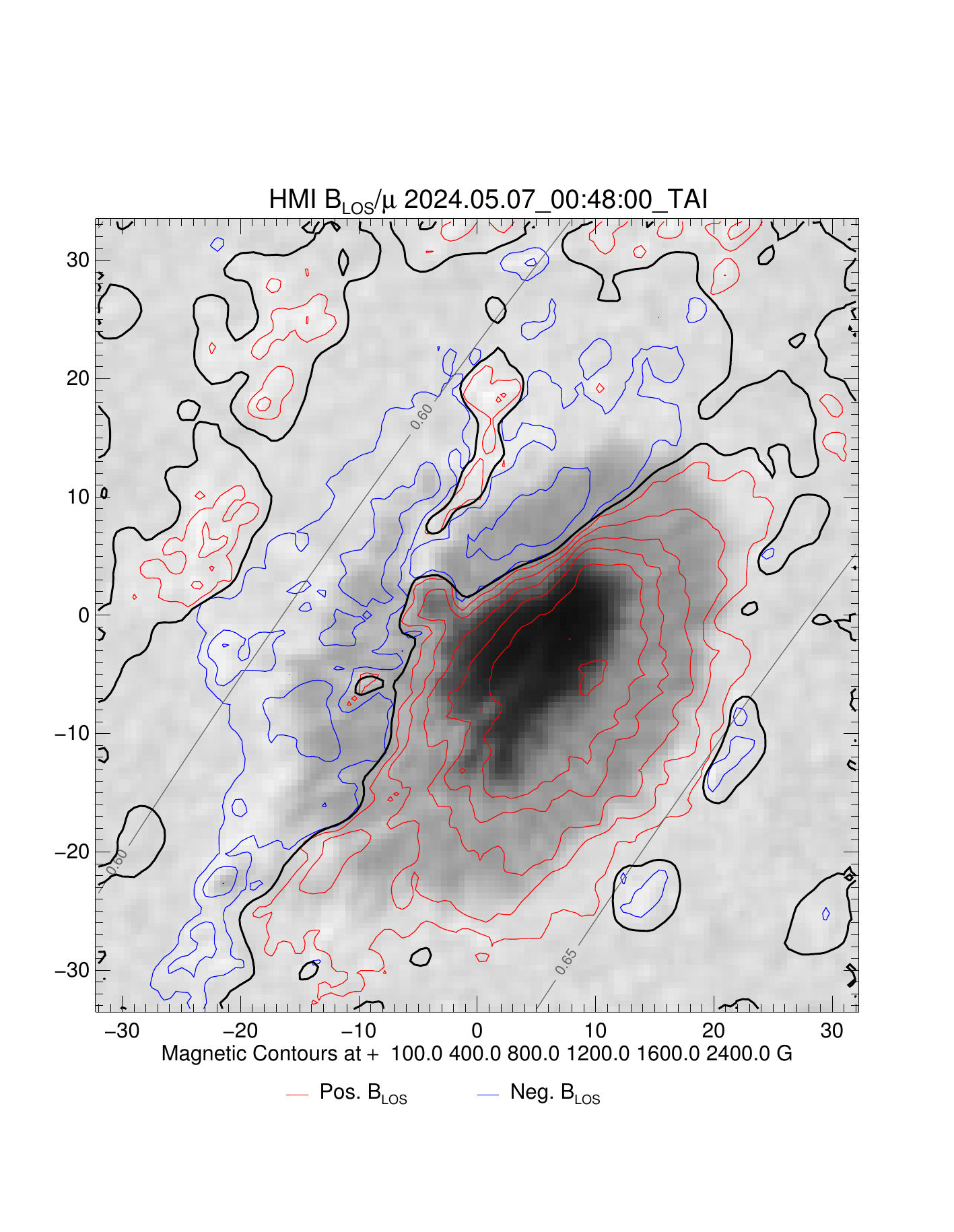}
\includegraphics[width=0.33\textwidth,clip, trim = 11mm 10mm 22mm 24mm]
{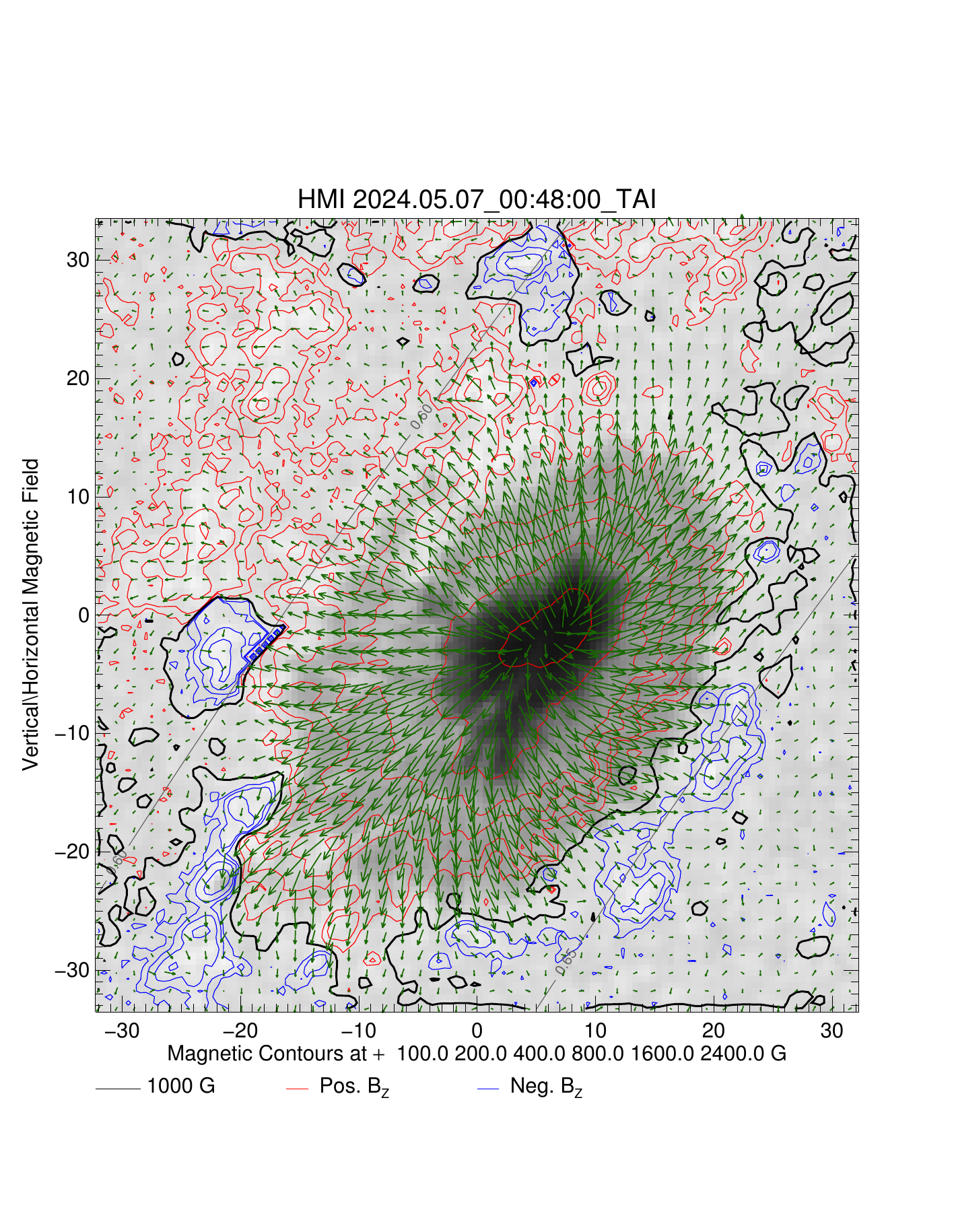}
\includegraphics[width=0.33\textwidth,clip, trim = 11mm 10mm 22mm 24mm]
{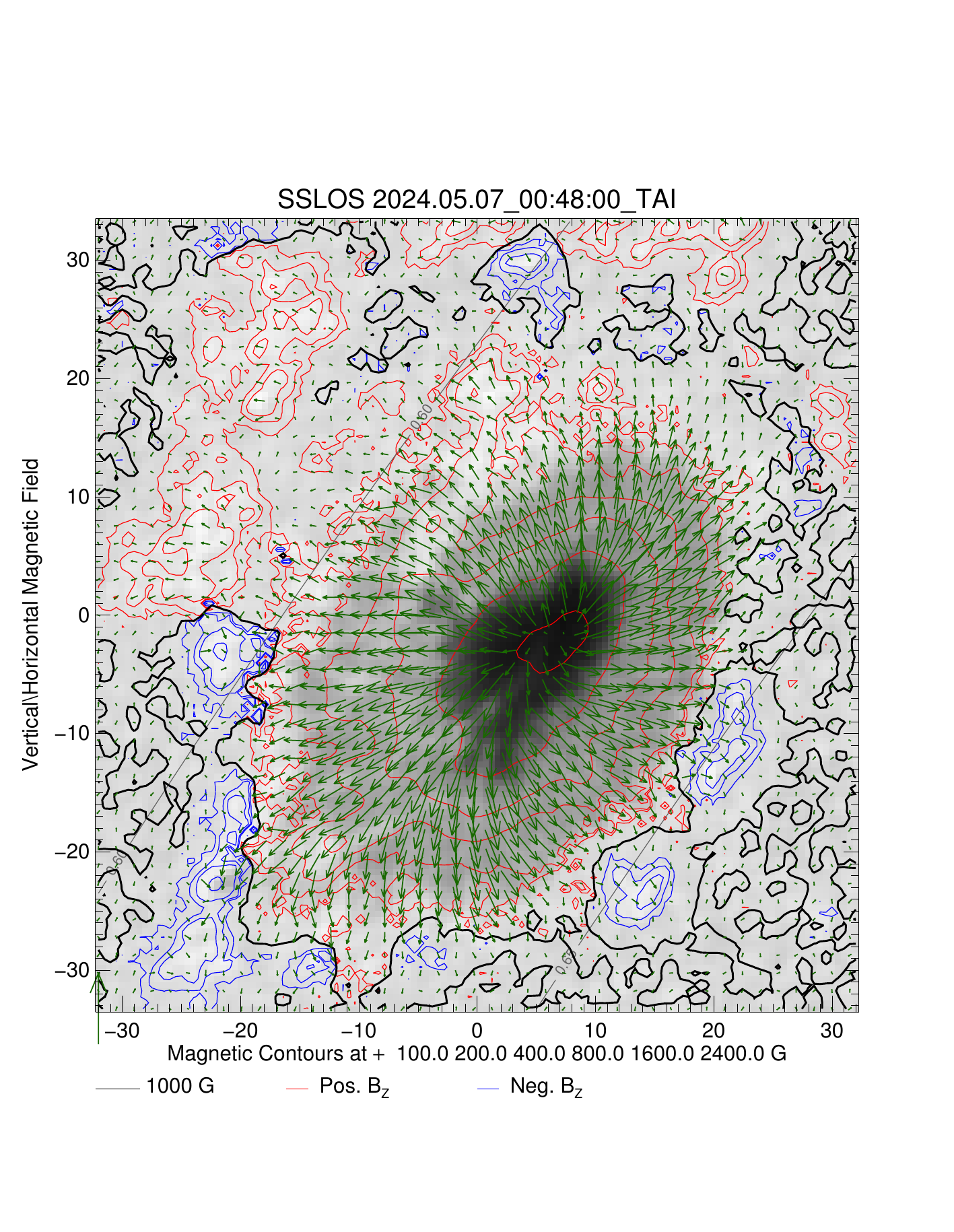}
}
\caption{Left: HMI \blos / $\mu$ from the {\tt hmi.M\_720s} series for NOAA AR\,13667, 2024 May 07.  Located in the far northeast
of the solar disk, the apparent polarity inversion line cuts across the middle of the sunspot.
Continuum image is from the {\tt hmi.Ic\_noLimbDark\_720s} series, isocontours indicate \textcolor{red}{positive} / \textcolor{blue}{negative} line-of-sight field including the ``$\mu$-correction''.
Middle: HMI vector field data from the {\tt hmi.B\_720s} showing isocontours of the \textcolor{red}{positive} / \textcolor{blue}{negative} radial field \br and the 
horizontal components \bp, \bt as \textcolor{ForestGreen}{vectors}, with every third vector displayed for visualization (sharp-eyed viewers may notice the bottom-left-pixel of \method has a strong vertically-oriented vector; this is a scaling reference, and all comparison plots will include this small artifact which forces the vector-lengths to be comparable between the two plots).  For all, axes are in arcseconds from image center, contours of $\mu$ are shown (as labeled), and the
grey contours indicate the smoothed ``polarity inversion line'' where the (line-of-sight or \br)
field changes from positive to negative.  The \method produces the full vector field on-par with
HMI vector field data, both being significantly different from the HMI \blos/$\mu$ field.
} 
\label{fig:vector_intro}

\end{figure*}
\begin{figure*}
    \centering
    \begin{tabular}{@{}c@{}c@{}c@{}c@{}}
    \includegraphics[width=0.24\linewidth]{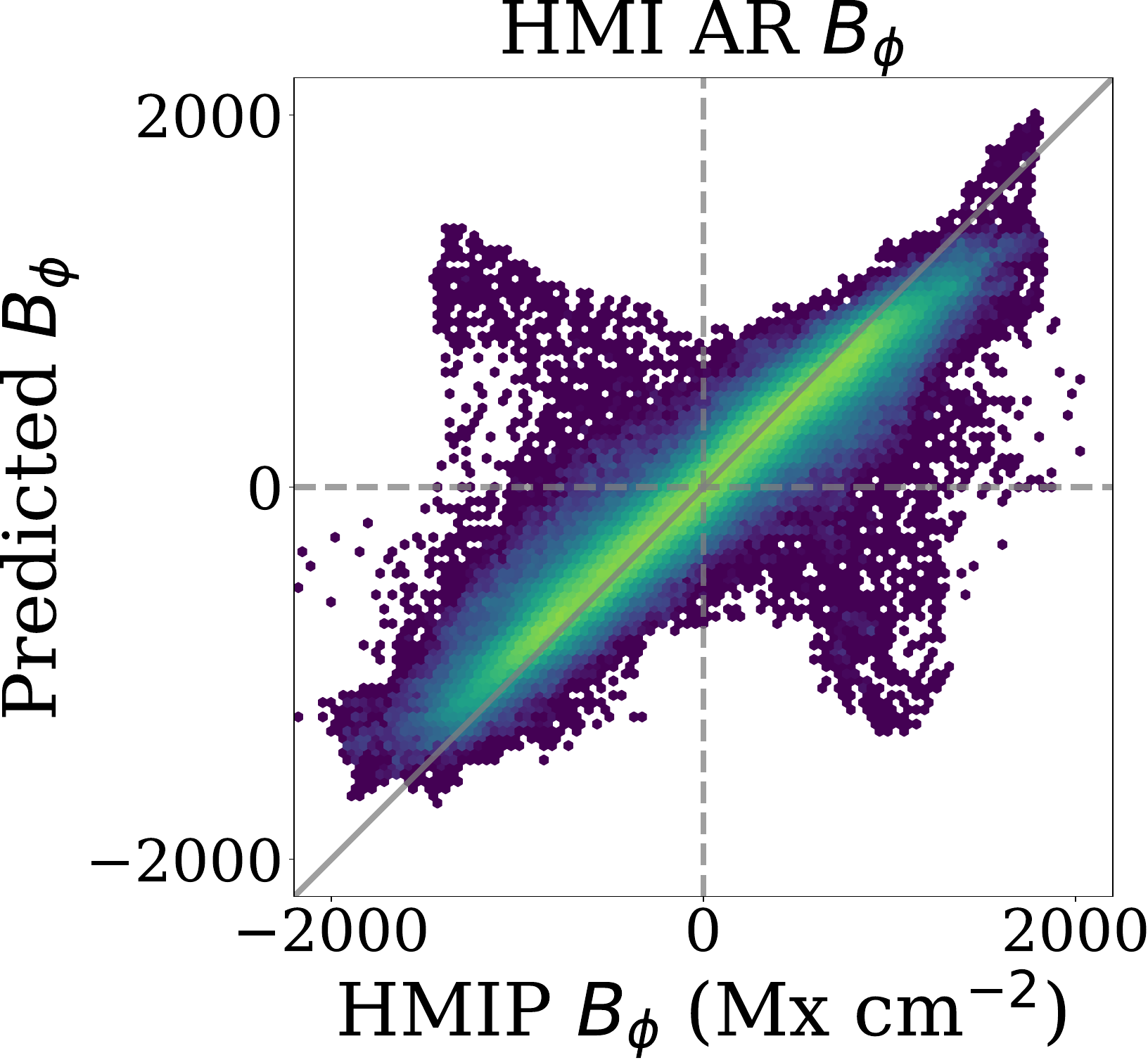} &
    \includegraphics[width=0.24\linewidth]{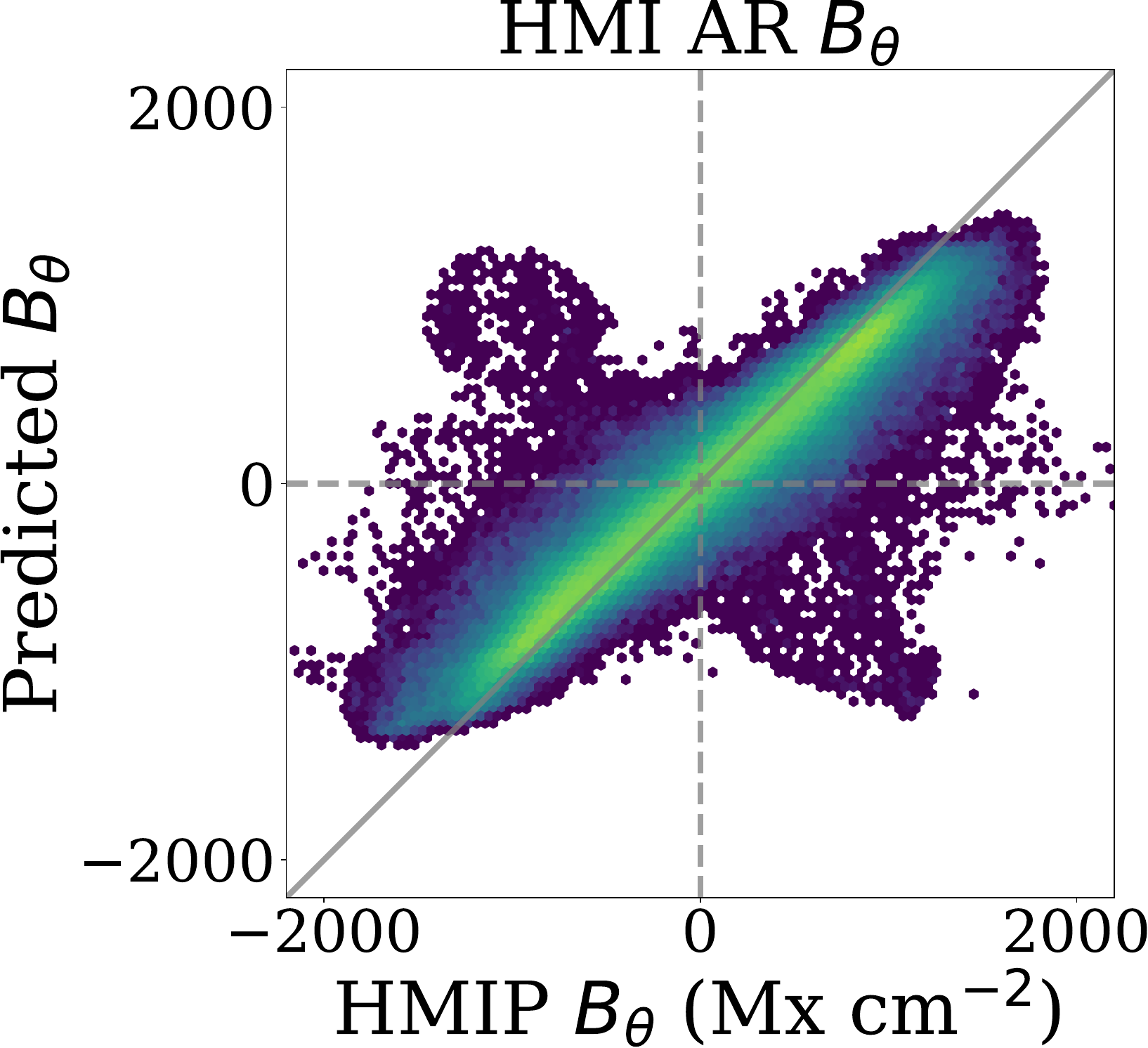} &
    \includegraphics[width=0.24\linewidth]{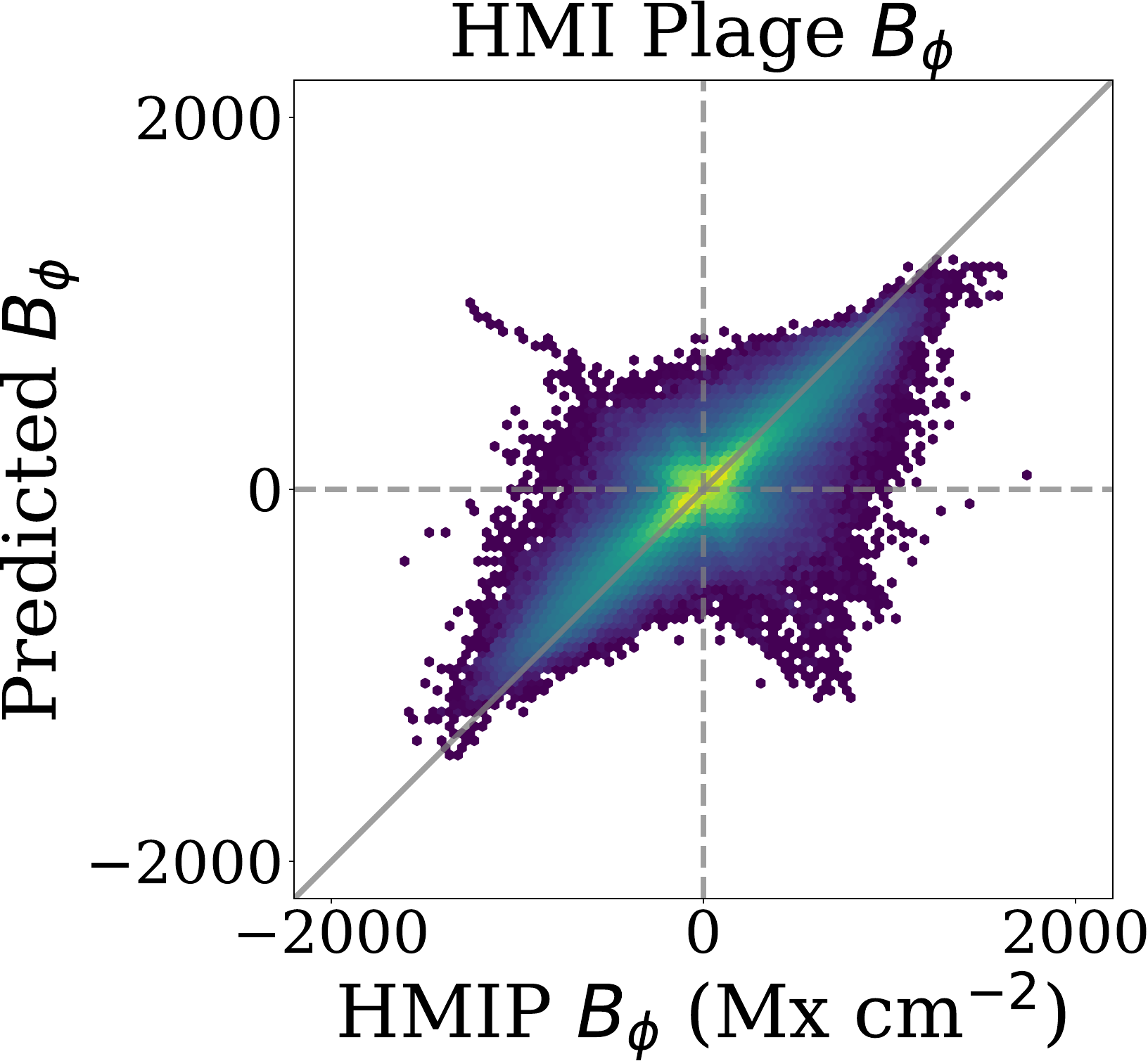} &
    \includegraphics[width=0.24\linewidth]{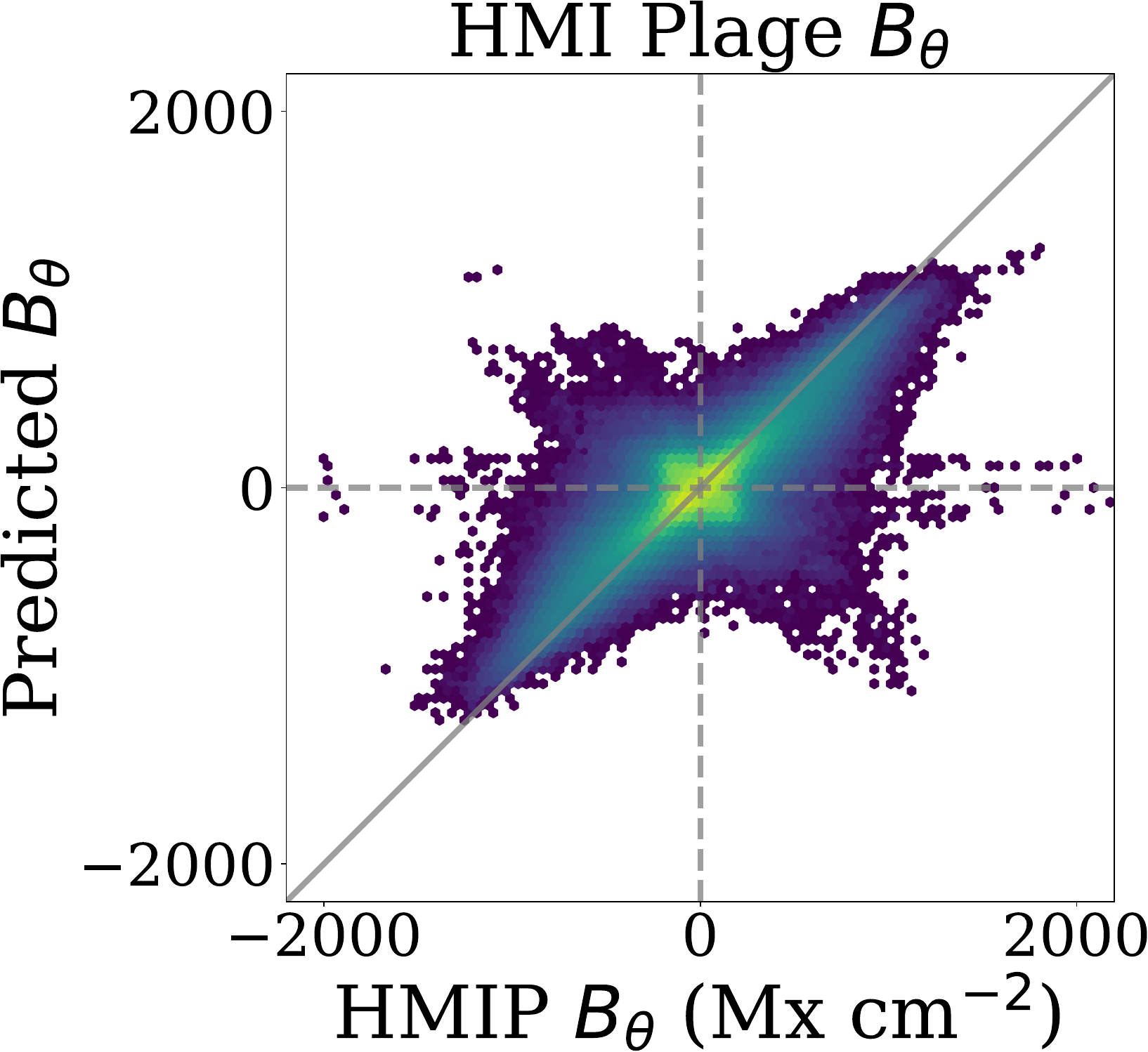} \\
    \includegraphics[width=0.24\linewidth]{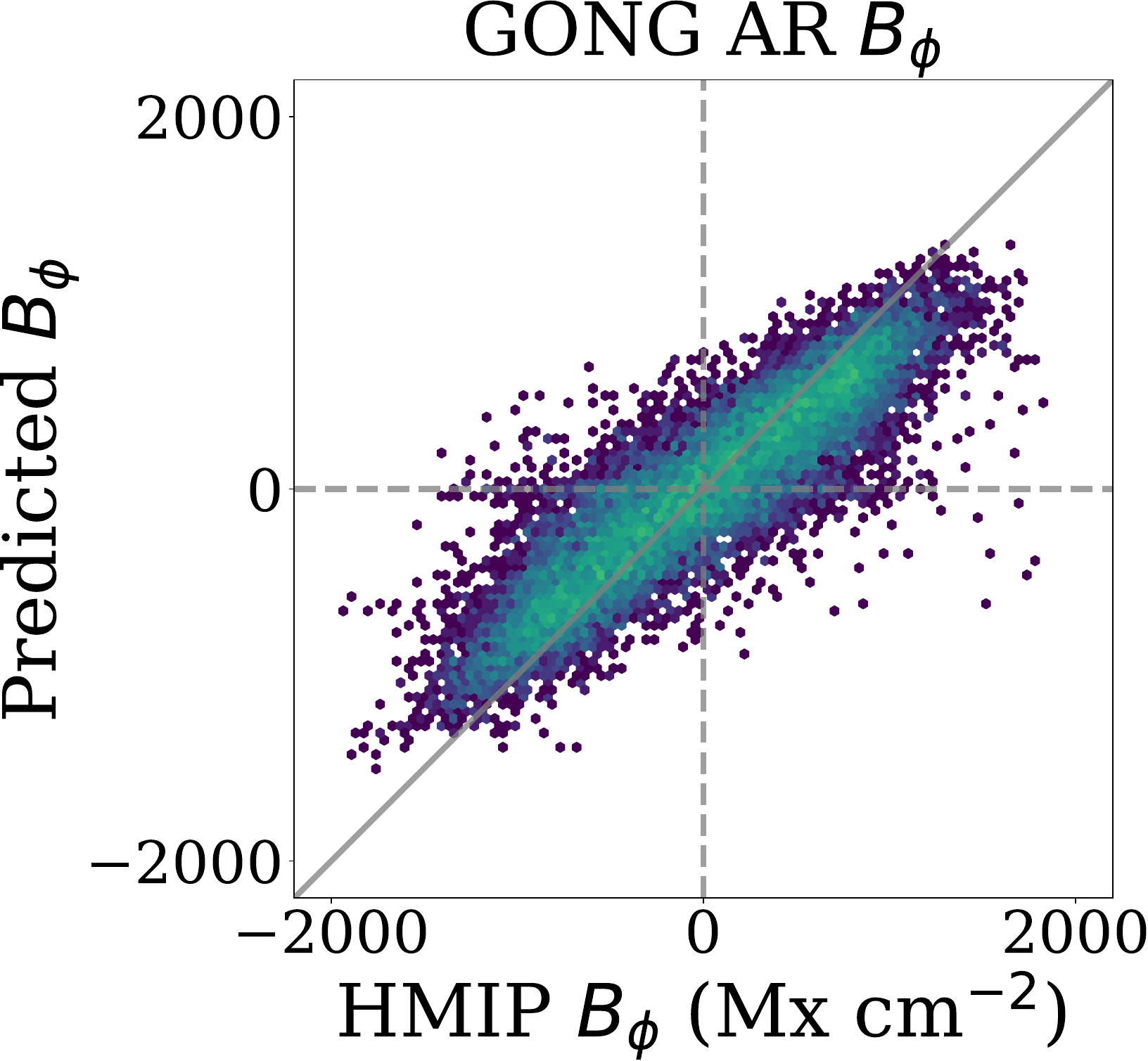} &
    \includegraphics[width=0.24\linewidth]{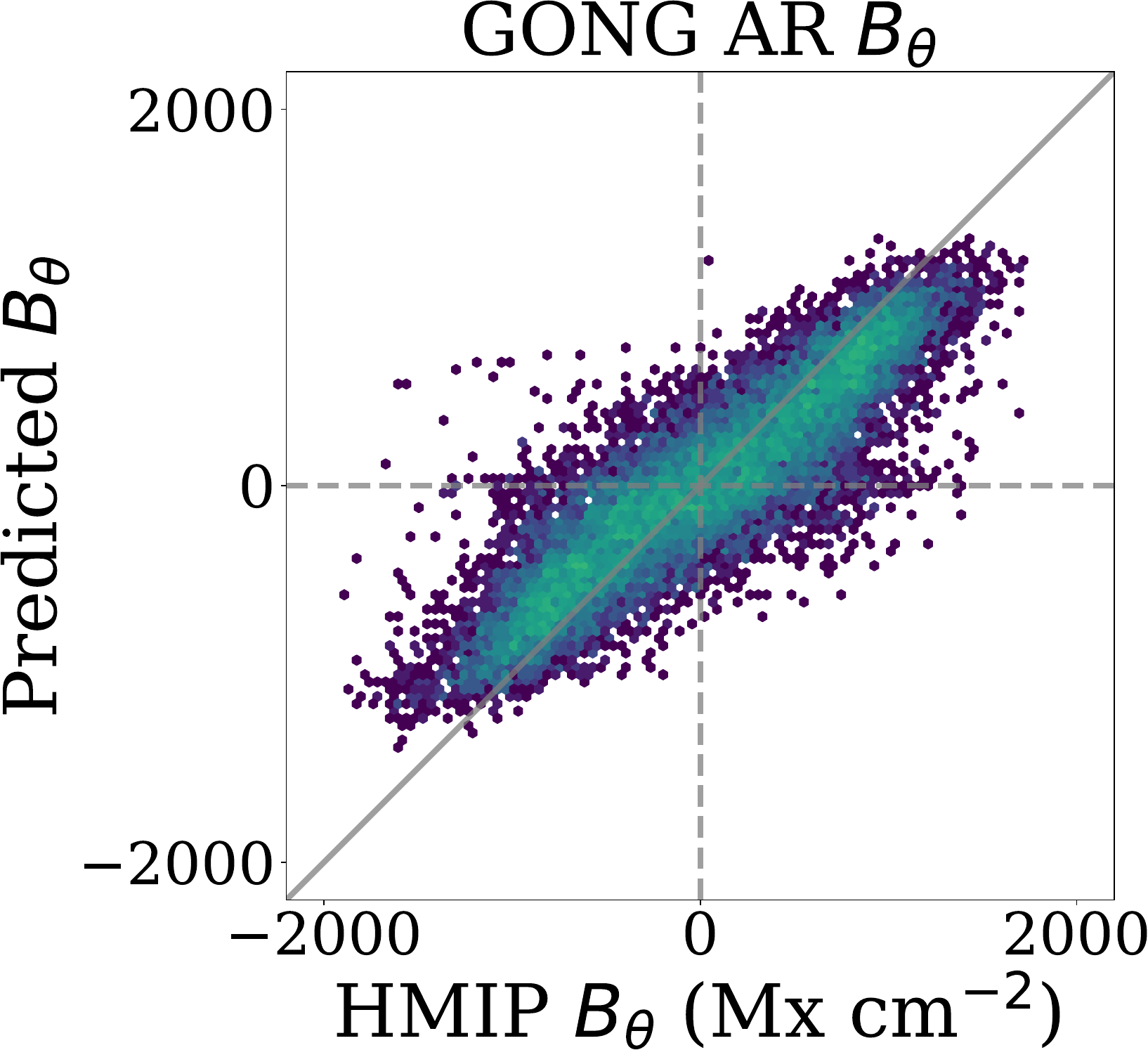} &
    \includegraphics[width=0.24\linewidth]{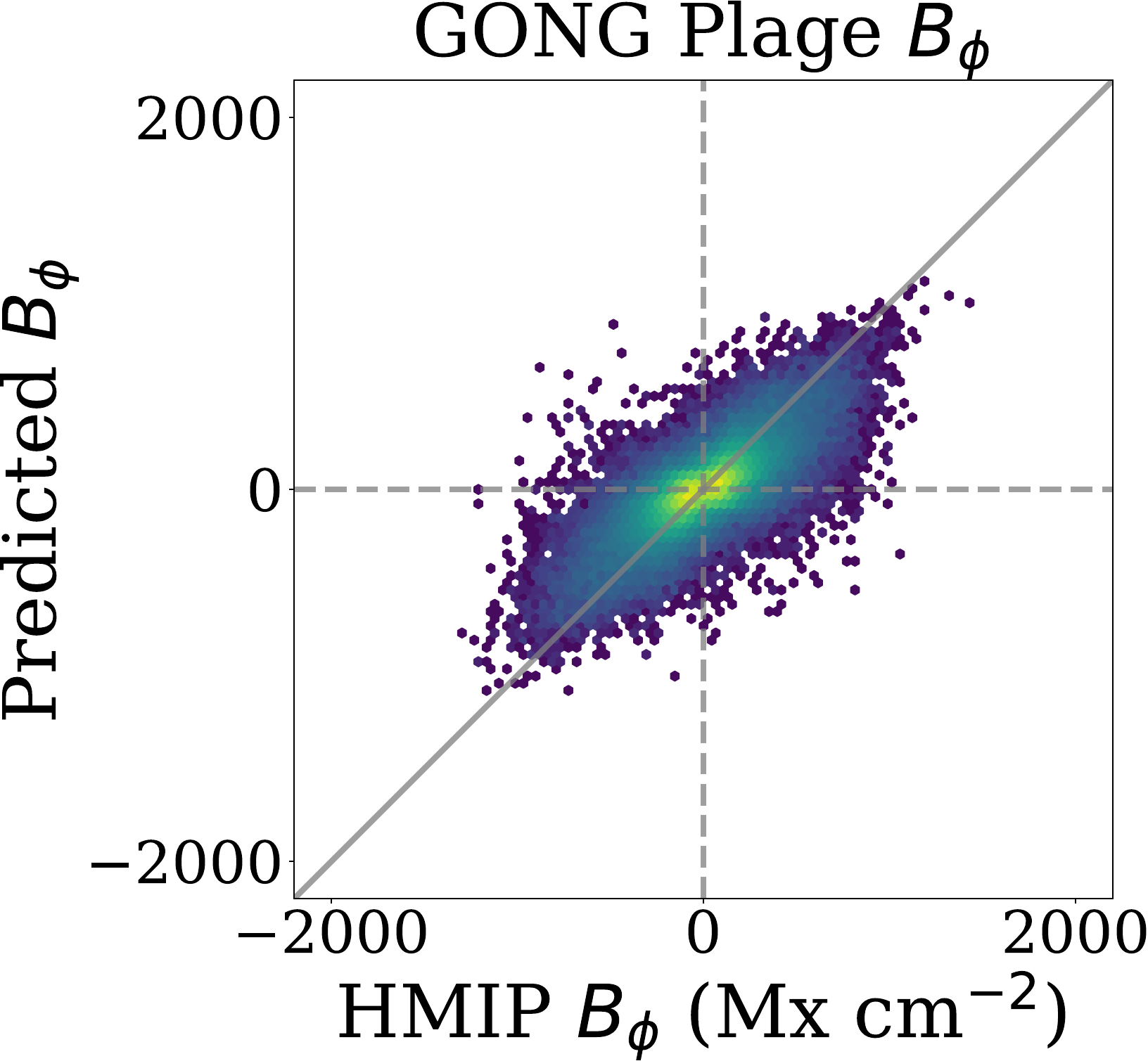} &
    \includegraphics[width=0.24\linewidth]{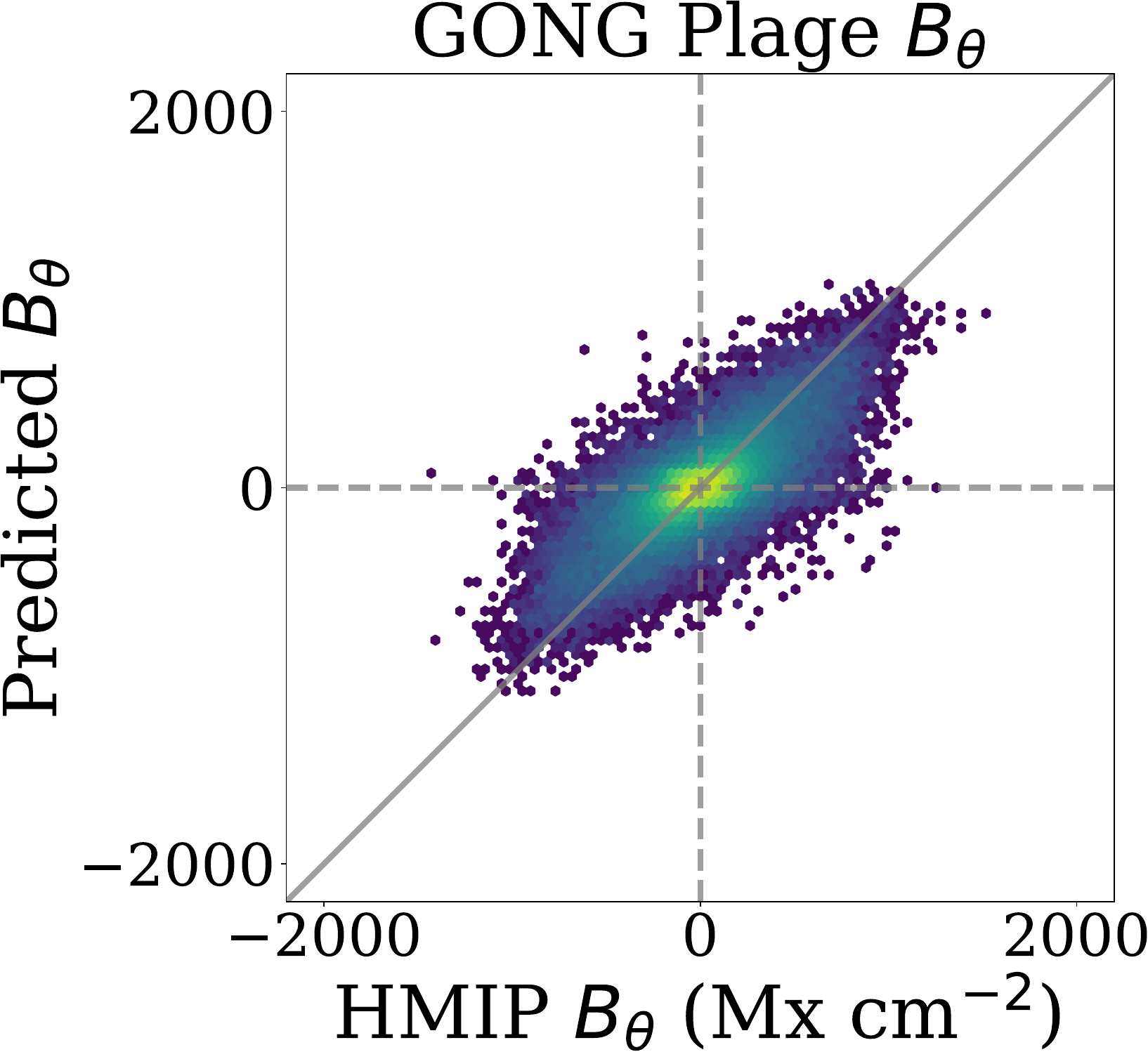} \\
    \end{tabular}
    \caption{{\bf Bivariate log-histograms of \bp and \bt produced by \method with input from \hmi (top row) and \gong (bottom row) \blos.} The method largely does a good job with relatively few polarity confusions. Colormap: empty bins are colored white; non-empty bins follow lowest \includegraphics[width=20pt,height=6pt]{viridis.png} highest density. Following Figure~\ref{fig:bivariateBr}, each condition (GONG AR/HMI Plage) is colormapped identically.}
    \label{fig:bivariabeBpt}
\end{figure*}

\begin{figure*}
\centerline{\includegraphics[width=0.4\textwidth,clip, trim = 10mm 10mm 24mm 24mm]{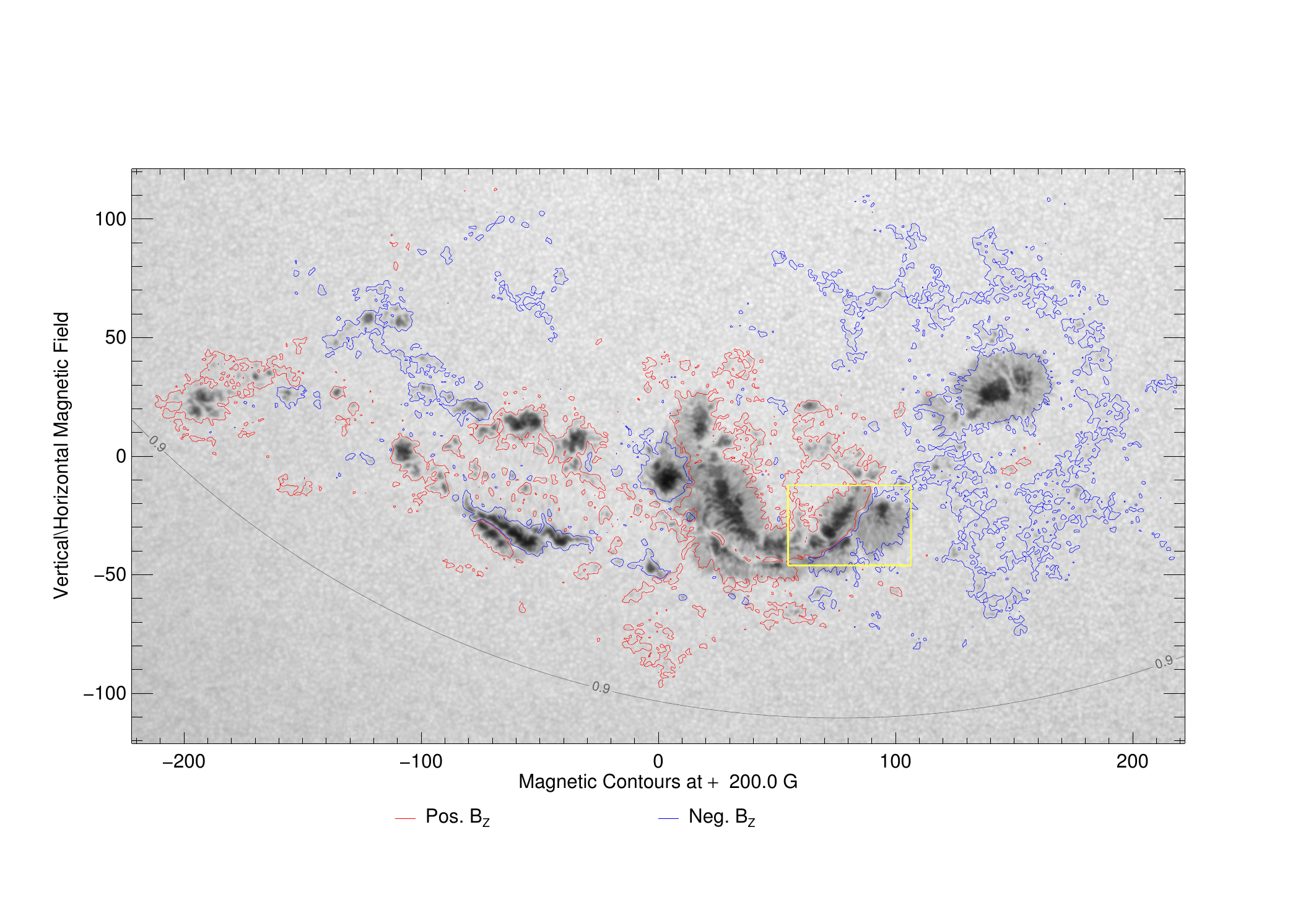}
\includegraphics[width=0.3\textwidth,clip, trim = 16mm 4mm 20mm 26mm]
{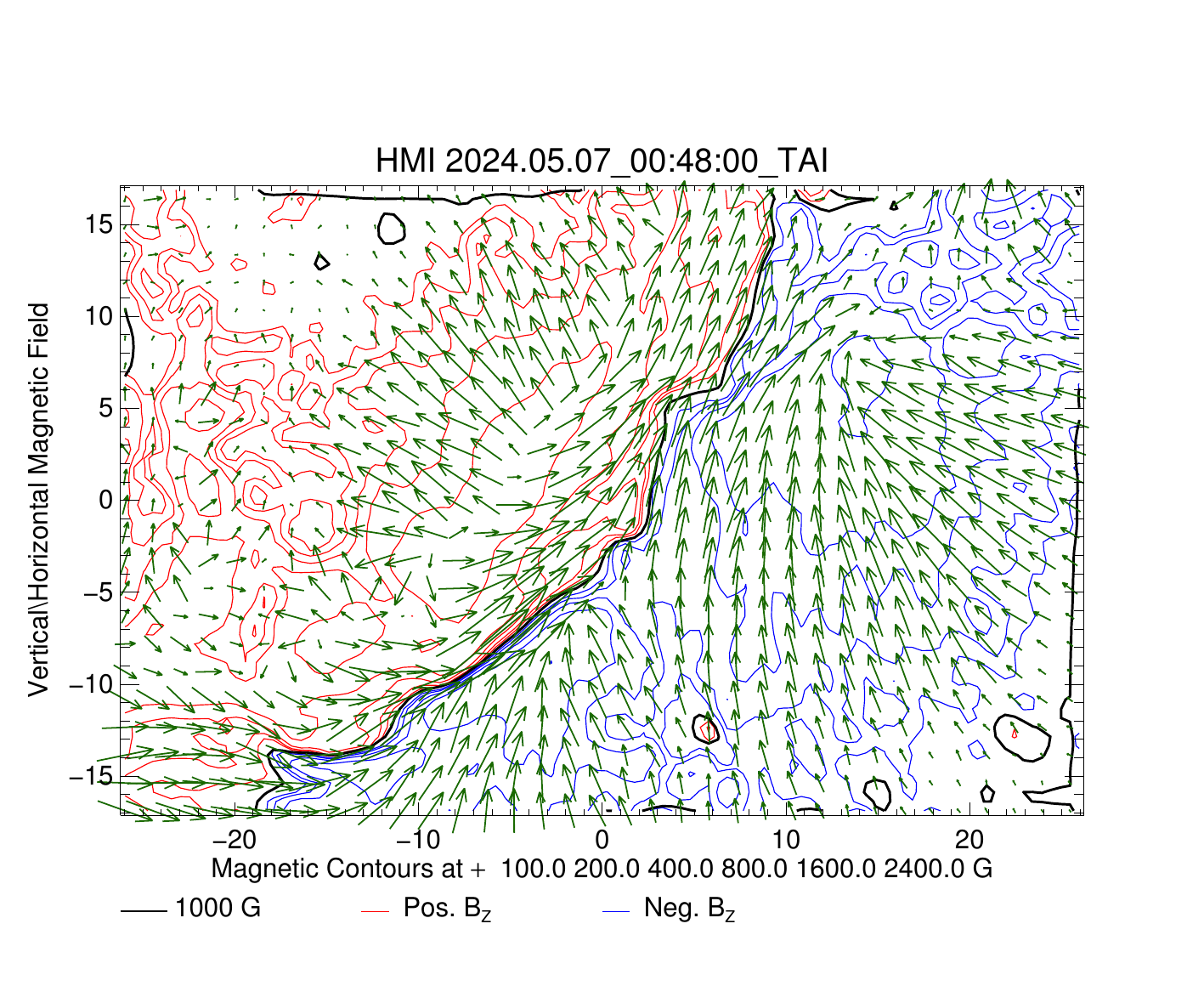}
\includegraphics[width=0.3\textwidth,clip, trim = 16mm 4mm 20mm 26mm]
{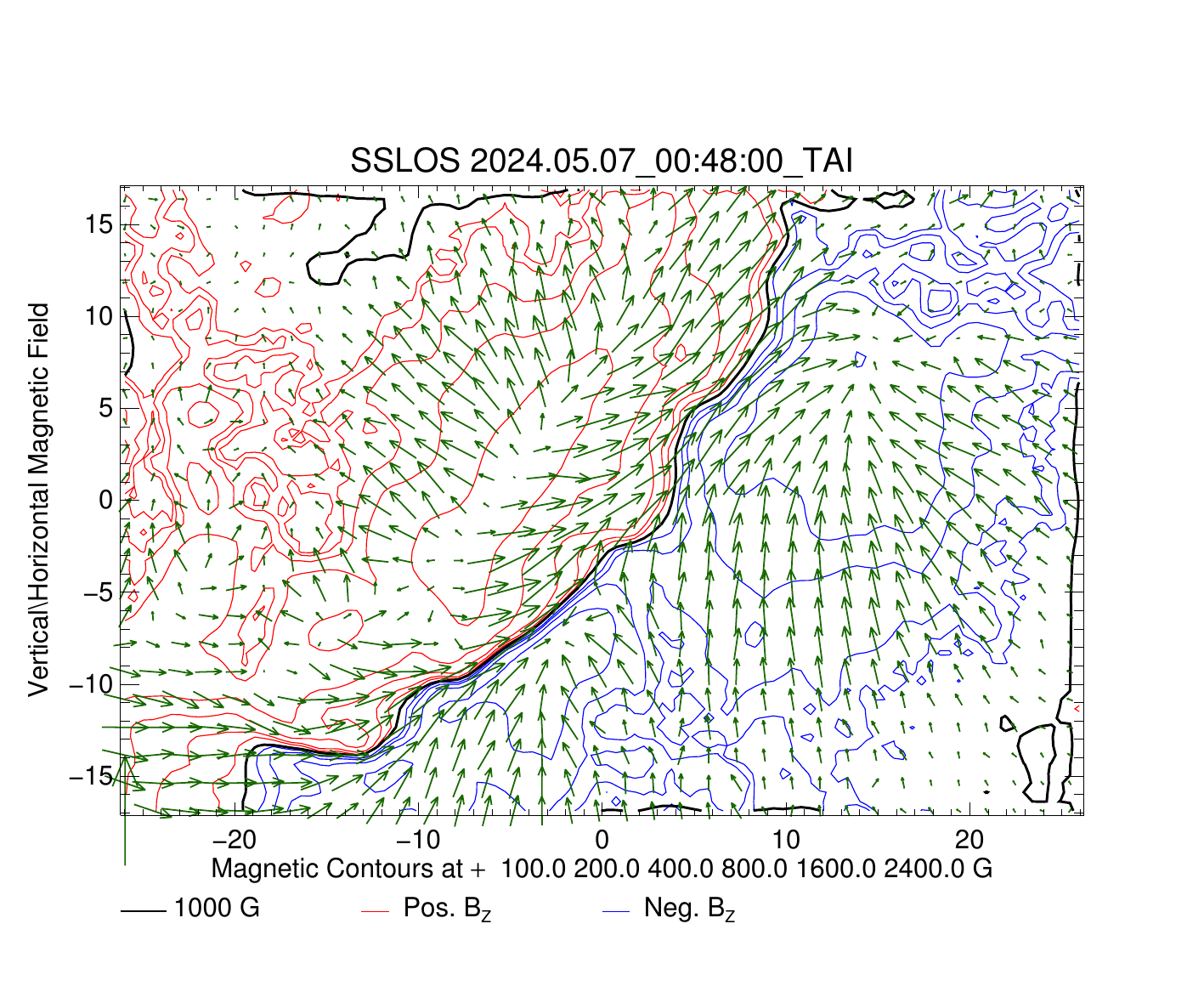}
}
\centerline{
\includegraphics[width=0.4\textwidth,clip, trim = 10mm 20mm 24mm 30mm]{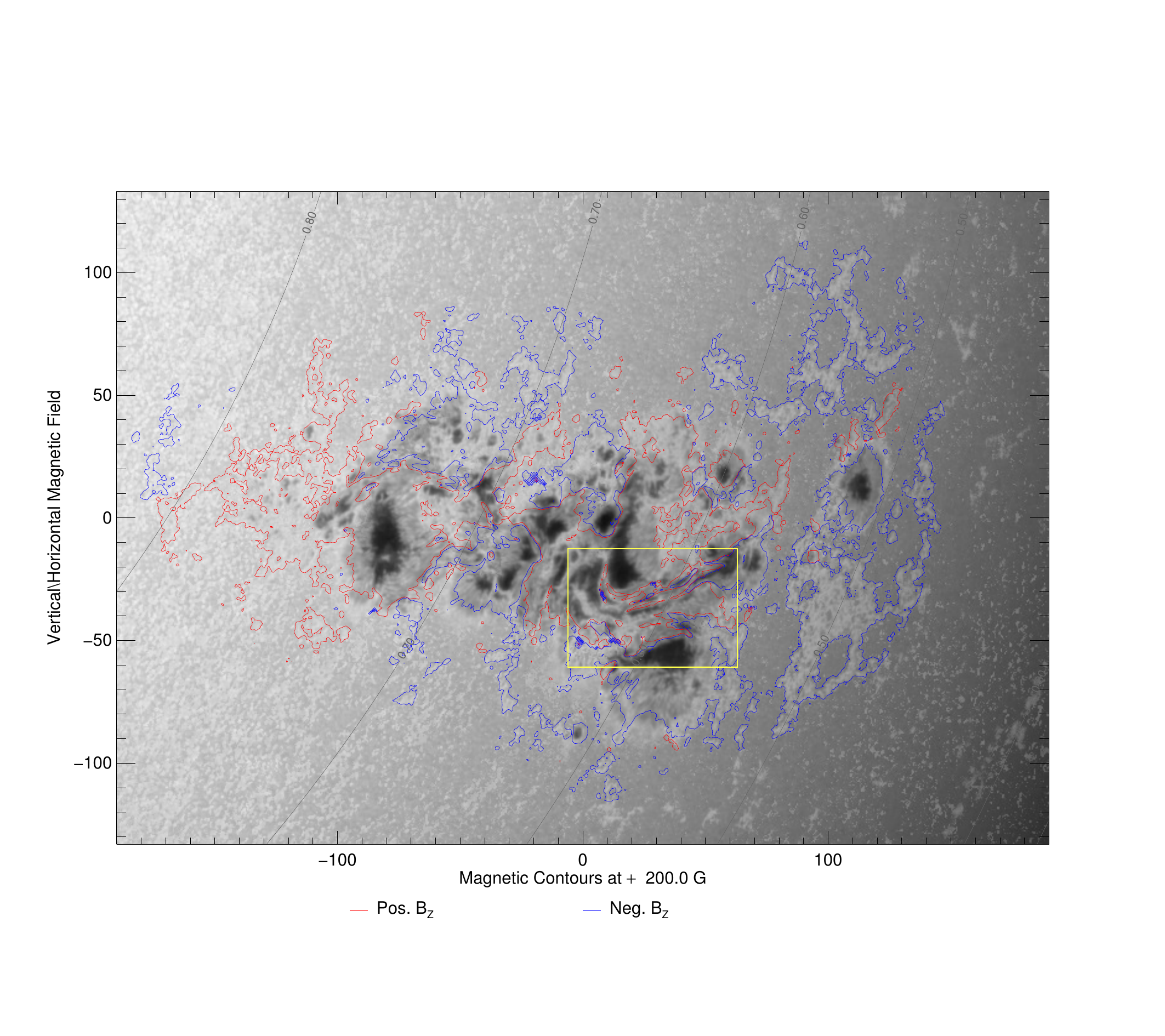}
\includegraphics[width=0.30\textwidth,clip, trim = 12mm -16mm 12mm 28mm]
{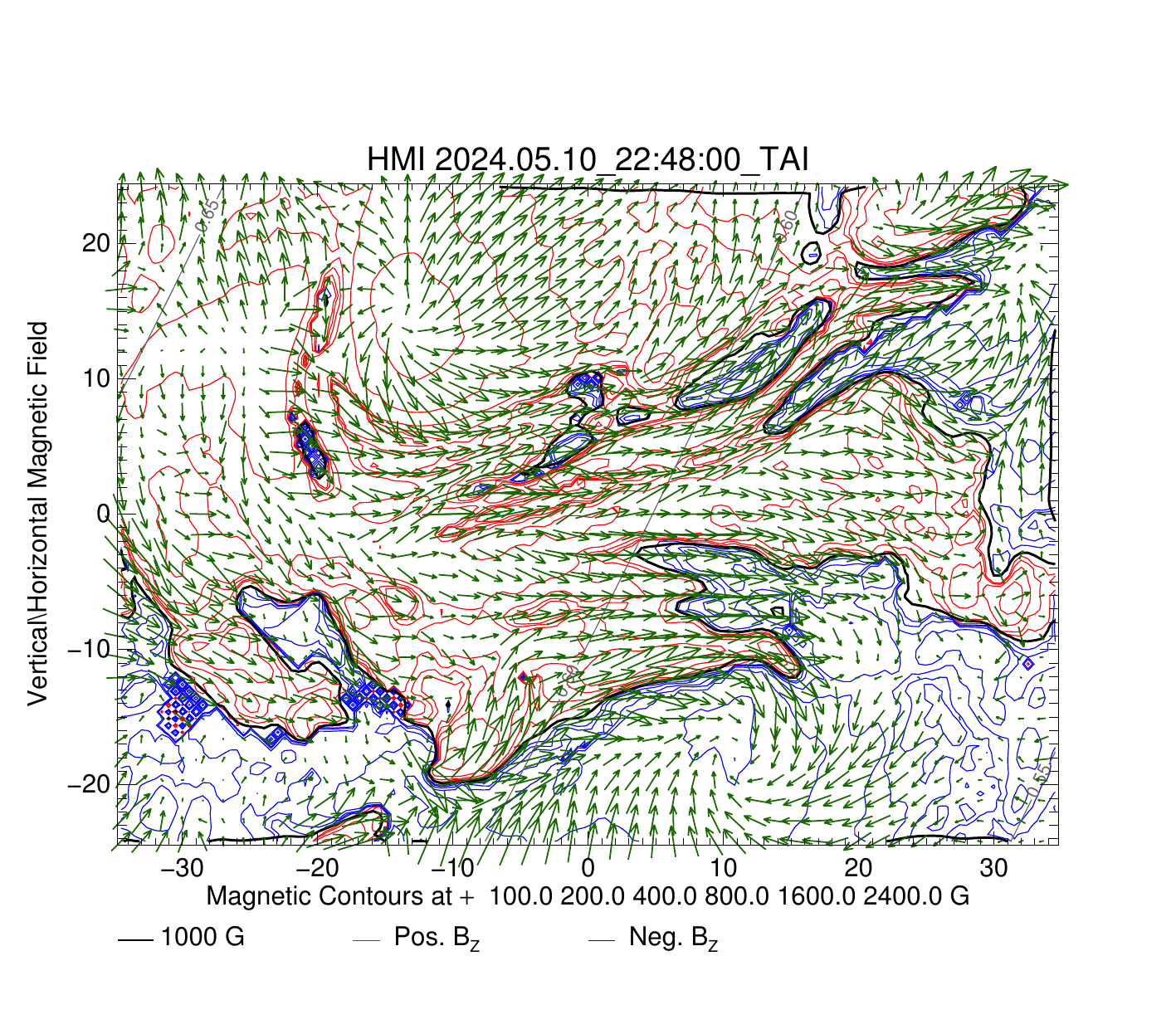}
\includegraphics[width=0.30\textwidth,clip, trim = 12mm -16mm 12mm 28mm]
{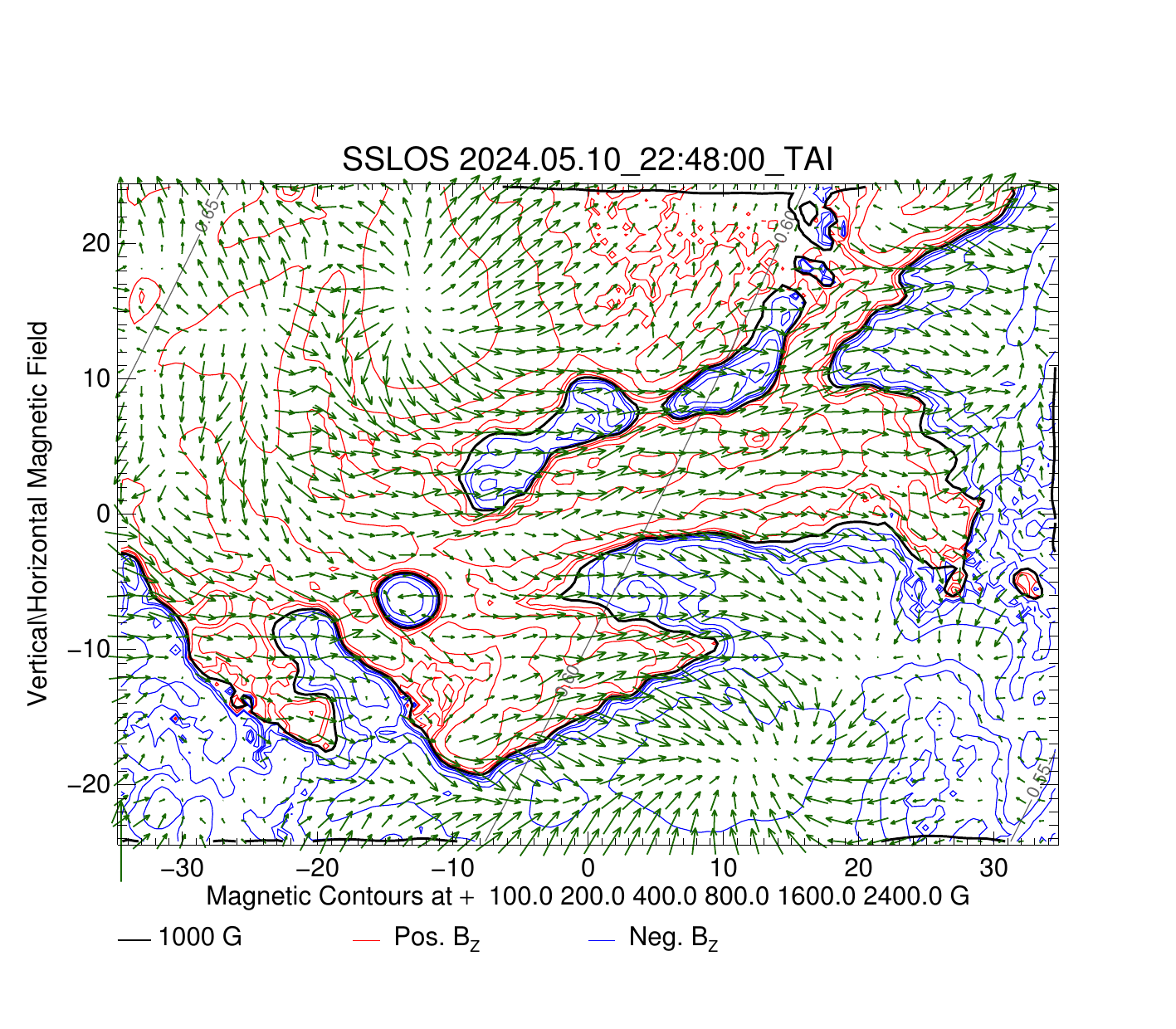}
}
\caption{{\bf Full vector magnetic field for the May 2024 storm-producing active regions}.  Top:  2024 May 07, and Bottom: 2024 May 10; Left: context figure showing NOAA ARs\,13664 and 13668, with isocontours of $B_z$ \textcolor{blue}{\bf blue (negative)} and \textcolor{red}{\bf red (positive)} over a continuum image (which includes limb-darkening), and contour(s) of $\mu$.
A small region of interest (ROI) is outlined in \textcolor{yellow}{\bf yellow}.  Middle: the full vector field
from HMI for the ROI, same format as Figure~\ref{fig:vector_intro} except that for clarity, we do 
not include the underlying continuum image.  Right: same, for \method.
 On May 07 (top), \method's reconstruction is largely correct, and recovers the general non-potential field structures. On May 10 (bottom), while \method does indicate the presence of complex magnetic structures, it underestimates the field magnitude as well as the degree of shear (deviation from potential). On the other hand, \method avoids the HMI disambiguation ``checkerboard'' artifacts visible in the bottom left of the \hmip data.}
\label{fig:brptfull}
\end{figure*}

Compared to $\ell_1$ regression, \method produces similar results but preserves total flux density far better, as shown in Tables~\ref{tab:hmi}, ~\ref{tab:totalFlux}, and Figure~\ref{fig:totalflux}. When examining \br in regions with well-defined signal (e.g., active region, plage), both approaches show similar performance. However, the $\ell_1$ regression handles regions with low polarization signal poorly. Flux density is dramatically underestimated as shown by regions with zero field, and there is a sudden transition in the output along with a clear ``boxing'' effect. The shape of the boxes resembles the isocontours of the disambiguation confidence masks used to determine whether full disambiguation is performed or not. We therefore hypothesize that the method has learned to identify which regions are likely to have been handled with the different disambiguation options.

Finally, we touch on how much information is lost by using \blos as opposed to Stokes data, evaluating on $||\BB||$.
As a reference point, we use the earlier Fast and Accurate Emulation paper by~\cite{higgins2021fast}, which uses \hmip as its ground-truth. Our full method trains on $>4000$ observations compared to \cite{higgins2021fast}'s ${\approx}180$ and so a direct comparison with a smaller scale model that saw ${\approx}300$ observations (documented in Appendix~\ref{sec:app_scale}) is more appropriate. In active regions, plage, and all pixels, the small-scale model obtains MAEs for $B_R$ of $147$, $38$, and $19$ \gauss respectively. In comparison, \cite{higgins2021fast} reports MAEs in the same populations of $108$, $19$, and $10$ \gauss, despite being trained half the data that spanned a smaller period. While the error rate of the proposed method is substantially better than the $\mu$-correction, there is clearly a considerable loss of information.  As such, we present \method as potentially an effective way to supplement existing missions, but not a substitute for acquiring a true vector magnetogram from Stokes spectropolarimetry.

\subsection{Recovering \hmip \br from \gong}
\label{sec:Q_gong}

Having demonstrated the ability to recover the \hmi radial field from its own \blos data, we turn to estimating \br (remapped from \hmi to \gong) from \gong \blos data. We compare the results with the \blos / $\mu$ estimate from \gong, including a variant that fits an additional scaling factor (to handle any instrument calibration). We consider \hmi \br, remapped to the \gong grid, as ground-truth and show results in 
in Table~\ref{tab:gong}. \method produces substantially better results than the $\mu$-correction, although unsurprisingly the overall performance is lower than earlier which relied on solely \hmi data, and is 
likely due to the ${\approx}4\times$ smaller grid, the ground-based \gong observatories, and stark differences in instrumentation. Together, these leave many \hmi features unresolved but (see Figure~\ref{fig:gong}), the false polarity inversion lines in \gong magnetograms are corrected by \method.

\subsection{Recovering the Full Vector Magnetic Field}
\label{sec:Q_fullfield}

\begin{deluxetable}{ccccccc}[t]
\caption{{\bf Evaluation of the full vector field 
from \blos}. We report errors for active regions and plage for \br, \bp, and \bt for \hmi and \gong. The proposed method is able to do a good job of recovering the \hmip vector field using \blos. Since \br, \bp, and \bt differ in distribution of values, one should not draw conclusions from the difference in their errors. Bivariate log-histograms appear in the Figure~\ref{fig:bivariabeBpt}.}
\label{tab:brpt}
\tablehead{
 & \multicolumn{3}{c}{Active Regions} & \multicolumn{3}{c}{Plage Regions} \\
 & MAE & ME & \%$<200$G & MAE & ME & \%$<200$G 
}
\startdata
 & \multicolumn{6}{c}{From \hmi \blos} \\ \midrule
\br  & 123 & 2.3 & 84.1 & 55 & -0.3 & 95.2\\
\bp & 133 & 2.0 & 81.0 & 48 & -0.5 & 97.7\\
\bt  & 152 & -5.9 & 75.8 & 65 & 0.2 & 94.9\\ \midrule
 & \multicolumn{6}{c}{From \gong \blos} \\ \midrule
\br  & 320 & 17.3 & 44.7 & 150 & -3.6 & 73.1\\
\bp  & 235 & 20.0 & 53.2 & 64 & -0.5 & 97.1\\
\bt  & 247 & -5.8 & 49.2 & 72 & 4.0 & 96.1\\
\bottomrule
\enddata
\end{deluxetable}

For the $\mu$-correction, it is assumed that there is no field perpendicular to the radial component. \method can estimate these components. 
Quantitative statistical results of the vector-field reconstruction are presented in in Table~\ref{tab:brpt}, per-component visualizations in Figure\,\ref{fig:teaser} (and Figure\,\ref{fig:gongStormQualitative} as part of Appendix\,\ref{sec:app_additional}, statistical properties in Figure\,\ref{fig:bivariabeBpt}, then vector representations of sample results in Figures~\ref{fig:vector_intro}, \ref{fig:brptfull}, \ref{fig:gongHinode}, and \ref{fig:ivm}.
Since \blos $/\mu$ {\it defines} the field as being purely radial (and thus its estimates of \bp and \bt are zero), the results from that approach for these particular experiments are not included.

When \method is tasked with recovering the vector field using \hmip \blos as input, the full vector field is well-reproduced for the most part.  The statistical differences from \hmip as ground-truth are easily within the overall uncertainties of \hmip data, and visual inspection confirms that \method's vector fields look reasonable especially for the simpler active regions (Fig.\,\ref{fig:vector_intro}).  In highly complex regions (e.g., NOAA AR\,13664, May 2024, Fig.\,\ref{fig:brptfull} ) we find that \method can reproduce not just simple vector fields but strongly-sheared polarity inversion lines, ``twist'', and other indicators of non-potential magnetic structures.  However, this is not always the case: as AR\,13664 approaches the solar limb and the fields become even more complex, \method still gives strong indication of highly complex magnetic fields with strong-gradient PILs and sheared horizontal field.  However, some of the extreme complexity in the \br is lost, the magnitude of the horizontal field ($\sqrt{B_{\phi}^2+B_{\theta}^2}$\,) is underestimated in places, and the (almost unprecedented degree of) shear and non-potentiality is not fully replicated.  Reassuringly, the uncertainties provided by \method do reflect these challenges, especially along the strongly-sheared PILs.  On the other hand, the well-known challenge of \hmip for near-limb vector field disambiguation \citep{Hoeksema2014} is 
present in this example as the distinct ``checkerboard'' pattern in the \hmip \br (Fig.\,\ref{fig:brptfull}, near $[x,y] = [-30,-14], [-16,-14]$ and $[-20,+5]$ ), and such artifacts are not present in the \method magnetogram.  These areas of disagreement, where \method in fact provides an improved solution by way of reduced artifacts, leads in part to the small population of points following the \mbox{$x = -y$} lines in Figure\,\ref{fig:bivariabeBpt} (top row). 

When the vector field is reconstructed by \method using \gong data as input, the results are still good but with additional weaknesses.  The statistical errors (using \hmip, binned to \gong $2.5^{\prime\prime}$ resolution, as ground truth) are larger, as expected (see discussion above) given the much lower spatial resolution and the \gong ground-based facilities.  When compared to the \method output using \hmip as input (Fig.\, \ref{fig:vector_intro} and \ref{fig:brptfull}) one can easily see the degradation.  Additional results from \method using \gong are discussed in Sec.\,\ref{sec:historic}, below.

\subsection{Applications to Historic Data}
\label{sec:historic}

Here,  \method is demonstrated on \gong data that predates the {\it Solar Dynamics Observatory} mission. We see this demonstration as important for two reasons. First, testing on past data serves as a proof of concept to obtaining quantitative vector results over a far longer period of time than was previously possible. Second, demonstrating outputs on data that could not {\it possibly} have been included in the training set is a strong demonstration of the generalization capabilities of the model.

\begin{figure*}[!t]
    \centering
    \begin{tabular}{cc}
    \includegraphics[trim=0.5in 0.0in 1.25in 0.5in,clip,width=0.48\linewidth]{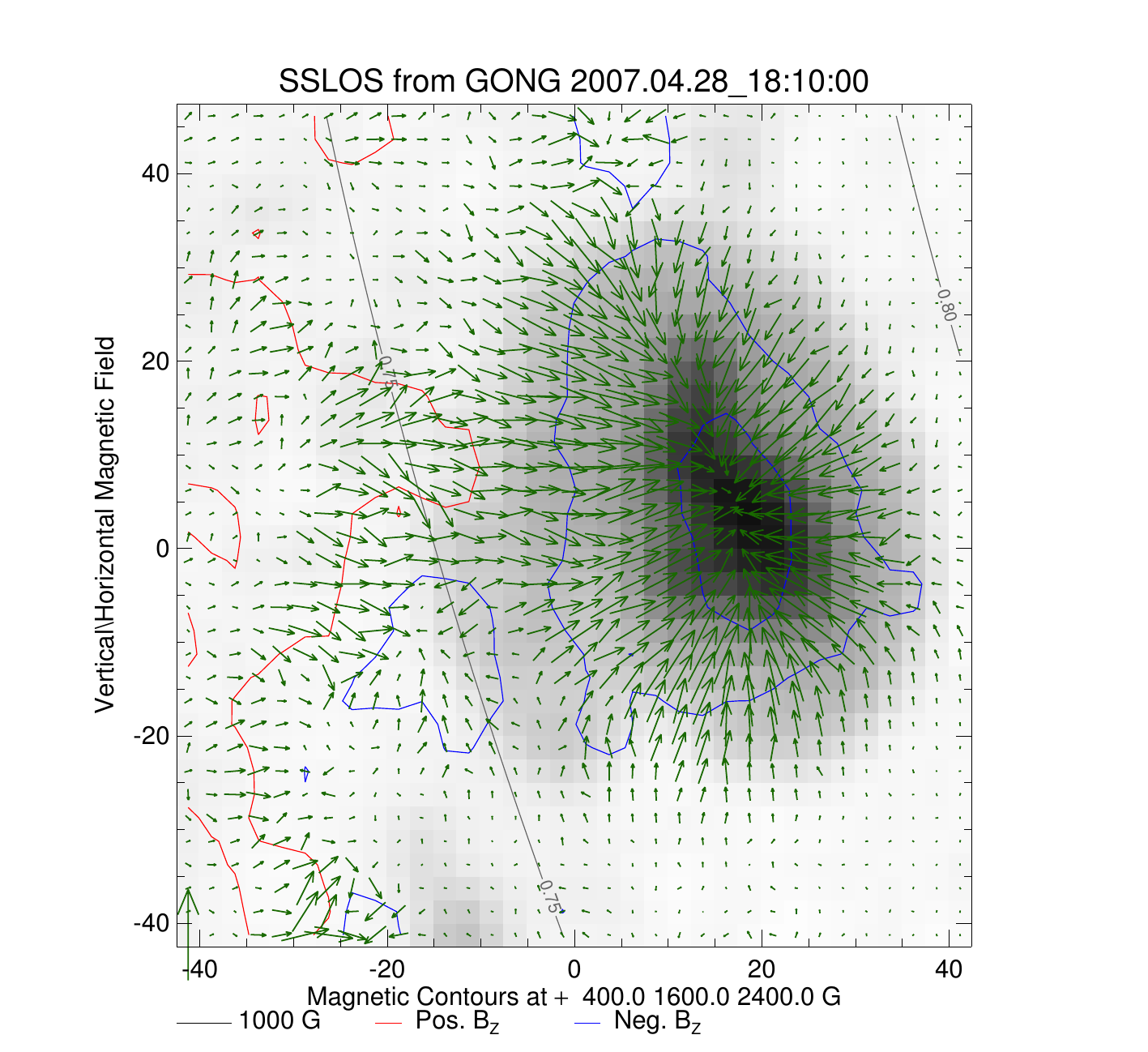}  & 
    \includegraphics[trim=0.5in 0.0in 1.25in 0.5in, clip,width=0.48\linewidth]{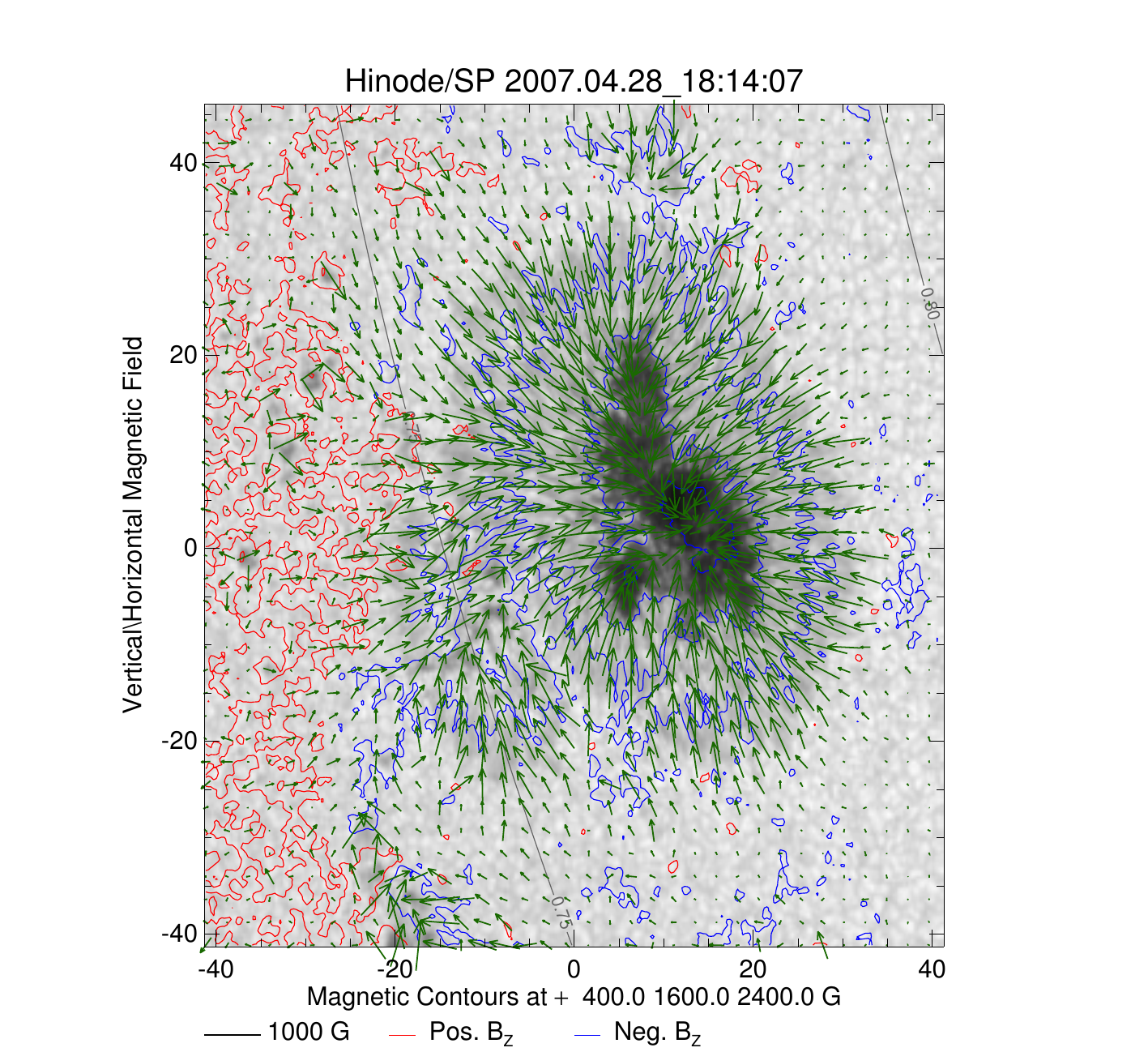} 
    \end{tabular}
    \caption{{\bf Comparison of \method with data from the {\it Hinode} Spectropolarimeter}, 28 April 2007 targeting NOAA AR\,10953. The image is continuum intensity from the respective sources, same format as Figure\,\ref{fig:vector_intro}. The {\it Hinode}/SP vectors are plotted every $8^{\rm th}$ pixel, and the \br and $\mu$ contours are as labeled.  \method output is substantially smoother since its inference is on an $8\times$ coarser grid from GONG, thus detailed structures visible in {\it Hinode}/SP are unresolved. Nonetheless, \method recovers the overall structure well.}
    \label{fig:gongHinode}
\end{figure*}

The first example uses a \gong network-merged magnetogram from 2007, near the start of the {\it Hinode} mission~\cite{kosugi2007hinode}, and targeting NOAA AR\,10953 (``the Japan sunspot'') on 28 April. In Figure~\ref{fig:gongHinode}, \method using this \gong input is compared with a Level 2.1 vector magnetogram from the {\it Hinode} Solar Optical Telescope-Spectro-Polarimeter~\cite[SOT-SP]{tsuneta2008solar}. \method recovers the general structure correctly, albeit at a significant reduction in spatial resolution. There are clearly some differences, including again a lack of magnetic shear (in the small negative-polarity area to the south/east of the large sunspot).  However, given the disparity of the information sources (SOT-SP being a high-resolution space-based vector polarimeter with far higher spectral information as compared to \gong, the \method output's overall structure is generally correct. 

A second example targets an active region from a far earlier time that produced large X-class flares, specifically NOAA AR~10486 (source of the Halloween Solar Storms of 2003). This epoch predates the site-merging data products from \gong, so we use site-specific data from Mauna Loa ({\tt mlbzi}), Big Bear ({\tt bbbzi}), Udaipur ({\tt udbzi}), and Learmonth ({\tt lebzi}). These data products are reported in m s$^{-1}$, which we convert to \gauss using $0.352$ \gauss m$^{-1}$\,s following~\cite{Sudol_2005}. There are differences in sign conventions between sites, which we fix manually; saturation effects in sunspot centers in the \blos data were also ``repaired'' ({\it c.f.} Section~\ref{sec:app_continuumfix}).

\begin{figure*}[!htp]
    \centering
    \begin{tabular}{cc}
    \includegraphics[trim=0.25in 0.0in 0.5in 0.25in, clip,width=0.47\linewidth]{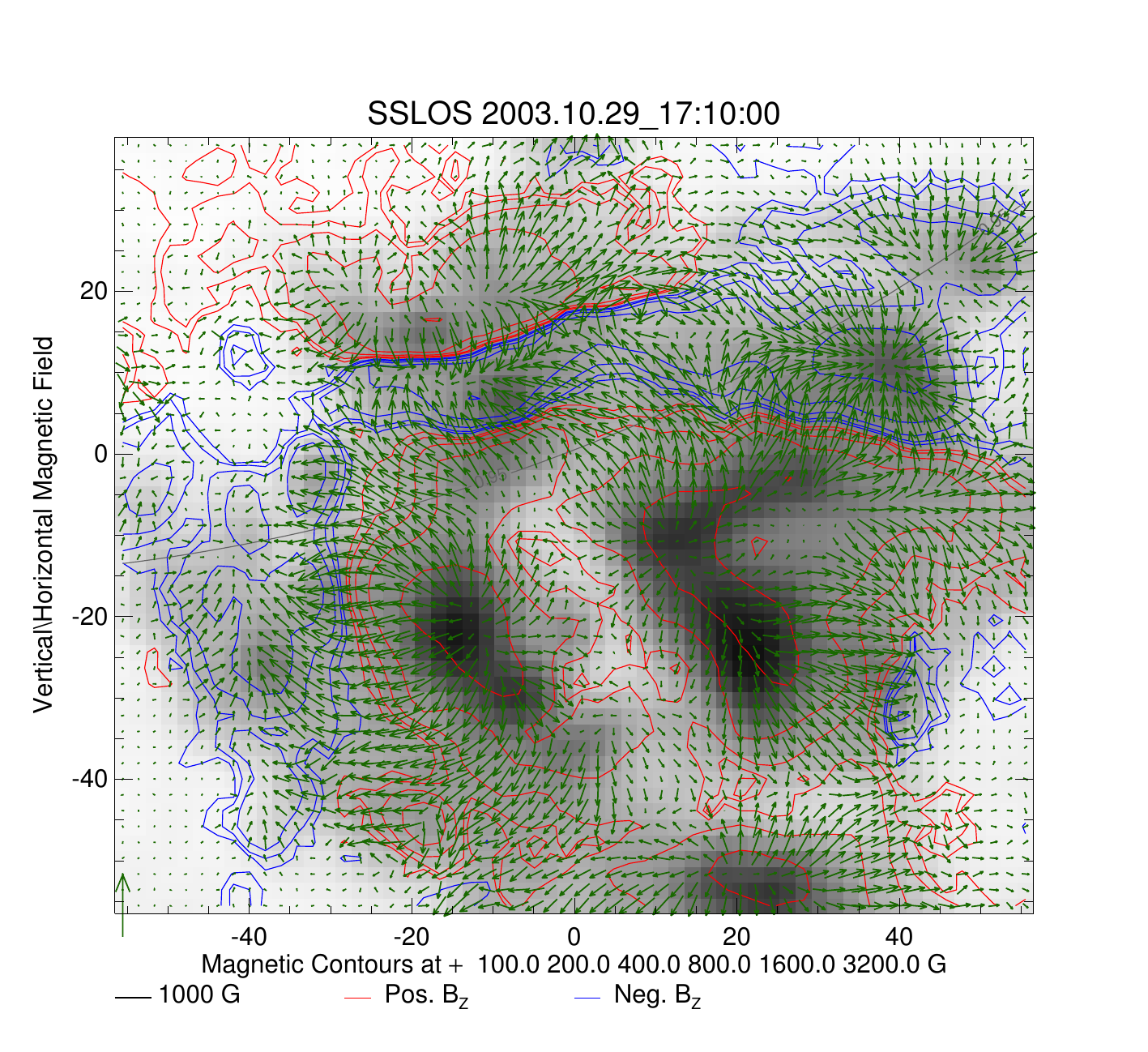}  & 
    \includegraphics[trim=0.25in 0.0in 0.5in 0.25in,clip,width=0.47\linewidth]{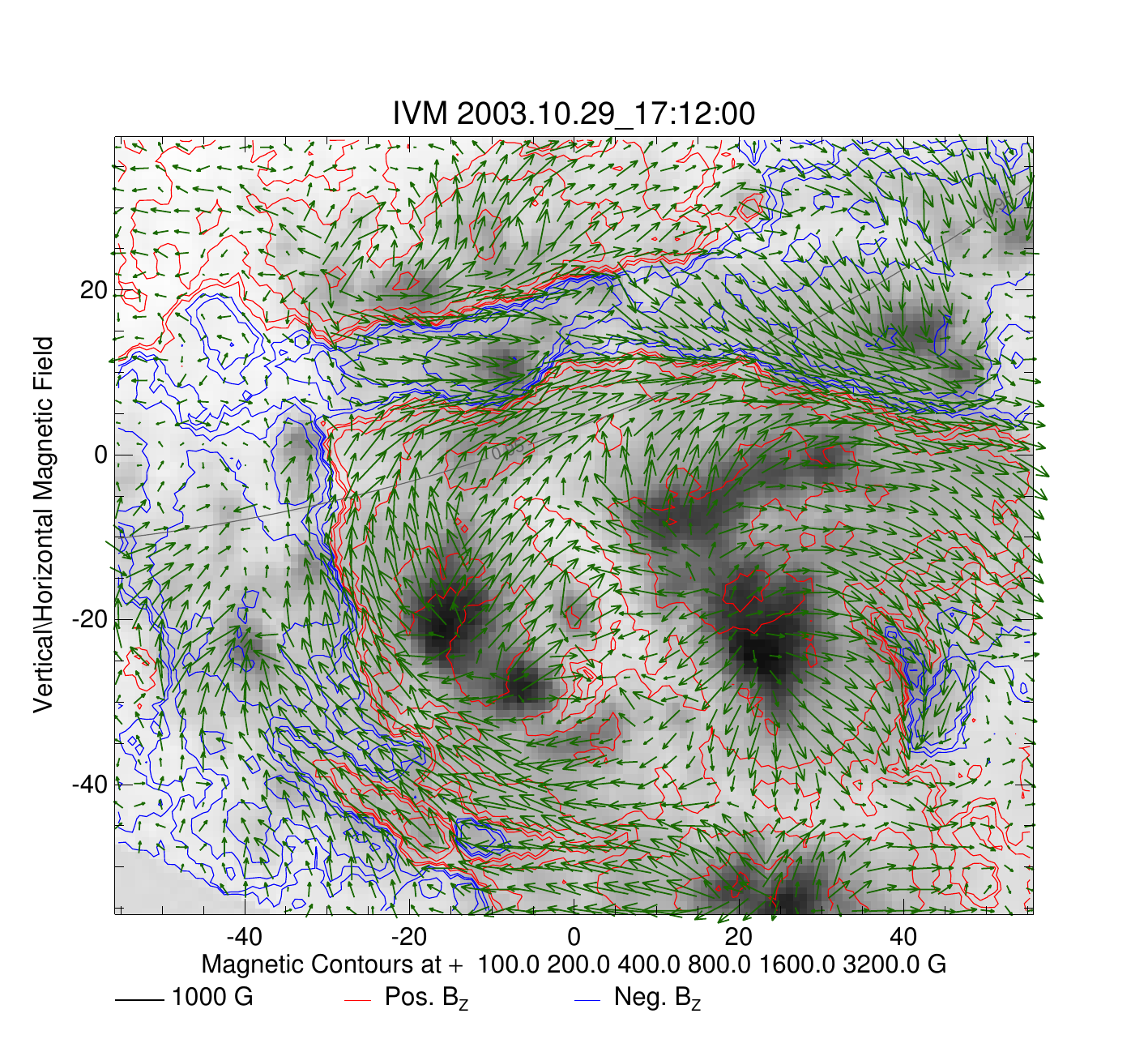} 
    \end{tabular}
    \caption{{\bf Comparison of \method with data from the U. Hawai`i Imaging Vector Magnetograph for NOAA AR\,10486 on 2003 October 29}, a few hours before the X10 flare (same format as Figure\,\ref{fig:vector_intro}).  Only the detail around the strongly sheared PILs is shown, with continuum
    image from each of the sources; for \method, every vector is plotted, for the IVM, every $3^{rd}$ vector is shown.  While \method does match the IVM \br well and there is good indication of complex field structure, there is less shear overall in the \method horizontal field, and in some areas it is in the incorrect direction. }
    \label{fig:ivm}
\end{figure*}

Vector magnetic field data from this time are of limited availability; we show a comparison between \method and a vector magnetogram from the U. Hawai`i / Mees Observatory Imaging Vector Magnetograph~\citep{mickey1996imaging} in Figure~\ref{fig:ivm}. \method produces reasonable global structures, and does an overall good job on estimating \br including very strong-gradient PILs. However, it underestimates the amount of shear at one of the polarity inversion lines, meaning that the vector field appears less twisted and more ``potential'' than the IVM vector field show, and in some sections of the PILs the direction of shear is opposite that seen in the IVM.  This particular IVM magnetogram is part of the ``quicklook'' dataset \citep{LekaBarnes2007} and may suffer from some  bias itself.  However, this active region produced some of the largest flares of solar cycle 23, and as such, indications of extreme shear and non-potential field structures would be expected.  A full statistical comparison between \method and all available IVM data is beyond the scope of this paper, but it is clear that \method provides additional information than available from just the \blos from \gong.  A video of a time-series of \method for AR\,10486 is available as supplemental material.

\subsection{Failure Modes}
\label{sec:failure_modes}

As with all approaches for generating of vector magnetograms, including optimization-based inversions from full-Stokes spectropolarimetry, our method is not perfect (and in fact relies on polarimetric data {\it ab initio}). Documenting known failure modes is important since different use-case scenarios will have different performance requirements.

One prominent short-coming is that the field is not as accurately recovered compared to methods using full Stokes data.
While the overall structure is generally reasonable, sometimes \method mis-estimates the magnitude and azimuthal direction of the horizontal field, tending toward an orientation that is more consistent with a simple or ``potential''-field configuration.  This effect is most noticeable in complex active regions.  The impact of this failure mode is an underestimation of properties such as apparent  ``twist'' in sunspots, and shear along polarity inversion lines.  We hypothesize that this failure mode is likely due to the relative infrequency of these sorts of structures available for training the model: sunspot area occupies at most only a few percent of the solar disk at any given time, and complex non-potential structures within active regions occupy only a few percent of all active regions. The post-processing smoothing may be exacerbating the problem.  \method's performance may be improved with more training, but the rarity of strongly-sheared non-potential magnetic field morphologies will continue to be a challenge.

Another failure mode is \method's tendency to flip-flop between polarities (in \br) or direction (in \bp and \bt) when it is uncertain, producing speckles of incorrect vectors that cannot be fixed without substantially heavier smoothing (which would negatively impact other regions). This effect is often seen in complex active regions (e.g., between points E and F in Figures~\ref{fig:uncertainty}). The origin is not the dithering operation, but rather the original network's estimate (e.g., as seen in the small tendrils of orange Figure~\ref{fig:ablation} in the ``No Logit Dither'' panel). This uncertainty was not seen in the output of neural networks for Stokes inversion and disambiguation
since the distribution of vector field conditioned on the full polarization state is substantially less ambiguous compared to conditioned on \blos, however it is reminiscent of the ``checkerboard'' outcome from the {\tt ME0} disambiguation when it is unable to find a global minimum \citep[see][Figure 15]{Hoeksema2014}. Currently, this failure mode is ameliorated (but not fully addressed) by a post-processing step that tries to ensure smoothness (again, similar to {\tt ME0}).  There may be alternate approaches {\it e.g.}, matching gradients \citep[as by][]{jungbluth2019single} or using an adversarial approach \citep[see][]{kim2019solar} which may produce slightly smoother results; the current approach, on the other hand, has the advantage of providing a way to identify uncertainty (via examining the estimated distribution over angles and norms). Formulating the problem as a conditional diffusion \citep[e.g.,][]{ramunno2024solar} may provide an alternate solution but at the cost of requiring multiple samples. in order to obtain uncertainty 

\section{Discussion}
\label{sec:discussion}

We have introduced a learning-based approach for recovering the vector magnetic field in their heliographic (physical) components simply from line of sight magnetograms. Our results demonstrate that the proposed method, \method, can often produce accurate results from \hmi and \gong input data. We have shown that \method can work on data predating the SDO era and produce vector magnetograms that are in agreement with those produced by other instruments such as the {\it Hinode}/SOT-SP and the {\it IVM}. Results in the Appendix suggest that \method has reasonable systematics including good (no worse than \hmi) center-to-limb bias in \br in both active regions and quiet Sun (Appendix~\ref{sec:app_CTL}). Temporally, \method does not exacerbate (but also does not fix) daily oscillations, and its average flux statistics match \hmi up to modest calibration constants (Appendix~\ref{sec:app_oscillations}). Moreover, the method provides uncertainties, like~\cite{SuperSynthIA} and thus can identify likely mistaken pixels, just as in Stokes inversion approaches. 

It is important to point out, from the start, that the proposed system is {\bf emphatically} not a replacement for full Stokes-vector capable instruments and ever-improving inversion capability.
Similarly, the recovered fields from \method do not have the precision, accuracy, and quality of ML-based techniques trained with Stokes data as input and similar architectures and losses, such as~\citep{higgins2021fast,Higgins2022}.  The gap in performance, nonetheless, however, understandable since there is vastly less information present in a single estimate of \blos as compared to the full polarization state in multiple passbands. 

We note that several of the issues (e.g., underestimates of shearing in extreme events) are in critical areas where magnetogram data is needed for research and monitoring. A particularly common problem was \method underestimating the magnitude of \bp and \bt. We hypothesize that the relative scarcity of training data for large and complex active regions contributes to these weaknesses, and that the smoothing may play a role.  

However, the combined results suggest that the learning can offer ways of obtaining additional information from archival \blos data, as we 
demonstrated by applying \method to data that predates the {\it Solar Dynamics Observatory} mission.  As such, \method could be a tool to aid efforts aimed at bringing past records to be consistent with current observations in order to better understand long-term solar trends \citep{lee2021generation,van2023probabilistic,dhuri2022deep}. Moreover, while Section~\ref{sec:method} describes a particular recipe that has success, it is likely that there are a variety of other effective models and methods (e.g., using conditional diffusion).

Compared to existing work in the area, our work is distinguished by focusing on the generation of {\it vector} magnetograms as opposed to LOS magnetograms or data products derived from magnetograms. For instance, \cite{dhuri2022deep} predicted SHARPs features~\cite{bobra2014helioseismic} from \blos data. The most similar work to ours is the complementary work of~\cite{jiang2023generating}, which uses H$\alpha$ data from BBSO as an additional signal to constrain the the orientation of the vector field. We see these approaches as complementary: our proposed approach uses the data from one instrument, rather than two, and~\cite{jiang2023generating} uses chromospheric signals to help constrain the vector.  An interesting future direction might be trying to use the additional H$\alpha$ data as an input signal, as well as other measurements in the higher layers of the photosphere (such as from \aia). 

Looking toward the future, machine learning may offer a way to extract additional value from constellations of instruments or produce new instruments. Within one mission, for instance, one could construct a combined instrument package that captures a small number of samples with high-quality full polarization information coupled with the acquisition of a large sample with limited polarization, using ML techniques to enhance the latter such that a large sample reflects the high-quality sample. Across instruments, AI may help provide surrogates to complete historical records across time and space. 

In the end, though, the results underscore the importance of {\it sustained} support for facilities that capture high-quality data of the Sun, especially in the particular case of vector magnetograms. The results presented here are made possible not just by the machine learning, but also by the considerable care that went into designing, calibrating, and maintaining the instruments that captured both the input data and other sources for comparisons. \method was only possible thanks to training on nearly 14 years of high-quality, high-cadence, high-availability, consistent data from the NASA {\it Solar Dynamics Observatory} mission. Our results suggest that this scale was important for producing strong results, as evidence by changes in performance as a function of data scale (Appendix~\ref{sec:app_scale}) as well as the starkest of our failure modes which occur in the presence of rare, highly sheared fields as shown in Figures~\ref{fig:brptfull}, \ref{fig:gongStormQualitative}.  
The solution for data scale in consumer computer vision  applications is often to simply gather more data, often in parallel and far-faster-than-real-time. Unfortunately, this conventional approach does not work for high-quality vector field data, which needs substantial infrastructure investment, and when an event is missed, there is no rewinding time.

\section{Acknowledgments}
\label{sec:acks}

This work was primarily supported by NASA/MIRO  80NSSC24M0174 and NASA/LWSSC grant 80NSSC22K0892. This work was supported in part through the NYU IT High Performance Computing resources, services, and staff expertise. DF thanks the ISSI team ``Quantitative Comparisons of Solar Surface Flux Transport Models'' for conversations that helped shape the framing of the paper.  This work makes use of data from NASA missions (specifically the {\it Solar Dynamics Observatory} and {\it Hinode})and NASA-supported instruments (the Helioseismic and Magnetic Imager and the Solar Optical Telescope, respectively), and we acknowledge the teams and support to make these data available.  This work utilizes GONG data obtained by the NSO Integrated Synoptic Program, managed by the National Solar Observatory, which is operated by the Association of Universities for Research in Astronomy (AURA), Inc. under a cooperative agreement with the National Science Foundation and with a contribution from the National Oceanic and Atmospheric Administration. The GONG network of instruments is hosted by the Big Bear Solar Observatory, High Altitude Observatory, Learmonth Solar Observatory, Udaipur Solar Observatory, Instituto de Astrof\'{\i}sica de Canarias, and Cerro Tololo Inter-American Observatory.

\appendix

Sections \ref{sec:app_trainingdetails}, \ref{sec:app_dither}, \ref{sec:app_continuumfix}, \ref{sec:app_uncertainty} present technical information that is beyond the scope of the main paper. Sections \ref{sec:app_CTL} and \ref{sec:app_oscillations} analyze systematics and trends in \method output that could cause issues for downstream tasks. Section~\ref{sec:app_scale} concludes with an analysis of how performance changes as a function of training data in terms of scale.  Section~\ref{sec:app_additional} shows additional samples of results using different sites for estimating vector fields from \gong data.

\section{Training Details}
\label{sec:app_trainingdetails}

In the main paper and analyses presented in the appendix that address data scale, we attempt to train as identically as possible. While there are some slight differences in parameter counts between models (e.g., with time/without time), these are negligible since they only change the U-Net's first layer filter count.

For \hmi data, we trained for approximately 1.5 million examples (or more accurately, the maximum number of epochs not exceeding 1.5M examples). For the main paper, this was 21 epochs. The learning rate started at $10^{-3}$ and was divided by 10 every 8 epochs. For \gong data, we initialized with the equivalent \hmi model and started training with a learning rate of $10^{-4}$. We trained the model for as many epochs as the \hmi data would have been trained for at the $10^{-4}$ or smaller learning rate -- 13 epochs in the main paper.

\section{Randomized Correction Details}
\label{sec:app_dither}

As described in Section~\ref{sec:method}, one option for producing corrections to the predicted bins is to randomize. Since the spacing between the bins is usually quite small, this randomization does not substantially increase the error (typically a few \gauss) but increases the apparent smoothness of the distribution. 

Since the bins are of differing magnitude and have variable gap, the distribution is per-bin. We define a per-bin distribution as a uniform distribution over the values that are closest to the bin. Specifically, let the sequence of bins be defined as $\{b_1, \ldots, b_K\}$. The distribution for the $i$th bin is a uniform distribution 
from $l = (b_{i} - b_{i-1})/2$ to $u=(b_{i+1} - b_{i})/2$. Thus, the distribution for bin $i$ touches the distributions for bins $i-1$ and $i+1$. For the first and last bin, this range of the distribution is clipped so that $0 \le b_i - l$ and $b_i + u \le 5000$ \gauss. 

The smoothness is illustrated in Section~\ref{sec:app_CTL} while showing distributions of \br as a function of viewing angle.

\section{Continuum Intensity Sunspot Fix Details}
\label{sec:app_continuumfix}

Stokes inversion methods occasionally fail at the center of sunspots where the intensity (photon count) becomes extremely low, and paradoxically produce an estimate of a field of $0$ \gauss. We found this in \gong data from the Halloween storm in 2003 and document the automatic fix that we apply. This is the same fix as used in the code for~\cite{SuperSynthIA} as a post-processing step when running in deployment. Given an intensity image $I$ and magnetogram $B$, we compute: (a) a limb-darkening-corrected continuum image $I_d = I / (1-0.7(1-\mu))$; (b) an estimated-from-field continuum image $I_B = 1.0 - B / 3000$. While the estimated-from-field continuum image is unlikely to precisely the same as the continuum images, pixels that have near-zero field strength but are actually dark (i.e., $I_B$ high and $I_d$ low) are implausible. To identify these, we look at the ratio between $I_B$ and $I_p$. Pixels with a ratio $I_B/I_d$ more than 2.5 over the median (i.e., field strength suggests would be dark, but are actually bright) are flagged and then interpolated from non-flagged pixels with a radial basis function interpolator.

\section{Estimating Uncertainties}
\label{sec:app_uncertainty}

To obtain uncertainty about the norm of the vector, we compute the standard deviation implied by the predicted distribution. We start with logits $\zB_n$ giving unnormalized probabilities over a sequence of bin values $\{b_1, \ldots, b_K\}$ as well as the final prediction from the inference procedure $n$. Then, if $\sigma$ is the softmax operator, $\sigma(\zB_n)$ is a probability distribution over the bin values that sums to unity. The norm uncertainty is then the standard deviation, or the square root of the expected squared distance to the mean, 
\begin{equation}
\left(\sum_{i=1}^K \sigma(\zB_n)[i] (b_i - n)^2\right)^{1/2}.
\end{equation}

Handling angular uncertainty requires a little more care. 
Given a sequence of unit vectors $\xB_1, \ldots, \xB_N$, the unnormalized sample mean is defined as $\muB_x = \frac{1}{N} \sum_{i=1}^N \xB_i$. This mean needs to be renormalized to the unit sphere to produce a direction, but the norm of the mean $||\muB_x||$ itself shows the concentration of the vectors. Note that $||\muB_x|| \le 1$ since each vector is unit norm. For instance, if the sequence of vectors is simply the same vector, $||\muB_x|| = 1$ and if the sequence of vectors is a vector $\xB$ and its opposite $-\xB$, then $||\muB_x|| = 0$. Finally, to obtain a variance (which correlates with dispersion rather than concentration), one subtracts the norm of the mean vector from unity or $1 - ||\frac{1}{N} \sum_{i=1}^N \xB_i||$.

In our case, we start with logits $\zB_a$ representing the un-normalized probabilities over the sequence of angular vectors $\SB_a \in \mathbb{R}^{K \times 3}$. For convenience, $\SB_a$ can be thought of a sequence of vectors $\sB_1, \ldots, \sB_K$. Recall that $\sigma(\zB_a)$ is a vector of the normalized probabilities and sums to unity. We then define the angular uncertainty as the circular variance of the bins under the distribution given by $\sigma(\zB_a)$, or
\begin{equation}
\label{eqn:circular}
 1 - ||\sigma(\zB_a)^\top \SB_a||
\textrm{~~or~~} 
1 - \left\Vert \sum_{j=1}^K \sigma(\zB_a)_j \sB_j\right\Vert
\end{equation} 
which is equivalent to the sample circular variance if one samples an infinite number of samples from the bin vectors according to the probability distribution $\sigma(\zB_n)$.

Finally, to produce per-component uncertainty, we compute a straightforward standard deviation. Note that the probability of the vector $b_i \sB_j$ (i.e., the $i$th norm and $j$th unit vector) is given by $\sigma(\zB_n)_i \sigma(\zB_a)_j$. If we let the prediction be $\pB$, then the standard deviation for a component $c$ is the square root of the sum of squared distances to the prediction's component, or: 
\begin{equation}
\label{eqn:componentStd}
\left( \sum_{i=1}^{K} \sum_{j=1}^K \sigma(\zB_n)_i \sigma(\zB_a)_j ((b_i \sB_j)[c] - \pB[c])^2 \right)^{1/2},
\end{equation}
which follows from the definition of the variance.

The \method's distribution is a distribution over the orientation of the vector and not just a scalar uncertainty around a best-prediction. Accordingly, we can examine the model's beliefs at varying points. We showed this in Figure~\ref{fig:uncertainty}.
For completeness, we also explain how we produced these distributions. We start with the non-uniformly distributed lattice points $\sB_1, \ldots, \sB_K$ and probabilities $\sigma(\zB_a)$ over the lattice. To visualize the distribution, we interpolate these probabilities onto the full sphere. Specifically, the probability density given at new coordinate is determined by  radial basis interpolation, using the default parameters for Scipy's {\tt RBFInterpolator} (specifically a thin plate spline). Probabilities are, by definition, strictly non-negative, and we found that interpolating in the original space produced ringing near positive bumps that led to negative probabilities. Accordingly, we interpolate the square roots of the probabilities and then square the results (which forces the data to be positive). Given the interpolation, the qualitative trends of the distributions are preserved, but one should not draw conclusions based on precise details or values. For instance, the smoothness is due to the RBF interpolation.

\begin{figure}[t]
\centering
\begin{tabular}{@{}ccc@{}}
\includegraphics[width=0.318\linewidth]{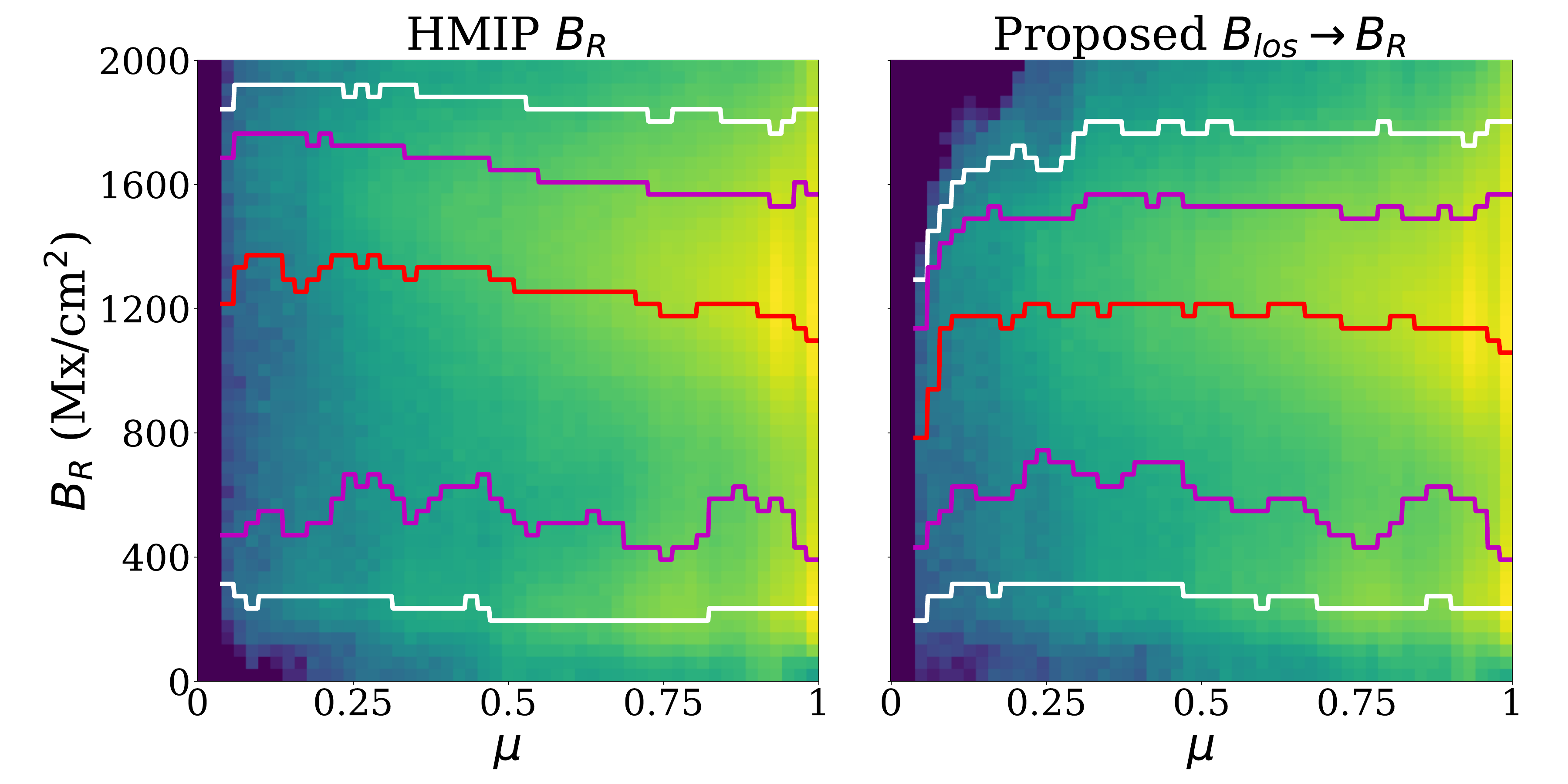} &
\includegraphics[width=0.318\linewidth]{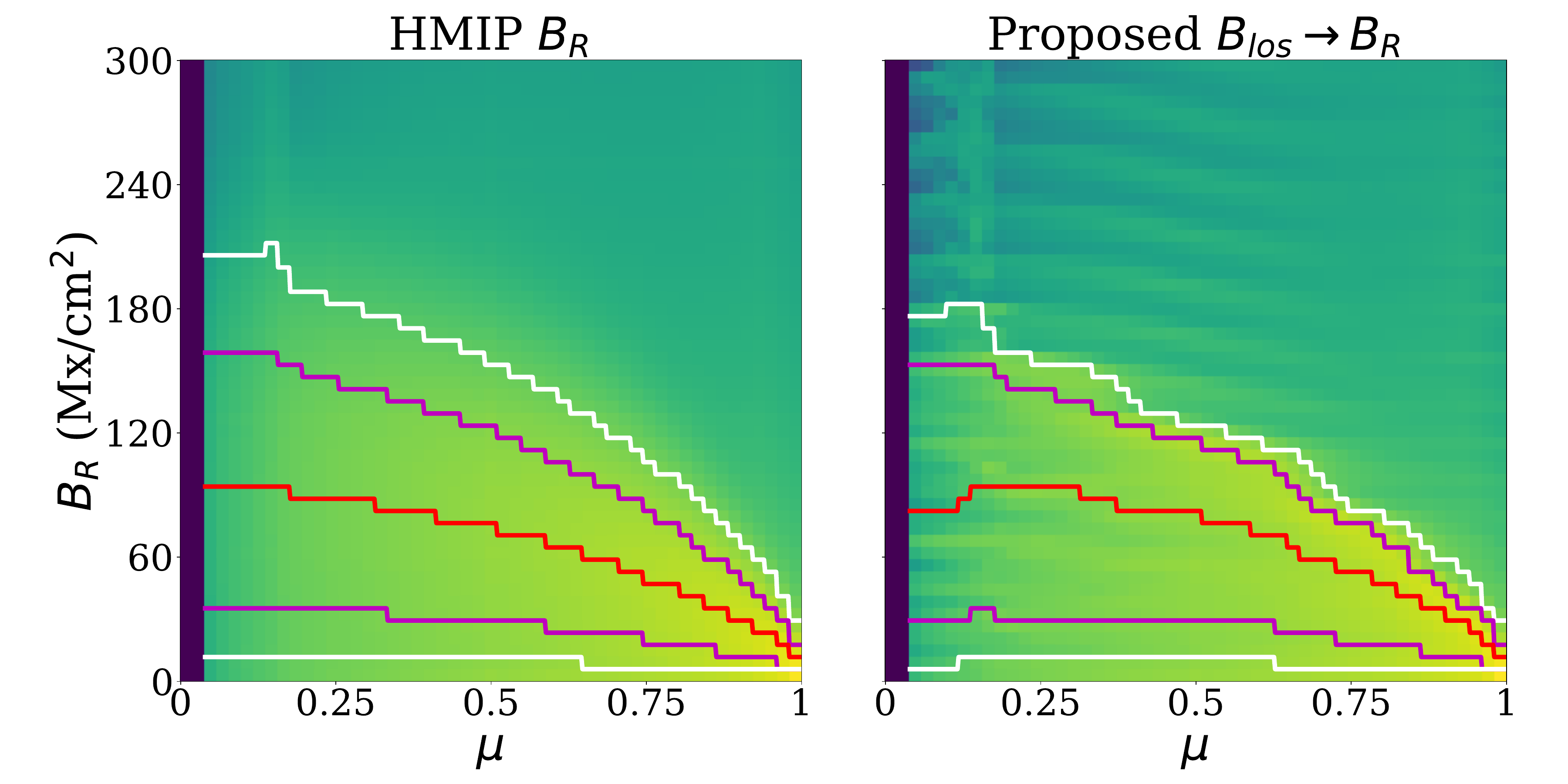} & 
\includegraphics[width=0.318\linewidth]{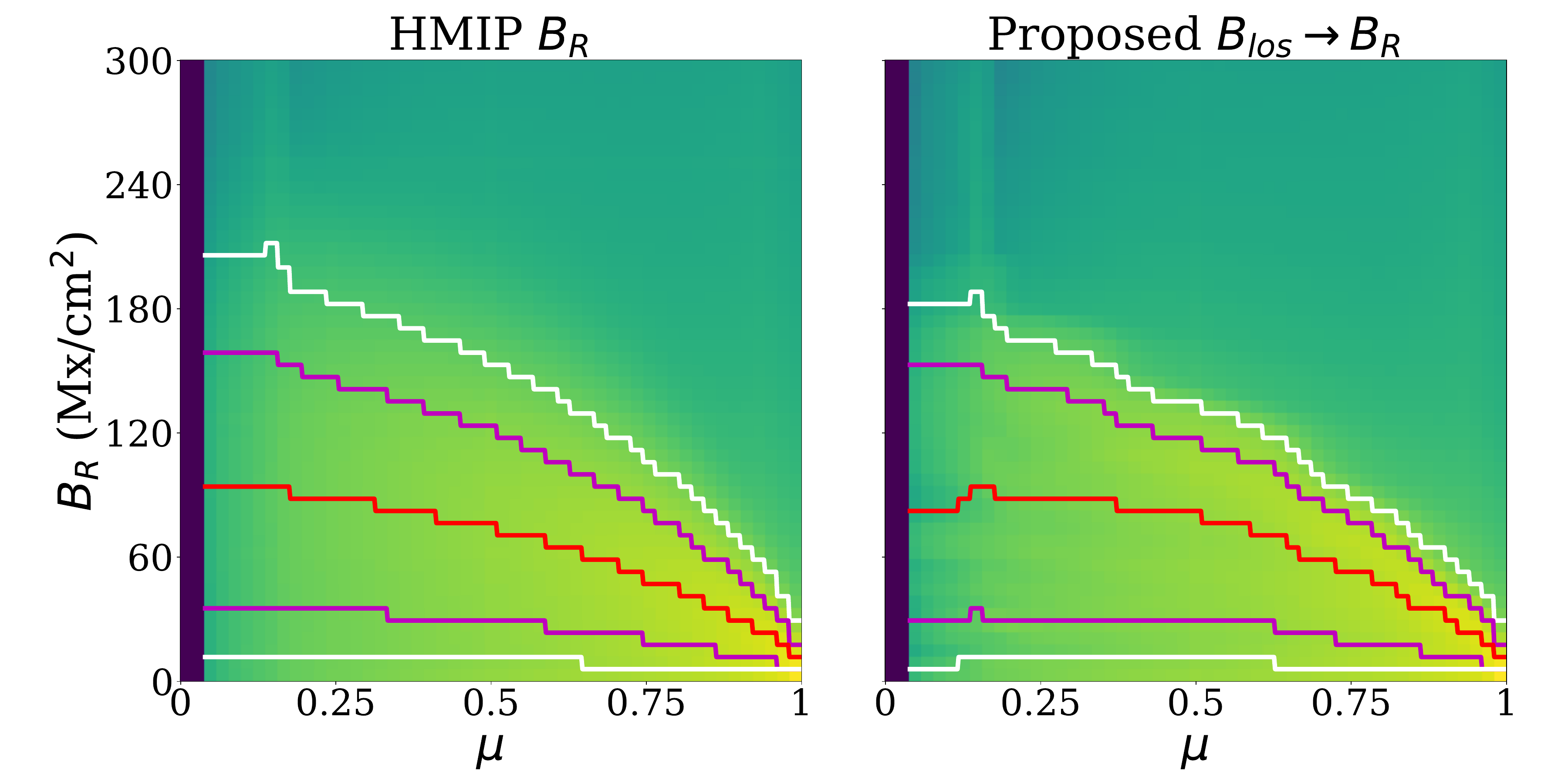} \\
(A) Active Regions Only & (B) All Pixels, Standard  & (C) All Pixels,  Randomized Norm \\
 &  Norm Estimate & Bin-Correction
\end{tabular}
\caption{{\bf The distribution of \br as a function of viewing angle $\mu$.} Like its target \hmip data, \method output (a) shows no (or at least, little) center-to-limb bias in active regions (b) shows a clear center-to-limb bias over all pixels. The CTL bias is driven by the large numbers of low-polarization signal pixels. We show two options: the standard method and the results obtained when the bin correction $\deltaB_n$ is randomized rather than predicted. Both closely track the \hmip results; randomization leads to a slight decrease in accuracy but mitigates banding artifacts in the output. Panels are identically colormapped so comparison within a panel is possible. 
}
\label{fig:app_muall}
\end{figure}

\section{Center-to-Limb Variations}
\label{sec:app_CTL}

The magnetic fields on the Sun should not vary with viewing angle from disk center toward the limb. \hmip data does show a viewing-angle dependence due to particulars of the inversion method \citep[e.g][]{leka2022identifying} although it is significantly less pronounced in active regions with good polarization signal. Ideally,\method should produce output with little-to-no viewing-angle bias even though the inputs do include a viewing angle dependence via both the \blos input and the use of $\mu$ itself. It is thus important to verify that \method does not strongly replicate viewing angle dependence in its outputs. 

We follow the definitions in~\cite{SuperSynthIA} and examine the distribution of \br as a function of $\mu$ for all data and then separately for active regions. Figure~\ref{fig:app_muall} shows that \method does not produce a viewing angle dependence in active regions, since the median $B_R$ as a function of viewing angle shows no substantial trends.
The slight dips near $\mu=0$ and $\mu=1$ in \method and the slight trend in \hmi are likely to be residual artifacts of pixel selection, since the definition of active regions is itself viewing-angle dependent. 

While there is no viewing-angle bias in active regions, \hmip data does have a viewing-angle bias when evaluated over {\it all} pixels. This bias happens because low polarization pixels have a predictable steady increase in flux density as a function of viewing angle, due to the additional noise in the component transverse to \blos. Unsurprisingly, since the relationship between viewing angle and flux density is predictable, \method reproduces it well in low polarization regions, duplicating \hmip's trends. We show both the standard norm prediction and the randomized correction described in Appendix~\ref{sec:app_dither}. The randomization helps prevents some banding artifacts. However, overall the same precautions used with \hmi (such as masking non-confident pixels) should be taken with \method's output, for downstream applications.

\begin{figure*}
\includegraphics[width=\linewidth]{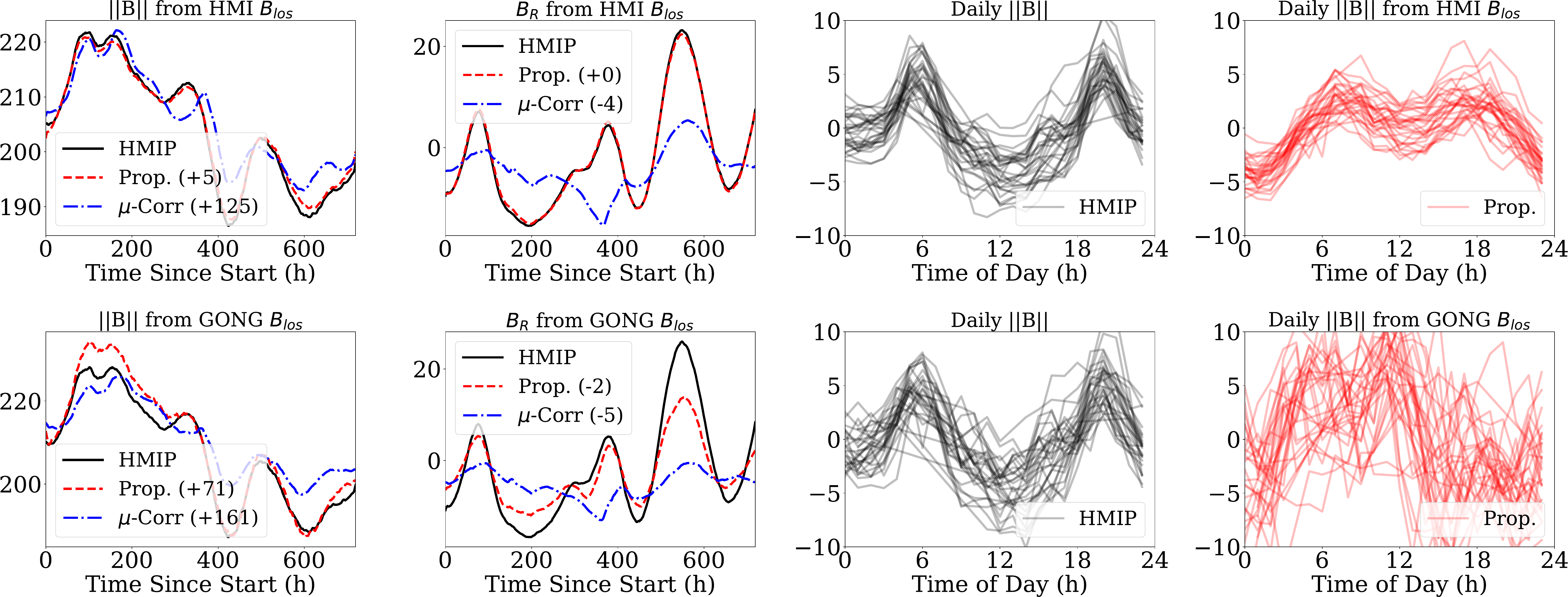}
\caption{{\bf Temporal trends, using \blos from either \hmi or \gong}. Following~\cite{higgins2021fast,Higgins2022}, we plot disk-averaged flux statistics since a start time of 2016/02/01, 00:36:00 TAI. We factor each time series into a long-term trends (shown in left two panels) and short-term trends to separately handle trends and the  known oscillations in \hmi's magnetic field data. For reference, we plot the \hmip result with a black line (or \hmip downsampled to \gong). The left two panels show plots of disk-averaged total flux $|B|$ and signed \br as a function of time. The plots co-align all methods to \hmip to facilitate comparison, reporting the additive correction in the legend. For both $|B|$ and \br, the proposed method follows \hmip far better and requires less correction. The right two panels show the residual high-frequency signal after detrending. \hmip's oscillation is clearly present. We report results only on confident confident pixels ({\tt conf\_disambig} $\ge 60$); results are similar using all pixels. }
\label{fig:app_time}
\end{figure*}

\begin{figure}
\centering
\includegraphics[width=\linewidth]{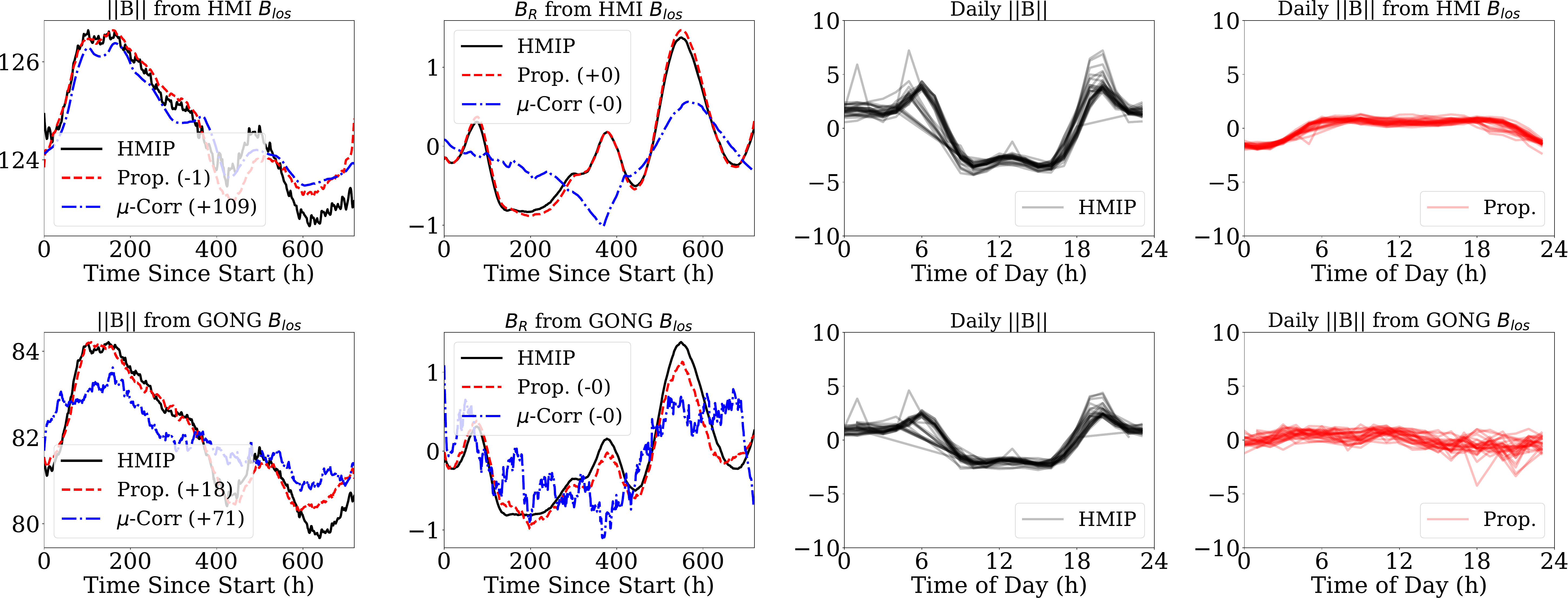}
\caption{{\bf Temporal trends over all pixels}. This figure follows Figure~\ref{fig:app_time} in terms of overall trends: the proposed method tracks \hmip well and far better than $\mu$-corrected \blos. The daily trends are slightly more concentrated, likely due to the far larger amount of data reducing variance. The difference in the \hmip is due to resampling the data to a smaller grid, which is done per-component, as opposed to by resampling the norm and angles. Accordingly, the average flux is reduced.}
\label{fig:app_timeall}
\end{figure}

\section{24-Hour and Month-Long Trends}
\label{sec:app_oscillations}

We next examine temporal trends of data produced by the method, looking at two factors of interest. The first factor is the long-term trends. \hmip is believed to capture long-term trends in magnetic flux density well and ideally the method's outputs should match \hmip. The second is short-term trends. \hmi's average magnetic flux density is known to oscillate with a 24-hour period plus harmonics~\citep{Hoeksema2014}. Ideally the proposed method should not make the results worse, but machine learning models have a tendency of exacerbating biases found in their training data~\citep{zhao2017men}.

We handle both by factoring a given time series of average flux densities into a long-term trend and short-term 24-hour trends. For each point in the time series, we fit a linear model mapping time to the time series value, using the adjacent 24-hour windows. We define the long-term trend as the linear model's value at the central point and the short-term trend as residual to the linear model. The original series is then the sum of the short and long term trends. We show both analysis with pixels with high polarization signal ({\tt conf\_disambig} $\ge 60$) in Figure~\ref{fig:app_time} as well as over all pixels (where the averages are driven by low-polarization pixels) in Figure~\ref{fig:app_timeall}. 

The proposed method's outputs match long-term trends for \hmip well using both \hmi and \gong \blos data. When using \hmi \blos data, the match is near-perfect up to a trivial few-Gauss additive calibration constant. For \gong data, there is strong alignment, albeit not as good as using \hmi \blos. The proposed method needs a larger additive recalibration and the peaks in the trend-lines are a little blunted. Nonetheless, in both cases, the proposed method shows much good agreement and much better compared to $\mu$-correcting \blos. These results are mirrored closely when using all pixels.

When we examine the short-term trends, we find that the proposed method has similar short-term oscillatory behavior. When using \blos data from \hmi, this is predictable: the 24-hour oscillation is believe to originate in calibration of Stokes profiles, and so it impacts the \blos data too. Thus with oscillations in both inputs and outputs the method is bound to replicate the oscillations. However, the method does not make it worse.

When using \blos data from \gong, the short-term trend pattern is different and the distinctive \hmi oscillation is gone. On the other hand, the short-term trend plot suggests that the $||\BB||$ recovered from \gong \blos data has a different bias: in the first half of the day, the signal is a few \gauss higher on average. This may be caused by the fact that \gong is a network of telescopes, and so different instruments are available at different times of the day. When the analysis is performed with all pixels, long-term trends match similarly, and oscillations are no worse, if not slightly better. Additionally, using a single magnetogram as input produces qualitatively similar magnetograms to using multiple magnetograms.

When using all pixels, the results are similar: the proposed method provides good estimates of $|\BB|$ and \br and also does much better than $\mu$-correcting \blos. There seems to be an improvement in the 24-hour oscillation; however, the more stringent test is whether there is an improvement in high-signal pixels. In either case, however, the lack of a {\it worse} oscillation in both cases shows the promise of the method.

\begin{deluxetable}{|cc|ccc|ccc|ccc|ccc|ccc|ccc|}[t]
\caption{{\bf Comparison of the impact of data scale and temporal signal.} Each row is a method parameterized by data scale from 32 scans to 4365 scans. We report results for \br, \bp, and \bt, reporting MAE (\gauss) and for all pixels, active regions (ARs), and limb active regions (Limb ARs), Plage, and Pixels over $250$G. Performance on all pixels are basically identical for all models. However, in active regions, there are substantial differences, especially on the limbs.  The results show that data scale helps, but that that looking at subpopulations of pixels, such as active regions, is important for evaluation.}
\label{tab:datascale}
\tablehead{
 & & \multicolumn{3}{c}{All} & \multicolumn{3}{c}{ARs} & \multicolumn{3}{c}{Limb ARs} & \multicolumn{3}{c}{Plage} & \multicolumn{3}{c|}{Over $250$G}  \\
 Data & Size & 
R & $\phi$ & $\theta$ & 
R & $\phi$ & $\theta$ & 
R & $\phi$ & $\theta$ & 
R & $\phi$ & $\theta$ & 
R & $\phi$ & $\theta$ 
}
\startdata
\hmi & 32 & 42 & 48 & 59 & 175 & 201 & 223 & 344 & 199 & 290 & 57 & 51 & 69 & 91 & 63 & 97\\
& 128  & 42 & 48 & 59 & 146 & 164 & 193 & 270 & 168 & 255 & 55 & 49 & 68 & 88 & 59 & 94\\
& 300 & 42 & 48 & 58 & 138 & 156 & 174 & 253 & 159 & 233 & 54 & 48 & 66 & 87 & 58 & 92\\
& 4365  & 42 & 48 & 58 & 123 & 133 & 152 & 229 & 128 & 193 & 55 & 48 & 65 & 88 & 56 & 90\\
\midrule \midrule
\gong & 300   & 43 & 45 & 55 & 331 & 253 & 264 & 451 & 309 & 347 & 157 & 68 & 74 & 119 & 70 & 94\\
& 4365  & 41 & 44 & 52 & 320 & 235 & 247 & 397 & 290 & 319 & 150 & 64 & 72 & 115 & 68 & 92\\
\enddata
\end{deluxetable}

\section{Data Scaling}
\label{sec:app_scale}

We start first by analyzing how changing the scale of data changes system performance, similar to understanding how studies in natural language processing~\citep{kaplan2020scaling}. This experiment is important since a current popular position has been that considerable progress in AI-based techniques has been driven {\it primarily} not by human ingenuity but instead by large amounts of data and compute -- an argument typically referred to as the Bitter Lesson, after the essay by~\cite{sutton2019bitter}.

We conduct experiments with identical models and vary the scale of data available. The smallest scale (XS -- 32 scans), is deliberately too small. The next (S -- 128 scans) is on par with the data scale used by~\cite{higgins2021fast}. This is then bumped up (M -- 300 scans), or about one scan a day for a year (sampled randomly from the training period). Finally, the largest scale (L -- 4365 scans) is approximately one scan a day for the {\it SDO} mission and corresponds to the main paper. We report results in Table~\ref{tab:datascale}, and reporting all four scales for \hmi and two of them for \gong.

\begin{figure*}[t]
    \centering
    \includegraphics[width=\linewidth]{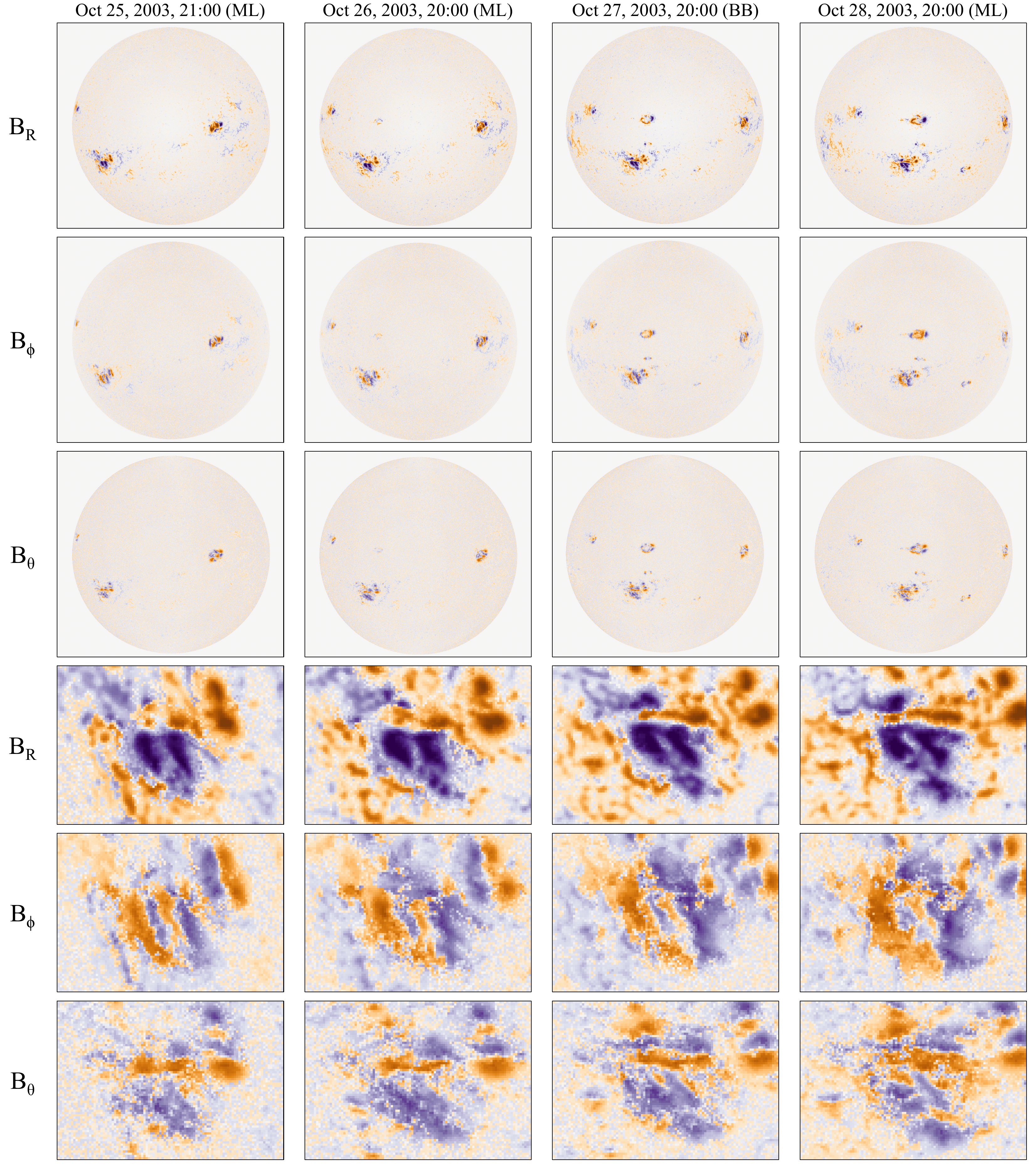}
    \caption{{\bf Results on \blos data predating   
    the Solar Dynamics Observatory from 2003 (best viewed in zoom on a screen in color)}. We show full disk images (top three rows) and close-ups of NOAA AR~10486 (bottom three rows), showing results from four dates, taken at Mauna Loa (ML) and Big Bear Observatory (BB). \method produces plausible vector fields. In the full disk image, one can see the fast emergence of NOAA AR~10488. Colormap: -2000 \includegraphics[width=20pt,height=6pt]{PuOrSqrt.png} 2000 \gauss.
    }
    \label{fig:gongStormQualitative}
\end{figure*}

We attempt to keep the number of training steps equivalent (the last epoch to not exceed 1.5M steps) and the recipes the same. With 300 scans, this took 312 epochs (as compared to 21 epochs). We took took the model with lowest validation loss during training. The learning rate started at $10^{-3}$ and was divided by 10 every 8 epochs for the large setting, and equivalently for the medium setting. In the small setting, we needed to train for reduce the learning rate earlier to train without instabilities. While the medium setting got fewer steps at the higher learning rate, the total number of steps was identical and the validation loss plateaued far before the learning rate drop. The small setting (S) was trained for 312 epochs; validation lass plateaued far before the conclusion of training. The extra-small (XS) case was trained for 1248 epochs, corresponding to ${\approx}638$K examples, but validation performance stopped improving between epoch 500 and 600 for both models.

Increasing data leads to a consistent, substantial improvement in active regions. The smallest data scale (XS) is too small, and performs clearly worse. If one starts at the next scale (S), performance improves considerably. Moving to the large data scale (L) leads to an improvement of $10-15\%$ in active regions for \hmi over S, with M in the middle of the two. \gong results show a similar trend -- there is a consistent gain from more data. 
These trends are in line with results in other fields, e.g., those of \cite{kaplan2020scaling}. More data consistently helps, and performance scales sublinearly with data: moving from XS to S increases the data by $4\times$ and reduces MAE in active regions by ${\approx}30$\gauss, and from S to L increases the data $34\times$ and reduces the data a similar amount. We note that at the L data size, it is likely that model capacity is a limiting factor.

On the other hand, if one looks at all pixels, increasing data has {\it absolutely} no effect. Simple thresholding helps a little, but not much. If one examines pixels over $250$ \gauss , the obviously too small (XS) dataset of 32 scans is slightly worse, and increasing the data scale by a factor of $34\times$ (S $\rightarrow$ L) has close to no effect -- at most 4 \gauss for \bp and \bt, but no improvement for \br. This surprising result underscores the critical importance of evaluation: if one looked at all pixels, as is common in many papers, one would conclude that data has no impact and might draw the incorrect conclusion that one gets as good results with 32 scans as one gets with $100\times$ the data. Additionally, the empirical results show that at least in this particular setting (estimating \brpt from \blos using HMIP data), estimating the quieter parts of the Sun from \blos data might only require 30 scans (for quiet regions) and 128 scans (for plage). 

\section{\method output variations with input variations}
\label{sec:app_additional}

Finally, we show additional samples of the Halloween Solar storms on four dates in Figure~\ref{fig:gongStormQualitative}.  There will necessarily be variations in \method output as the input changes between \gong sites, visible (subtley) in this figure but more obviously in the supplemental movie.  The fact that \method provide full-disk vector data is highlighted here (and in the movie), an aspect not well emphasized in the region-focused examples provided in the main paper.

\bibliography{ms}{}
\bibliographystyle{aasjournal}

\end{document}